\documentclass[pdftex,twocolumn,epjc3_preprint,runningheads]{svjour3}

\usepackage[T1]{fontenc}
\usepackage{lmodern}
\usepackage{calc}
\usepackage{graphicx}
\usepackage{booktabs}
\usepackage{textcomp}
\usepackage{xspace}
\usepackage{relsize}
\usepackage{amssymb}
\usepackage{amsmath}
\usepackage{listings}
\usepackage{microtype}
\usepackage{multirow}
\usepackage{tabularx}
\usepackage{array}
\usepackage{placeins}
\usepackage{cuted}
\usepackage{soul} 
\usepackage{fixltx2e}
\usepackage[numbers,sort&compress]{natbib}
\usepackage[labelfont=bf,font=small]{caption}
\usepackage[skip=-2pt]{subcaption}
\usepackage[colorlinks,citecolor=blue,urlcolor=blue,linkcolor=blue,breaklinks=breakall]{hyperref}
\usepackage{breakurl}
\usepackage[dvipsnames]{xcolor}
\usepackage[clockwise,figuresright]{rotating}
\usepackage{siunitx}
\usepackage{tikz}

\usepackage{etoolbox}
\AfterEndEnvironment{strip}{\leavevmode}

\allowdisplaybreaks

\newcolumntype{L}{>{\raggedright\let\newline\\\arraybackslash\hspace{0pt}}X}
\newcolumntype{R}{>{\raggedleft\let\newline\\\arraybackslash\hspace{0pt}}X}
\newcolumntype{C}{>{\centering\let\newline\\\arraybackslash\hspace{0pt}}X}

\newcommand{\imperial}{Department of Physics, Imperial College London, Blackett Laboratory, Prince Consort Road, London SW7 2AZ, UK}
\newcommand{\nordita}{NORDITA, Roslagstullsbacken 23, SE-10691 Stockholm, Sweden}
\newcommand{\oslo}{Department of Physics, University of Oslo, N-0316 Oslo, Norway}
\newcommand{\adelaide}{Department of Physics, University of Adelaide, Adelaide, SA 5005, Australia}
\newcommand{\glasgow}{SUPA, School of Physics and Astronomy, University of Glasgow, Glasgow, G12 8QQ, UK}
\newcommand{\monash}{School of Physics and Astronomy, Monash University, Melbourne, VIC 3800, Australia}
\newcommand{\coepp}{Australian Research Council Centre of Excellence for Particle Physics at the Tera-scale}
\newcommand{\okc}{Oskar Klein Centre for Cosmoparticle Physics, AlbaNova University Centre, SE-10691 Stockholm, Sweden}
\newcommand{\su}{Department of Physics, Stockholm University, SE-10691 Stockholm, Sweden}
\newcommand{\mcgill}{Department of Physics, McGill University, 3600 rue University, Montr\'eal, Qu\'ebec H3A 2T8, Canada}
\newcommand{\ucla}{Physics and Astronomy Department, University of California, Los Angeles, CA 90095, USA}
\newcommand{\annecy}{LAPTh, Universit\'e de Savoie, CNRS, 9 chemin de Bellevue B.P.110, F-74941 Annecy-le-Vieux, France}
\newcommand{\harvard}{Department of Physics, Harvard University, Cambridge, MA 02138, USA}
\newcommand{\grappa}{GRAPPA, Institute of Physics, University of Amsterdam, Science Park 904, 1098 XH Amsterdam, Netherlands}
\newcommand{\sydney}{Centre for Translational Data Science, Faculty of Engineering and Information Technologies, School of Physics, The University of Sydney, NSW 2006, Australia}

\newcommand{\desy}{DESY, Notkestra\ss e 85, D-22607 Hamburg, Germany}

\newcommand{\cernth}{Theoretical Physics Department, CERN, CH-1211 Geneva 23, Switzerland}
\newcommand{\lyon}{Univ Lyon, Univ Lyon 1, ENS de Lyon, CNRS, Centre de Recherche Astrophysique de Lyon UMR5574, F-69230 Saint-Genis-Laval, France}
\newcommand{\iuf}{Institut Universitaire de France, 103 boulevard Saint-Michel, 75005 Paris, France}
\newcommand{\zurich}{Physik-Institut, Universit\"at Z\"urich, Winterthurerstrasse 190, 8057 Z\"urich, Switzerland}
\newcommand{\krakow}{H.~Niewodnicza\'nski Institute of Nuclear Physics, Polish Academy of Sciences, 31-342  Krak\'ow, Poland}

\newcommand{\gambitacknosplus}{We warmly thank the Casa Matem\'aticas Oaxaca, affiliated with the Banff International Research Station, for hospitality whilst part of this work was completed, and the staff at Cyfronet, for their always helpful supercomputing support.  \GB has been supported by STFC (UK; ST/K00414X/1, ST/P000762/1), the Royal Society (UK; UF110191), Glasgow University (UK; Leadership Fellowship), the Research Council of Norway (FRIPRO 230546/F20), NOTUR (Norway; NN9284K), the Knut and Alice Wallenberg Foundation (Sweden; Wallenberg Academy Fellowship), the Swedish Research Council (621-2014-5772), the Australian Research Council (CE110001004, FT130100018, FT140100244, FT160100274), The University of Sydney (Australia; IRCA-G162448), PLGrid Infrastructure (Poland), Polish National Science Center (Sonata UMO-2015/17/D/ST2/03532), the Swiss National Science Foundation (PP00P2-144674), European Commission Horizon 2020 (Marie Sk\l{}odowska-Curie actions H2020-MSCA-RISE-2015-691164, European Research Council Starting Grant ERC-2014-STG-638528), the ERA-CAN+ Twinning Program (EU \& Canada), the Netherlands Organisation for Scientific Research (NWO-Vidi 016.149.331), the National Science Foundation (USA; DGE-1339067), the FRQNT (Qu\'ebec) and NSERC/The Canadian Tri-Agencies Research Councils (BPDF-424460-2012).}

\makeatletter

\newcommand{\preprintnumber}[1]{\gdef\@preprintnumber{\begin{flushright}{#1}\end{flushright}}}

\g@addto@macro\bfseries{\boldmath}
\makeatother

\newcommand{\subparagraph}{} 
\usepackage{titlesec}
\titleformat*{\paragraph}{\bfseries}

\journalname{Eur. Phys. J. C}
\smartqed
\let\underscore\_
\renewcommand{\_}{\discretionary{\underscore}{}{\underscore}}

\makeatletter
\let\orgdescriptionlabel\descriptionlabel
\renewcommand*{\descriptionlabel}[1]{%
  \let\orglabel\label
  \let\label\@gobble
  \phantomsection
  \protected@edef\@currentlabel{#1}%
  \let\label\orglabel
  \orgdescriptionlabel{#1}%
}
\makeatother

\newcommand\postnewlinemarker{\hbox{\ensuremath{\hookrightarrow}}}
\newcommand\cpp[1]{{\lstinline!#1!}}  
\newcommand\cpppragma[1]{{\CPPcommentstyle#1}}
\newcommand\yaml[1]{{\lstset{style=yaml}\lstinline!#1!\lstset{style=cpp}}}
\newcommand\yamlvalue[1]{{\YAMLvaluestyle\ttfamily#1}}
\newcommand\term[1]{{\lstset{style=terminal}\lstinline!#1!\lstset{style=cpp}}}
\newcommand\fortran[1]{{\lstset{style=fortran}\lstinline!#1!\lstset{style=cpp}}}
\newcommand\py[1]{{\lstset{style=python}\lstinline!#1!\lstset{style=cpp}}}
\newcommand\customtilde{{\raisebox{0.2ex}{\scalebox{0.6}{\boldmath$\sim$}}}}
\newcommand\mathematica[1]{{\lstset{style=Mathematica}\lstinline!#1!\lstset{style=cpp}}}

\lstnewenvironment{lstlistingyaml}{\lstset{style=yaml}}{\lstset{style=cpp}}
\lstnewenvironment{lstlistingterm}{\lstset{style=terminal}}{\lstset{style=cpp}}
\lstnewenvironment{lstlistingfortran}{\lstset{style=fortran}}{\lstset{style=cpp}}
\lstnewenvironment{lstcpp}{\lstset{style=cpp}}{\lstset{style=cpp}}
\lstnewenvironment{lstcppalt}{\lstset{style=cppalt}}{\lstset{style=cpp}}
\lstnewenvironment{lstcppnum}{\lstset{style=cppnum}}{\lstset{style=cpp}}
\lstnewenvironment{lstyaml}{\lstset{style=yaml}}{\lstset{style=cpp}}
\lstnewenvironment{lstterm}{\lstset{style=terminal}}{\lstset{style=cpp}}
\lstnewenvironment{lsttermalt}{\lstset{style=terminalalt}}{\lstset{style=cpp}}
\lstnewenvironment{lsttext}{\lstset{style=text}}{\lstset{style=cpp}}
\lstnewenvironment{lstfortran}{\lstset{style=fortran}}{\lstset{style=cpp}}
\lstnewenvironment{lstpy}{\lstset{style=python}}{\lstset{style=cpp}}
\lstnewenvironment{lstmathematica}{\lstset{style=mathematica}}{\lstset{style=cpp}}

\newcommand{\tmpname}{}
\newcommand{\tmplistingname}{}
\makeatletter
\newif\ifATOlabelname
\lst@Key{labelname}{Listing}{\def\ATOlabelname{#1}\global\ATOlabelnametrue}
\makeatother
\lstnewenvironment{lstcpplabel}[1][]{
  \lstset{style=cpp,#1} 
  \ifATOlabelname
    \renewcommand{\tmpname}{\lstlistingname}
    \renewcommand{\tmplistingname}{\lstlistlistingname}
    \renewcommand{\lstlistingname}{\ATOlabelname}
    \renewcommand{\lstlistlistingname}{List of \lstlistingname s}
  \fi
}{
  \renewcommand{\lstlistingname}{\tmpname}
  \renewcommand{\lstlistlistingname}{\tmplistingname}
  \lstset{style=cpp}
}
\definecolor{solarized@base03}{HTML}{002B36}
\definecolor{solarized@base02}{HTML}{073642}
\definecolor{solarized@base01}{HTML}{586e75}
\definecolor{solarized@base00}{HTML}{657b83}
\definecolor{solarized@base0}{HTML}{839496}
\definecolor{solarized@base1}{HTML}{93a1a1}
\definecolor{solarized@base2}{HTML}{EEE8D5}
\definecolor{solarized@base3}{HTML}{FDF6E3}
\definecolor{solarized@yellow}{HTML}{B58900}
\definecolor{solarized@orange}{HTML}{CB4B16}
\definecolor{solarized@red}{HTML}{DC322F}
\definecolor{solarized@magenta}{HTML}{D33682}
\definecolor{solarized@violet}{HTML}{6C71C4}
\definecolor{solarized@blue}{HTML}{268BD2}
\definecolor{solarized@cyan}{HTML}{2AA198}
\definecolor{solarized@green}{HTML}{859900}
\definecolor{darkred}{HTML}{550003}
\definecolor{darkgreen}{HTML}{00AA00}

\newcommand\YAMLstringstyle{\footnotesize\color{solarized@green}\mdseries}
\newcommand\YAMLkeystyle{\footnotesize\color{solarized@blue}\ttfamily}
\newcommand\YAMLvaluestyle{\footnotesize\color{blue}\mdseries}
\newcommand\ProcessThreeDashes{\llap{\color{cyan}\mdseries-{-}-}}

\newcommand\CPPidentifierstyle{\color{solarized@blue}\footnotesize\ttfamily}
\newcommand\CPPcommentstyle{\color{solarized@violet}\footnotesize\ttfamily}
\newcommand\CPPdirectivestyle{\color{solarized@magenta}\footnotesize\ttfamily}
\newcommand\termplainstyle{\footnotesize\ttfamily}

\newcommand\processLongMacroDelimiter
{%
\CPPdirectivestyle%
\#define%
}

\lstdefinestyle{cpp}
{
  language=C++,
  basicstyle=\footnotesize\ttfamily,
  basewidth={0.53em,0.44em}, 
  numbers=none,
  tabsize=2,
  breaklines=true,
  escapeinside={@}{@},
  showstringspaces=false,
  numberstyle=\tiny\color{solarized@base01},
  keywordstyle=\color{solarized@orange},
  stringstyle=\color{solarized@red}\ttfamily,
  identifierstyle=\color{solarized@blue},
  commentstyle=\CPPcommentstyle,
  directivestyle=\CPPdirectivestyle,
  emphstyle=\color{solarized@green},
  frame=single,
  rulecolor=\color{solarized@base2},
  rulesepcolor=\color{solarized@base2},
  literate={~} {\customtilde}1,
  moredelim=*[directive]\ \ \#,
  moredelim=*[directive]\ \ \ \ \#
}

\lstdefinestyle{cppalt}
{
  language=C++,
  basicstyle=\footnotesize\ttfamily,
  basewidth={0.53em,0.44em}, 
  numbers=none,
  tabsize=2,
  breaklines=true,
  escapeinside={*@}{@*},
  showstringspaces=false,
  numberstyle=\tiny\color{solarized@base01},
  keywordstyle=\color{solarized@orange},
  stringstyle=\color{solarized@red}\ttfamily,
  identifierstyle=\color{solarized@blue},
  commentstyle=\CPPcommentstyle,
  directivestyle=\CPPdirectivestyle,
  emphstyle=\color{solarized@green},
  frame=single,
  rulecolor=\color{solarized@base2},
  rulesepcolor=\color{solarized@base2},
  literate={~}{\customtilde}1,
  moredelim=**[is][\processLongMacroDelimiter]{BeginLongMacro}{EndLongMacro} 
}

\lstdefinestyle{cppnum}
{
  language=C++,
  basicstyle=\footnotesize\ttfamily,
  basewidth={0.53em,0.44em}, 
  numbers=none,
  tabsize=2,
  breaklines=true,
  escapeinside={@}{@},
  numberstyle=\tiny\color{solarized@base01},
  showstringspaces=false,
  numberstyle=\tiny\color{solarized@base01},
  keywordstyle=\color{solarized@orange},
  stringstyle=\color{solarized@red}\ttfamily,
  identifierstyle=\color{solarized@blue},
  commentstyle=\CPPcommentstyle,
  directivestyle=\CPPdirectivestyle,
  emphstyle=\color{solarized@green},
  frame=single,
  rulecolor=\color{solarized@base2},
  rulesepcolor=\color{solarized@base2},
  literate={~} {\customtilde}1,
  moredelim=*[directive]\ \ \#,
  moredelim=*[directive]\ \ \ \ \#
}

\lstdefinestyle{python}
{
  language=Python,
  basicstyle=\footnotesize\ttfamily,
  basewidth={0.53em,0.44em},
  numbers=none,
  tabsize=2,
  breaklines=true,
  escapeinside={@}{@},
  showstringspaces=false,
  numberstyle=\tiny\color{solarized@base01},
  keywordstyle=\color{blue},
  stringstyle=\color{orange}\ttfamily,
  identifierstyle=\color{darkred},
  commentstyle=\color{purple},
  emphstyle=\color{green},
  frame=single,
  rulecolor=\color{solarized@base2},
  rulesepcolor=\color{solarized@base2},
  literate = {~}{\customtilde}1
             {\ as\ }{{\color{blue}\ as\ \color{black}}}3
}

\lstdefinestyle{fortran}
{
  language=Fortran,
  basicstyle=\footnotesize\ttfamily,
  basewidth={0.53em,0.44em},
  numbers=none,
  tabsize=2,
  breaklines=true,
  escapeinside={@}{@},
  showstringspaces=false,
  numberstyle=\tiny\color{solarized@base01},
  keywordstyle=\color{blue},
  stringstyle=\color{orange}\ttfamily,
  identifierstyle=\color{Periwinkle},
  commentstyle=\color{purple},
  emphstyle=\color{green},
  morekeywords={and, or, true, false},
  frame=single,
  rulecolor=\color{solarized@base2},
  rulesepcolor=\color{solarized@base2},
  literate={~}{\customtilde}1
}

\lstdefinestyle{terminal}
{
  language=bash,
  basicstyle=\termplainstyle,
  numbers=none,
  tabsize=2,
  breaklines=true,
  escapeinside={@}{@},
  frame=single,
  showstringspaces=false,
  numberstyle=\tiny\color{solarized@base01},
  keywordstyle=\color{solarized@orange},
  stringstyle=\color{solarized@red}\ttfamily,
  identifierstyle=\color{black},
  commentstyle=\color{solarized@violet},
  emphstyle=\color{solarized@green},
  frame=single,
  rulecolor=\color{solarized@base2},
  rulesepcolor=\color{solarized@base2},
  morekeywords={gambit, cmake, make, mkdir},
  deletekeywords={test},
  literate = {\ gambit}{{\ }{\color{black}}gambit}7
             {/gambit}{{/}{\color{black}}gambit}6
             {gambit/}{{\color{black}}gambit{/}}6
             {/include}{{/}{\color{black}}include}8
             {cmake/}{{\color{black}}cmake/}6
             {.cmake}{{.}{\color{black}}cmake}6
             {~}{\customtilde}1
}

\lstdefinestyle{terminalalt}
{
  language=bash,
  basicstyle=\footnotesize\ttfamily,
  numbers=none,
  tabsize=2,
  breaklines=true,
  escapeinside={*@}{@*},
  frame=single,
  showstringspaces=false,
  numberstyle=\tiny\color{solarized@base01},
  keywordstyle=\color{solarized@orange},
  stringstyle=\color{solarized@red}\ttfamily,
  identifierstyle=\color{black},
  commentstyle=\color{solarized@violet},
  emphstyle=\color{solarized@green},
  frame=single,
  rulecolor=\color{solarized@base2},
  rulesepcolor=\color{solarized@base2},
  morekeywords={gambit, cmake, make, mkdir},
  deletekeywords={test},
  literate = {\ gambit}{{\ }{\color{black}}gambit}7
             {/gambit}{{/}{\color{black}}gambit}6
             {gambit/}{{\color{black}}gambit{/}}6
             {/include}{{/}{\color{black}}include}8
             {cmake/}{{\color{black}}cmake/}6
             {.cmake}{{.}{\color{black}}cmake}6
             {~}{\customtilde}1
}

\lstdefinestyle{text}
{
  language={},
  basicstyle=\footnotesize\ttfamily,
  identifierstyle=\color{black},
  numbers=none,
  tabsize=2,
  breaklines=true,
  escapeinside={*@}{@*},
  showstringspaces=false,
  frame=single,
  rulecolor=\color{solarized@base2},
  rulesepcolor=\color{solarized@base2},
  literate={~}{\customtilde}1
}

\lstdefinestyle{yaml}
{
  language=bash,
  escapeinside={@}{@},
  keywords={true,false,null},
  otherkeywords={},
  keywordstyle=\color{solarized@base0}\bfseries,
  basicstyle=\footnotesize\color{black}\ttfamily,
  identifierstyle=\YAMLkeystyle,
  sensitive=false,
  commentstyle=\color{solarized@orange}\ttfamily,
  morecomment=[l]{\#},
  morecomment=[s]{/*}{*/},
  stringstyle=\YAMLstringstyle\ttfamily,
  moredelim=**[s][\YAMLkeystyle]{,}{:},   
  moredelim=**[l][\YAMLvaluestyle]{:},    
  morestring=[b]',
  morestring=[b]",
  literate =    {---}{{\ProcessThreeDashes}}3
                {>}{{\textcolor{solarized@red}\textgreater}}1
                {|}{{\textcolor{solarized@red}\textbar}}1
                {\ -\ }{{\mdseries\color{black}\ -\ \negmedspace}}3
                {\}}{{{\color{black} \}}}}1
                {\{}{{{\color{black} \{}}}1
                {[}{{{\color{black} [}}}1
                {]}{{{\color{black} ]}}}1
                {~}{\customtilde}1,
  breakindent=0pt,
  breakatwhitespace,
  columns=fullflexible
}

\lstdefinestyle{mathematica}
{
  language={Mathematica},
  basicstyle=\footnotesize\ttfamily,
  basewidth={0.53em,0.44em},
  numbers=none,
  tabsize=2,
  breaklines=true,
  escapeinside={@}{@},
  numberstyle=\tiny\color{black},
  showstringspaces=false,
  numberstyle=\tiny\color{solarized@base01},
  keywordstyle=\color{solarized@orange},
  stringstyle=\color{solarized@red}\ttfamily,
  identifierstyle=\color{solarized@orange}\ttfamily,
  commentstyle=\color{solarized@gray}\ttfamily,
  directivestyle=\color{solarized@orange}\ttfamily,
  emphstyle=\color{solarized@green},
  frame=single,
  rulecolor=\color{solarized@base2},
  rulesepcolor=\color{solarized@base2},
  literate={~} {\customtilde}1,
  moredelim=*[directive]\ \ \#,
  moredelim=*[directive]\ \ \ \ \#,
  mathescape=true
}

\newcommand{\cross}[1]{\ref{#1}}
\newcommand{\doublecross}[2]{\hyperref[#2]{\textbf{#1}}}
\newcommand{\doublecrosssf}[2]{\hyperref[#2]{\textbf{\textsf{#1}}}}
\newcommand{\gitem}[1]{\item[\textbf{#1}\label{#1}]}
\newcommand{\gsfitem}[1]{\item[\textbf{\textsf{#1}}\label{#1}]}
\newcommand{\gsfitemc}[1]{\item[\textbf{\textsf{#1}}\label{#1}:]}

\newcommand{\startglossary}{\section{Glossary}\label{glossary}Here we explain some terms that have specific technical definitions in \GB.\begin{description}}
\newcommand{\finishglossary}{\end{description}}

\newcommand{\bcode}{\begin{lstlisting}}
\newcommand{\ecode}{\end{lstlisting}}
\newcommand{\metavarf}[1]{\textit{\color{darkgreen}\footnotesize\textrm{#1}}}

\newcommand{\metavar}{\metavarf}

\DeclareMathOperator\erf{erf}

\newcommand{\sss}{\scriptscriptstyle}

\newcommand{\lhs}{\lambda_{h\sss S}}
\newcommand{\ls}{\lambda_{\sss S}}

\newcommand{\DR}{$\overline{DR}$\xspace}
\newcommand{\DRbar}{\DR}
\newcommand{\MSbar}{$\MSBar$\xspace}
\newcommand{\MSBar}{\overline{MS}}

\newcommand{\ie}{i.e.\ }

\newcommand{\gambit}{\textsf{GAMBIT}\xspace}

\newcommand{\darkbit}{\textsf{DarkBit}\xspace}
\newcommand{\colliderbit}{\textsf{ColliderBit}\xspace}
\newcommand{\flavbit}{\textsf{FlavBit}\xspace}
\newcommand{\specbit}{\textsf{SpecBit}\xspace}
\newcommand{\decaybit}{\textsf{DecayBit}\xspace}
\newcommand{\precisionbit}{\textsf{PrecisionBit}\xspace}
\newcommand{\scannerbit}{\textsf{ScannerBit}\xspace}
\newcommand{\examplebita}{\textsf{ExampleBit\_A}\xspace}

\newcommand{\BOSS}{\textsf{BOSS}\xspace}
\newcommand{\GB}{\gambit}

\newcommand{\omp}{\textsf{OpenMP}\xspace}
\newcommand{\mpi}{\textsf{MPI}\xspace}
\newcommand{\posix}{\textsf{POSIX}\xspace}
\newcommand{\buckfast}{\textsf{BuckFast}\xspace}

\newcommand{\pythiaeight}{\textsf{Pythia\,8}\xspace}

\newcommand{\madgraph}{\textsf{MadGraph}\xspace}

\newcommand{\higgsbounds}{\textsf{HiggsBounds}\xspace}

\newcommand{\higgssignals}{\textsf{HiggsSignals}\xspace}
\newcommand{\ds}{\textsf{DarkSUSY}\xspace}
\newcommand{\darksusy}{\ds}

\newcommand{\micromegas}{\textsf{micrOMEGAs}\xspace}

\newcommand{\feynhiggs}{\textsf{FeynHiggs}\xspace}
\newcommand{\FH}{\feynhiggs}

\newcommand\flexiblesusy{\FlexibleSUSY}
\newcommand\FlexibleSUSY{\textsf{FlexibleSUSY}\xspace}

\newcommand\SOFTSUSY{\textsf{SOFTSUSY}\xspace}

\newcommand\SUSYHIT{\textsf{SUSY-HIT}\xspace}

\newcommand\susyhit{\SUSYHIT}
\newcommand\gmtwocalc{\textsf{GM2Calc}\xspace}

\newcommand\SPheno{\textsf{SPheno}\xspace}
\newcommand\superiso{\textsf{SuperIso}\xspace}

\newcommand\Mathematica{\textsf{Mathematica}\xspace}

\newcommand\nulike{\textsf{nulike}\xspace}
\newcommand\gamLike{\textsf{gamLike}\xspace}
\newcommand\gamlike{\gamLike}

\newcommand\pippi{\textsf{pippi}\xspace}
\newcommand\MultiNest{\textsf{MultiNest}\xspace}
\newcommand\multinest{\MultiNest}
\newcommand\great{\textsf{GreAT}\xspace}
\newcommand\twalk{\textsf{T-Walk}\xspace}
\newcommand\diver{\textsf{Diver}\xspace}
\newcommand\ddcalc{\textsf{DDCalc}\xspace}

\newcommand\xx{\raisebox{0.2ex}{\smaller ++}\xspace}
\newcommand\Cpp{\textsf{C\xx}\xspace}
\newcommand\Cppeleven{\textsf{C\raisebox{0.2ex}{\smaller ++}11}\xspace}
\newcommand\plainC{\textsf{C}\xspace}
\newcommand\Python{\textsf{Python}\xspace}
\newcommand\python{\Python}
\newcommand\Fortran{\textsf{Fortran}\xspace}
\newcommand\YAML{\textsf{YAML}\xspace}

\newcommand\beq{\begin{equation}}
\newcommand\eeq{\end{equation}}

\renewcommand{\url}[1]{\href{#1}{#1}}

\newcommand{\TODO}[1]{\textbf{\textcolor{red}{#1}}}

\usepackage{pdfpages}

\begin{document}

\preprintnumber{CERN-TH-2017-166, CoEPP-MN-17-6, NORDITA 2017-074}

\title{GAMBIT: The Global and Modular Beyond-the-Standard-Model Inference Tool}

\author
{
The GAMBIT Collaboration:
Peter Athron\thanksref{inst:a,inst:b} \and
Csaba Balazs\thanksref{inst:a,inst:b} \and
Torsten Bringmann\thanksref{inst:c} \and
Andy Buckley\thanksref{inst:d} \and
Marcin Chrz\k{a}szcz\thanksref{inst:e,inst:f} \and
Jan Conrad\thanksref{inst:g,inst:h} \and
Jonathan M.~Cornell\thanksref{inst:i} \and
Lars A.~Dal\thanksref{inst:c} \and
Hugh Dickinson\thanksref{inst:w} \and
Joakim Edsj\"o\thanksref{inst:g,inst:h} \and
Ben Farmer\thanksref{inst:g,inst:h,e1} \and
Tom\'as E.\ Gonzalo\thanksref{inst:c} \and
Paul Jackson\thanksref{inst:k,inst:b} \and
Abram Krislock\thanksref{inst:c} \and
Anders Kvellestad\thanksref{inst:m,e2} \and
Johan Lundberg\thanksref{inst:g,inst:h} \and
James McKay\thanksref{inst:q} \and
Farvah Mahmoudi\thanksref{inst:n,inst:o,e5} \and
Gregory D.\ Martinez\thanksref{inst:p} \and
Antje Putze\thanksref{inst:r} \and
Are Raklev\thanksref{inst:c} \and
Joachim Ripken\thanksref{inst:x} \and
Christopher~Rogan\thanksref{inst:s} \and
Aldo~Saavedra\thanksref{inst:t,inst:b} \and
Christopher Savage\thanksref{inst:m} \and
Pat Scott\thanksref{inst:q,e3} \and
Seon-Hee Seo\thanksref{inst:y} \and
Nicola Serra\thanksref{inst:e} \and
Christoph Weniger\thanksref{inst:u,e4} \and
Martin White\thanksref{inst:k,inst:b} \and
Sebastian Wild\thanksref{inst:v}
}

\institute{%
  \monash\label{inst:a} \and
  \coepp\label{inst:b} \and
  \oslo\label{inst:c} \and
  \glasgow\label{inst:d} \and
  \zurich\label{inst:e} \and
  \krakow\label{inst:f} \and
  \okc\label{inst:g} \and
  \su\label{inst:h} \and
  \mcgill\label{inst:i} \and
  Minnesota Institute for Astrophysics, University of Minnesota, Minneapolis, MN 55455, USA \label{inst:w} \and
  \adelaide\label{inst:k} \and
  \nordita\label{inst:m} \and
  \imperial\label{inst:q} \and
  \lyon\label{inst:n} \and
  \cernth\label{inst:o} \and
  \ucla\label{inst:p} \and
  \annecy\label{inst:r} \and
  Max Planck Institute for Solar System Research, Justus-von-Liebig-Weg 3, D-37077 G\"ottingen, Germany \label{inst:x} \and
  \harvard\label{inst:s} \and
  \sydney\label{inst:t} \and
  Department of Physics and Astronomy, Seoul National University, 1 Gwanak-ro, Gwanak-gu, Seoul 08826, Korea \label{inst:y} \and
  \grappa\label{inst:u} \and
  \desy\label{inst:v}
}

\thankstext{e1}{benjamin.farmer@fysik.su.se}
\thankstext{e2}{anders.kvellestad@nordita.org}
\thankstext{e3}{p.scott@imperial.ac.uk}
\thankstext{e4}{c.weniger@uva.nl}
\thankstext[*]{e5}{Also \iuf.}

\titlerunning{GAMBIT: The Global and Modular BSM Inference Tool}
\authorrunning{The GAMBIT Collaboration}

\date{Received: date / Accepted: date}

\maketitle

\begin{abstract}
We describe the open-source global fitting package \GB: the Global And Modular Beyond-the-Standard-Model Inference Tool.  \GB combines extensive calculations of observables and likelihoods in particle and astroparticle physics with a hierarchical model database, advanced tools for automatically building analyses of essentially any model, a flexible and powerful system for interfacing to external codes, a suite of different statistical methods and parameter scanning algorithms, and a host of other utilities designed to make scans faster, safer and more easily-extendible than in the past.  Here we give a detailed description of the framework, its design and motivation, and the current models and other specific components presently implemented in \GB.  Accompanying papers deal with individual modules and present first \GB results. \GB can be downloaded from \href{http://gambit.hepforge.org}{gambit.hepforge.org}.
\end{abstract}

\tableofcontents

\section{Introduction}

The search for physics Beyond the Standard Model (BSM) is a necessarily multidisciplinary effort, as evidence for new physics could appear in any observable in particle, astroparticle or nuclear physics.  Strategies include producing new particles at high-energy colliders \cite{ATLAS_diphoton,ATLAS15,CMS_SMS}, hunting for their influences on rare processes and precision measurements \cite{gm2exp,BelleII,CMSLHCb_Bs0mumu}, directly detecting dark matter (DM) in the laboratory \cite{XENON2013,PICO60,LUX2016}, carefully examining cosmological observations for the influence of new physics \cite{Planck15cosmo,Slatyer15a,keVsterile_whitepaper}, and detecting high-energy particles from DM annihilation or decay \cite{BringmannWeniger,LATdwarfP8,IC79_SUSY}.  In this context, exclusions have so far been just as valuable as apparent detections; many purported signals of new physics have appeared \cite{Bernabei08, Goodenough09, Pamelapositron, CoGeNTAnnMod11, Bringmann12, Bulbul14, Boyarsky14, Vincent14, ATLAS_diboson}, often only to be disproven by a lack of correlated signals in other experiments \cite{CDMSLowE12, Bartels15, Lee:2015fea, Jeltema15, CRESST_momdep,LATdwarfP8}.

Properly and completely weighing the sum of data relevant to a theory, from many disparate experimental sources, and making rigorous statistical statements about which models are allowed and which are not, has become a challenging task for both theory and experiment.  This is the problem addressed by global fits: simultaneous predictions of a raft of different observables from theory, coupled with a detailed combined statistical analysis of the various experimental searches sensitive to them.  Several attempts to address this problem have been made in particle physics, from the characterisation of electroweak physics with \textsf{ZFitter} \cite{ZFitter} and later \textsf{GFitter} \cite{GFitter11} to CKM fits \cite{CKMFitter}, neutrino global fits \cite{Bari13,Tortola14,NuFit15} and global analyses of supersymmetry (SUSY) \cite{Baltz04,Allanach06,SFitter}, in particular with packages like \textsf{SuperBayeS} \cite{Ruiz06,2007NewAR..51..316T,2007JHEP...07..075R,Roszkowski09a,Trotta08,Martinez09,Roszkowski09b,Roszkowski10,Scott09c,BertoneLHCDD,SBCoverage,SuperbayesHiggs,Nightmare,BertoneLHCID,IC22Methods,Strege13,SuperbayesXENON100,SuperBayesGC}, \textsf{Fittino} \cite{Fittino06,Fittino,Fittino12}, \textsf{MasterCode} \cite{Buchmueller08,Buchmueller09,MasterCodemSUGRA,MasterCode11,MastercodeXENON100,Mastercode12,MastercodeHiggs,Mastercode12b,MastercodeCMSSM,Buchmueller:2014yva,Bagnaschi:2016afc,Bagnaschi:2016xfg} and others \cite{Allanach:2007qk,Abdussalam09a,Abdussalam09b,Allanach11b,Allanach11a,arXiv:1111.6098,Roszkowski12,Farmer13,arXiv:1212.4821,Fowlie13,Henrot14,Kim:2013uxa,arXiv:1402.5419,arXiv:1405.4289,arXiv:1503.08219,arXiv:1604.02102,Han:2016gvr}.

BSM global fits today remain overwhelmingly focused on SUSY, specifically lower-dimensional subspaces of the minimal supersymmetric standard model (MSSM) \cite{Strege15,Fittinocoverage,Catalan:2015cna,MasterCodeMSSM10,Mastercode15}, or, more rarely, the next-to-minimal variant (NMSSM) \cite{Balazs08,LopezFogliani:2009np,Kowalska:2012gs,Fowlie14}.  There are only a handful of notable exceptions for non-SUSY models \cite{BertoneUED,Cheung:2012xb,Arhrib:2013ela,Sming14,Chowdhury15,Liem16,LikeDM,Banerjee:2016hsk,Matsumoto:2016hbs,Cuoco:2016jqt} and none for SUSY beyond the NMSSM.  These studies, and their underlying software frameworks, were each predicated on one specific theoretical framework, relying on the phenomenologist-as-developer to hardcode the relevant observables and theory definitions.  This inflexibility and the correspondingly long development cycle required to recode things to work with a new theory, are two of the primary reasons that global fits have been applied to so few other models. The unfortunate result has been that proper statistical analyses have not been carried out for most of the theories of greatest current interest.  This is in spite of the fact that the LHC and other experiments have yet to discover any direct evidence for SUSY, heavily constraining the minimal variant \cite{Strege15,Fittinocoverage,Catalan:2015cna,MasterCodeMSSM10,Mastercode15}.  It is therefore essential that as many new ideas as possible are rigorously tested with global fits.

\begin{table*}[tb]
\centering
\caption{A \GB reading list \label{readinglist}}
\begin{tabular}{lll}
\hline
User/reader: & Wants: & Should read Sections / References: \\
\hline
Cheersquad     & to get an overview of features                   &  \ref{design}, referring to \ref{glossary} \\
Playtime       & to run \GB                                       &  ~~$\hookrightarrow$ \ref{quickstart}, \ref{cmake_backends_scanners}--\ref{config_options}, \ref{examples++} \\
Runtime        & to run \GB without causing injury                &  ~~~~$\hookrightarrow$ \ref{inventory}, \ref{interface}, \ref{stats}, \ref{output_overview}, \ref{output_specific}, \ref{logs}, \ref{diagnostics}, \ref{component_databases}, \ref{resume} \\
Dev            & to add observables, backends, etc.               &  ~~~~~~$\hookrightarrow$ all of \ref{cmake} $\rightarrow$ \ref{modules}, \ref{declaration_be_func}, \ref{backend_types}, all of \ref{utils} \\
Model-Dev      & to add new theories                              &  ~~~~~~~~$\hookrightarrow$ all of \ref{models} \\
Guru           & \doublecrosssf{BOSS}{BOSS} the world             &  ~~~~~~~~~~$\hookrightarrow$ \ref{boss}, \ref{backendinfo}, \ref{depresolver}, \ref{print_overloads} \\
\hline
Physicist      & details of physics implemented              &  \cite{ColliderBit,FlavBit,DarkBit,SDPBit} \\
Stats/CompSci  & details of scanning algorithms \& framework &  \cite{ScannerBit} \\
\hline
\end{tabular}
\end{table*}

Even working within the limited theoretical context for which they were designed, existing global fits do not offer a public framework that can be easily extended to integrate new observables, datasets and likelihood functions into the fit as they become available.  Neither do they provide any standardised or streamlined way to deal with the complex interfaces to external codes for calculating specific observables or experimental likelihoods.  Of the major SUSY global fit codes, only one (the now-discontinued \textsf{SuperBayeS} \cite{SBWeb}) has seen a public code release, in stark contrast to many of the public phenomenological tools that they employ.  Public code releases improve the reproducibility, accessibility, development and, ultimately, critique, acceptance and adoptance of methods in the community.

Another difficulty is that carrying out detailed joint statistical analyses in many-dimensional BSM parameter spaces is technically hard.  It requires full understanding of many different theory calculations and experiments, considerable coding experience, large amounts of computing time, and careful attention to statistical and numerical methods \cite{Akrami09,SBSpike,Akrami11coverage,Strege12,Fittinocoverage}.  Outside of global fits, the response has been to focus instead on individual parameter combinations or a limited, not-necessarily-representative part of the parameter space, e.g.\ \cite{ATLAS15,CMS_SMS,Berger09}. Making concrete statements across ranges of parameters requires adopting either the Bayesian or frequentist statistical framework.  These each impose specific mathematical conditions on how one discretely samples the parameter space and then combines the samples to make statements about continuous parameter ranges.  The choice of statistical framework therefore has a strong bearing upon the appropriateness and efficiency of the scanning algorithm one employs \cite{Akrami09,SBSpike}; random sampling is rarely adequate.  Most global fits have so far assumed either Bayesian or frequentist statistics, discarding the additional information available from the other.  They have also employed only a single parameter sampling algorithm each, despite the availability and complementarity of a wide range of relevant numerical methods.

Here we introduce \GB, the Global And Modular BSM Inference Tool.  \GB is a global fitting software framework designed to address the needs listed above: theory flexibility, straightforward extension to new observables and external interfaces, code availability, statistical secularism and computational speed.  In this paper we describe the \GB framework itself in detail.  First results for SUSY and the scalar singlet DM model can be found in accompanying papers \cite{CMSSM,MSSM,SSDM}, as can detailed descriptions of the constituent physics and statistics modules \cite{ColliderBit,FlavBit,DarkBit,SDPBit,ScannerBit}.  The \GB codebase is released under the standard 3-clause BSD license\footnote{\href{http://opensource.org/licenses/BSD-3-Clause}{http://opensource.org/licenses/BSD-3-Clause}.  Note that \textsf{fjcore} \cite{Cacciari:2011ma} and some outputs of \flexiblesusy \cite{Athron:2014yba} (incorporating routines from \SOFTSUSY \cite{Allanach:2001kg}) are also shipped with \GB \textsf{1.0}.  These code snippets are distributed under the GNU General Public License (GPL; \href{http://opensource.org/licenses/GPL-3.0}{http://opensource.org/licenses/GPL-3.0}), with the special exception, granted to \GB by the authors, that they do not require the rest of \GB to inherit the GPL.}, and can be obtained from \href{http://gambit.hepforge.org}{gambit.hepforge.org}.

This paper serves three purposes.  It is:
\begin{enumerate}
\item An announcement of the public release of \GB,
\item A generally-accessible presentation of the novel and innovative aspects of \GB, along with the possibilities it provides for future particle phenomenology,
\item A reference manual for the framework and associated code.
\end{enumerate}
Goals 2 and 3 imply slightly different things for the structure and content of this paper.  Here we begin by specifically addressing Goal 2, in Sec.\ \ref{design}. This section provides an extended synopsis of the flexible and modular design concept of \GB, describing its main features and advances compared to previous global fits.  Sec.\ \ref{design} provides something of a teaser for the more extended `manual', which can be found in Secs.~\ref{modules}--\ref{summary}.  These sections describe how the features of \GB are actually implemented, used and extended.  A quick start guide can be found in Appendix \ref{quickstart}, library dependencies and supported compiler lists in Appendix \ref{dependencies}, specific SM parameter definitions in Appendix \ref{SMdefs}, and a glossary of \GB-specific terms in Appendix \ref{glossary}.  When terms make their first or other crucial appearances in the text, we \doublecross{cross-link}{glossary} them to their entries in the glossary.  To the end of this arXiv submission, we also append the published \hyperlink{addendum}{Addendum} to this paper, which describes the changes implemented in \GB \textsf{1.1} compared to \GB \textsf{1.0}, including support for calling external \Mathematica codes from \GB.

\begin{figure*}[t]
\centering
\includegraphics[width=0.9\textwidth]{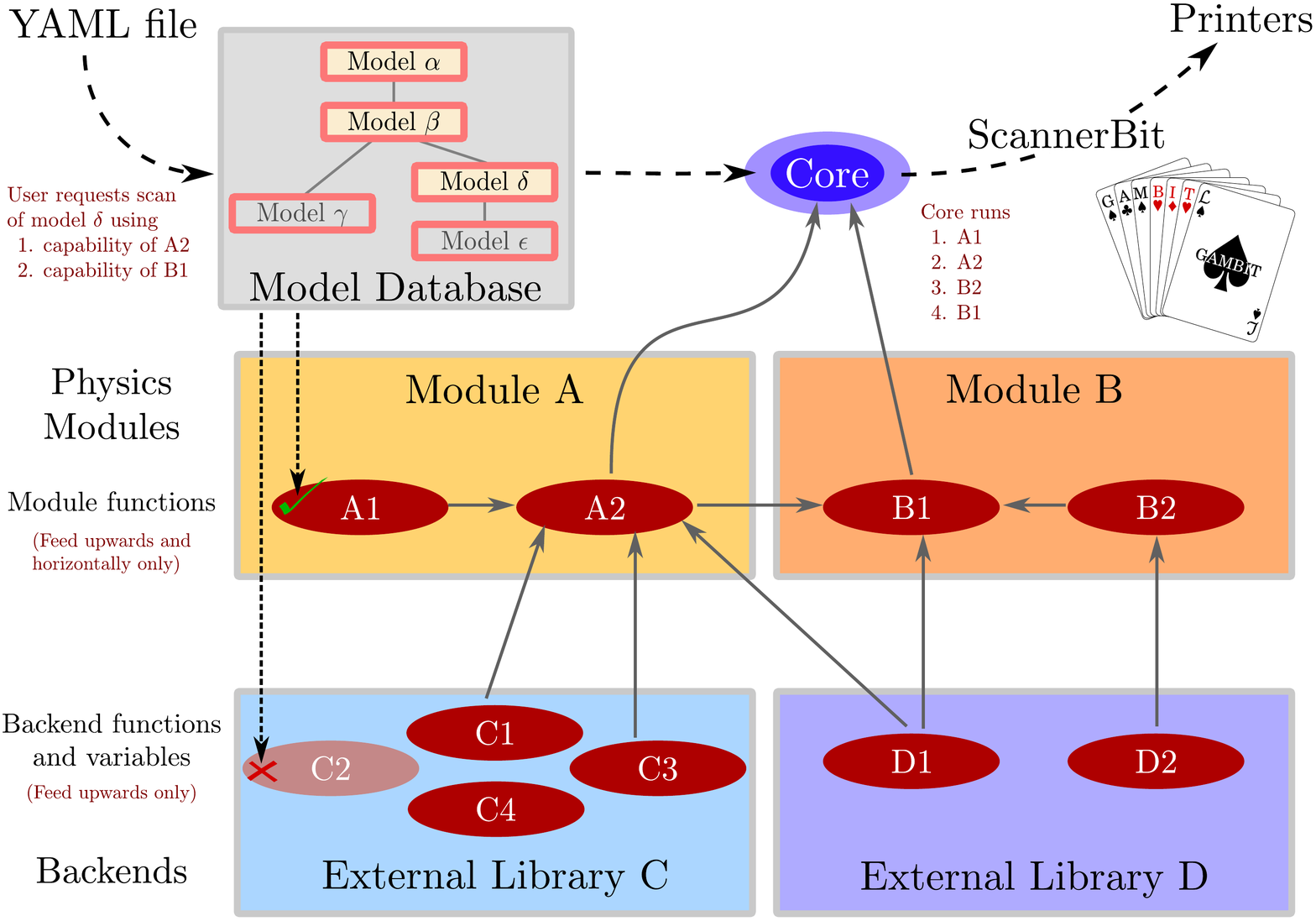}
\caption[A schematic representation of the basic elements of a \GB scan]{A schematic representation of the basic elements of a \GB scan.  The user provides a \YAML input file (see \href{http://www.yaml.org/}{www.yaml.org}), which chooses a model to scan and some observables or likelihoods to calculate.  The requested model $\delta$ and its ancestor models (see text for definition) $\beta$ and $\alpha$ are activated.  All model-dependent module and backend functions/variables are tested for compatibility with the activated models; incompatible functions are disabled (C2 in the example).  Module functions are identified that can provide the requested quantities (A2 and B1 in the example), and other module functions are identified to fulfil their dependencies.  More are identified to fulfil those functions' dependencies until all dependencies are filled.  Backend functions and variables are found that can fulfil the backend requirements of all chosen module functions.  The \textsf{Core} determines the correct module function evaluation order.  It passes the information on to \scannerbit, which chooses parameter combinations to sample, running the module functions in order for each parameter combination.  The requested quantities are output by the printer system for each parameter combination tested.}
\label{fig::design}
\end{figure*}

Within the `manual' part of the paper, Sec.\ \ref{modules} describes in detail how a physics module in \GB works, Sec.\ \ref{backends} details the system \GB uses for interfacing with external codes, and Sec.\ \ref{models} covers the internal model database and its influence on analyses and the rest of the code.  Sec.\ \ref{interface} explains the user interface to \GB and documents the available settings in the master initialisation file.  Sec.~\ref{depresolver} details the \GB system for instigating scans by automatically activating different calculations, depending on the models scanned and the observables requested by the user.  Sec.\ \ref{stats} explains how \GB deals with statistical and parameter scanning issues; further details of the specific methods and optimisation options in the scanning module can be found in Ref.\ \cite{ScannerBit}.  Sec.\ \ref{printers} describes the system for outputting results from \GB.  Sec.\ \ref{utils} covers other assorted utility subsystems.  Section \ref{cmake} discusses the build and automatic component registration system, including a crash course in adding new models, observables, likelihoods, scanners and other components to \GB.  Sec.\ \ref{examples++} describes some minimal examples included in the base distribution, and provides information about releases and support.

A code like \GB and a paper such as this have multiple levels of user and reader.  The relevant sections of this paper for each are summarised in Table \ref{readinglist}. Those more interested in understanding what \GB offers than actually running or extending it need only this introduction, Sec.\ \ref{design} and the glossary (Appendix \ref{glossary}).  Users interested in running scans without modifying any code should find more than enough to get started in Appendix \ref{quickstart}, Secs.\ \ref{cmake_backends_scanners}--\ref{config_options} and \ref{examples++}.  To get the most out of the code, such users should then move progressively on to Secs.\ \ref{inventory}, \ref{interface}, \ref{stats}, \ref{output_overview}, \ref{output_specific}, \ref{logs}, \ref{diagnostics}, \ref{component_databases} and \ref{resume}.  Those interested in adding new observables, likelihoods or interfaces to external codes should also read Secs.\ \ref{modules}, \ref{declaration_be_func}, \ref{backend_types}, and the rest of Secs.\ \ref{utils} and \ref{cmake}.  Users wanting to extend \GB to deal with new models and theories should add the remainder of Sec.\ \ref{models} to this tally.  Power users and developers wanting to have a complete understanding of the software framework should also familiarise themselves with Secs.\ \ref{boss}, \ref{backendinfo}, \ref{depresolver} and \ref{print_overloads}.  Readers and users with specific interests in particular physical observables, experiments or likelihoods should also add the relevant physics module paper(s) \cite{ColliderBit,FlavBit,DarkBit,SDPBit} to this list, and those interested in details of parameter scanning or statistics should likewise add Ref.\ \cite{ScannerBit}.

\section{Design overview}
\label{design}

\GB consists of a number of \doublecross{modules}{module} or `Bits', along with various Core components and utilities.  Fig.\ \ref{fig::design} is a simplified representation of how these fit together. \GB modules are each either \doublecross{physics modules}{physics module} (\darkbit, \colliderbit, etc.) or the scanning module, \scannerbit. \scannerbit is responsible for parameter sampling, prior transformations, interfaces to external scanning and optimisation packages and related issues; it is discussed in more detail in Sec.\ \ref{stats} and Ref.\ \cite{ScannerBit}.

\subsection{Modularity}

\subsubsection{Physics modules, observables and likelihoods}

The first version of \GB ships with six physics modules:

\begin{description}
\item[\textbf{\colliderbit}\!]calculates particle collider observables and likelihoods.  It includes detailed implementations of LEP, ATLAS and CMS searches for new particle production, and measurements of the Higgs boson.  The LEP likelihoods are based on direct cross-section limits on sparticle pair production from ALEPH, OPAL and L3.  Fast Monte Carlo simulation of signals at ATLAS and CMS can be performed with a specially parallelised version of \pythiaeight \cite{Sjostrand:2014zea}.  \colliderbit offers the option to carry out detector simulation with \buckfast, a fast smearing tool, or the external package \textsf{Delphes} \cite{Ovyn:2009tx,deFavereau:2013fsa}.  We have validated all likelihoods and limits via extensive comparison to experimental limits and cutflows.  Higgs likelihoods in the first version of \colliderbit are provided exclusively by communication with \higgsbounds \cite{Bechtle:2008jh,Bechtle:2011sb,Bechtle:2013wla} and \higgssignals \cite{HiggsSignals}. Supersymmetic models are presently supported natively by the LEP and LHC likelihoods.  The Higgs likelihoods are model-independent in as much as they require only Higgs couplings as inputs.  Other models can be supported in LHC calculations by reading matrix elements into \pythiaeight, e.g.\ from \madgraph \cite{Alwall:2011uj,Alwall:2014hca}.  For a detailed description, see \cite{ColliderBit}.
\item[\textbf{\flavbit}\!]calculates observables and likelihoods from flavour physics, in particular $B$, $D$ and $K$ meson decays as observed by LHCb, including angular observables and correlations.  Possibilities for inter-code communication exist with \superiso \cite{Mahmoudi:2007vz,Mahmoudi:2008tp,Mahmoudi:2009zz} and \FH \cite{Bahl:2017aev,Bahl:2016brp,Hahn:2013ria,Frank:2006yh,Degrassi:2002fi,Heinemeyer:1998np,Heinemeyer:1998yj}.  Supersymmetry is supported directly.  A broad range of other models is supported, via the use of effective field theory. Likelihoods and observables have been validated by comparison to existing flavour fits \cite{Mahmoudi:2014mja,Altmannshofer:2013foa,Descotes-Genon:2015uva}.  See \cite{FlavBit}.
\item[\textbf{\darkbit}\!]calculates DM observables and likelihoods, from the relic abundance to direct and indirect searches.  It includes an on-the-fly cascade decay spectral yield calculator, and a flexible, model-independent relic density calculator capable of mixing and matching aspects from existing backends, including \darksusy \cite{darksusy} and \micromegas \cite{Belanger:2001fz,Belanger:2004yn,Belanger:2006is,Belanger:2010gh,Belanger:2013oya,Belanger:2014vza}.  Direct detection likelihoods in \darkbit are based on calls to the \ddcalc package \cite{DarkBit}.  Indirect detection can be carried out with the help of \nulike \cite{IC79_SUSY} (neutrinos) and \gamlike \cite{DarkBit} (gamma rays).  Validation of relic density calculations is based on extensive comparison with results from standalone versions of \darksusy and \micromegas.  Direct and indirect limits are validated by comparison with exclusion curves from the relevant experiments.  All calculations support MSSM neutralinos and all other WIMPs (in particular, this includes Higgs portal models such as scalar singlet dark matter).  See \cite{DarkBit} for details.
\item[\textbf{\specbit}\!]interfaces to one of a number of possible external spectrum generators in order to determine pole masses and running parameters, and provides them to the rest of \GB in a standardised spectrum container format.  Spectrum generators currently supported include \flexiblesusy \cite{Athron:2014yba} and \SPheno \cite{Porod:2003um,Porod:2011nf}. Models include MSSM models defined at arbitrary scales and the scalar singlet model.  Support for additional spectrum generators and models is straightforward for users to add.  Results of the existing code have been validated by comparison to standalone versions of \flexiblesusy, \SPheno and \SOFTSUSY \cite{Allanach:2001kg,Allanach:2009bv,Allanach:2011de,Allanach:2013kza,Allanach:2014nba}. \specbit also carries out vacuum stability calculations and perturbativity checks, which have been validated against existing results in the literature.  See \cite{SDPBit} for full details.
\item[\textbf{\decaybit}\!]calculates decay rates of all relevant particles in the BSM theory under investigation, and contains decay data for all SM particles.  Theory calculations can make use of interfaces to \FH \cite{Bahl:2017aev,Bahl:2016brp,Hahn:2013ria,Frank:2006yh,Degrassi:2002fi,Heinemeyer:1998np,Heinemeyer:1998yj} and an improved version of \susyhit \cite{Djouadi:2006bz,Muhlleitner:2003vg,Djouadi:2002ze,Djouadi:1997yw}, validated against direct SLHA communication with the same codes.  \decaybit supports the MSSM and scalar singlet models.  See \cite{SDPBit}.
\item[\textbf{\precisionbit}\!]calculates model-dependent precision corrections to masses, couplings and other observables, as well as precision nuisance likelihoods for e.g. Standard Model (SM) parameters.  BSM calculations are presently limited to the MSSM, and can call on \gmtwocalc \cite{gm2calc}, \FH \cite{Bahl:2017aev,Bahl:2016brp,Hahn:2013ria,Frank:2006yh,Degrassi:2002fi,Heinemeyer:1998np,Heinemeyer:1998yj} and \superiso \cite{Mahmoudi:2007vz,Mahmoudi:2008tp,Mahmoudi:2009zz}.  See \cite{SDPBit}.
\end{description}

Physics modules are collections of \doublecross{module functions}{module function}, each capable of calculating a single physical or mathematical quantity. This may be an observable, likelihood component or any intermediate quantity required for computing one or more observables or likelihoods.

Each module function is tagged with a \cross{capability}, which together with the associated \cross{type} describes exactly what \cross{quantity} it is capable of calculating.  Module functions, rather than modules themselves, are the main building blocks of \GB.  The capability-type pairs associated with module functions are the threads that allow \GB to automatically stitch together multiple functions into arbitrarily complicated physics calculations.

Individual module functions may have one or more \doublecross{dependencies}{dependency} on quantities that their own calculations depend on.  At runtime, \GB selects an appropriate module function to fulfil each dependency, by matching the declared capabilities of module functions with the declared dependencies of other module functions.  This process also requires matching the declared return types of module functions with the types requested in each dependency.

A simple example is the $W$ mass likelihood function in \precisionbit, which has capability \cpp{lnL_W_mass}. This function calculates a basic $\chi^2$ likelihood for the $W$ mass, and is correspondingly named \cpp{lnL_W_mass_chi2}.  To do its job, \cpp{PrecisionBit::lnL_W_mass_chi2} must be provided with a predicted value for the $W$ mass, by some other module function in \GB.  These aspects are declared
\begin{lstcpp}
#define CAPABILITY lnL_W_mass
START_CAPABILITY
  #define FUNCTION lnL_W_mass_chi2
  START_FUNCTION(double)
  DEPENDENCY(mw, triplet<double>)
  #undef FUNCTION
#undef CAPABILITY
\end{lstcpp}
Here the \cpp{DEPENDENCY} on the $W$ mass \cpp{mw} is explicitly declared, and the declaration demands that it must be provided as a set of three real numbers, corresponding to a central value with upper and lower uncertainties (a \cpp{triplet<double>}).  \cpp{lnL_W_mass_chi2} accesses these values in its actual source via a pointer named \cpp{mw} placed in a special \cpp{namespace} reserved for dependencies (\cpp{Pipes::lnL_W_mass_chi2::Dep}).  It then uses the values to compute the likelihood, which it returns as its result:
\begin{lstcpp}
/// W boson mass likelihood
const double mw_central_observed = 80.385;
const double mw_err_observed = 0.015;
void lnL_W_mass_chi2(double &result)
{
  using namespace Pipes::lnL_W_mass_chi2;
  double theory_uncert = std::max(Dep::mw->upper,
   Dep::mw->lower);
  result = Stats::gaussian_loglikelihood(
   Dep::mw->central, mw_central_observed,
   theory_uncert, mw_err_observed);
}
\end{lstcpp}

This module function has no concern for precisely \textit{where} or \textit{how} the $W$ mass has been determined.  This allows \GB to choose for itself at runtime, on the basis of the model being scanned, whether it should provide e.g.\ an MSSM-corrected prediction (for an MSSM scan), or a different prediction (for a scan of a different model).  This serves to illustrate the power of the modular design of \GB, allowing different calculations to be automatically reused in myriad different physics scenarios, with essentially zero user intervention.

Section \ref{modules} covers declaring and writing module functions in detail.

\subsubsection{Backends}

External software packages that might be useful for calculating specific quantities are referred to in \GB as \doublecross{backends}{backend}.  Examples of these might be \textsf{DarkSUSY}~\cite{darksusy} (for, e.g., relic density calculations), or \textsf{FeynHiggs}~\cite{Heinemeyer:1998yj,Heinemeyer:1998np,Degrassi:2002fi,Frank:2006yh,Hahn:2013ria} (for, e.g., Higgs mass calculations).  A full list of existing codes with which \GB can communicate via the backend system, along with all relevant references, can be found in the file \term{README.md} included in the main distribution.  All studies that make use of \GB with a backend must cite all the literature associated with that backend, along with all relevant \GB literature.

Although \GB itself is written in \Cpp, with a small admixture of \python for build abstraction, backends can in principle be written in any language.  Module functions can directly call \doublecross{backend functions}{backend function} and access global \doublecross{backend variables}{backend variable} from these codes.  To do this, a module function must declare that it has a \cross{backend requirement}, which is then matched at runtime to the declared capability of a function or variable from some backend. This mirrors the manner in which \GB fills dependencies from amongst the available module functions.

Whilst module functions can have both dependencies (resolvable with other module functions) and backend requirements (resolvable with backend functions or variables), backend functions and variables cannot themselves have either dependencies nor backend requirements.  This is illustrated in the example in Fig.\ \ref{fig::design}: backend functions and variables feed into module functions, but nothing feeds into the backend functions nor variables themselves.

A simple example is the calculation in \darkbit of the rate at which DM is gravitationally captured by the Sun:
\begin{lstcpp}
#define CAPABILITY capture_rate_Sun
START_CAPABILITY
  #define FUNCTION capture_rate_Sun_const_xsec
  START_FUNCTION(double)
  DEPENDENCY(mwimp, double)
  DEPENDENCY(sigma_SI_p, double)
  DEPENDENCY(sigma_SD_p, double)
  BACKEND_REQ(cap_Sun_v0q0_isoscalar, (), double,
   (const double&, const double&, const double&))
  #undef FUNCTION
#undef CAPABILITY
\end{lstcpp}
Here \cpp{DarkBit::capture_rate_Sun_const_xsec} depends on the DM mass and scattering cross-sections, and explicitly declares that it requires access to a function from a backend.  It demands that the backend function be tagged with capability \cpp{cap_Sun_v0q0_isoscalar}, that it take three \cpp{const double&} arguments, and that it must return a \cpp{double} result.  The declaration of a matching backend function (taken in this example from the interface to \darksusy\ \textsf{5.1.3}) would then look like:
\begin{lstcpp}
BE_FUNCTION(dsntcapsuntab, double, (const double&,
 const double&, const double&), "dsntcapsuntab_",
 "cap_Sun_v0q0_isoscalar")
\end{lstcpp}
The function \cpp{DarkBit::capture_rate_Sun_const_xsec} then accesses the backend function from its source via a similar alias system to the one used for dependencies:
\begin{lstcpp}
// @\cpppragma{Capture rate in s$^{-1}$ of regular DM in the Sun}@
// @\cpppragma{($\implies$ $\sigma$ is neither $v$-dependent nor $q$-dependent),}@
// @\cpppragma{assuming isoscalar interactions ($\sigma_p = \sigma_n$)}@.
void capture_rate_Sun_const_xsec(double &result)
{
  using namespace
   Pipes::capture_rate_Sun_const_xsec;
  result =
   BEreq::cap_Sun_v0q0_isoscalar(*Dep::mwimp,
   *Dep::sigma_SI_p, *Dep::sigma_SD_p);
}
\end{lstcpp}
Typically, the requirement \cpp{cap_Sun_v0q0_isoscalar} will be fulfilled by \darksusy, a \Fortran code -- but there is nothing about this particular example function nor its declaration that forces such a pairing.  The only conditions are that the selected backend function fulfils the requirements laid out in the \cpp{BACKEND_REQ} declaration.  This is another example of the power of the modular design of \GB, allowing it to attach any matching function from any backend at runtime, and to adapt to the presence or absence of different versions of different backends present on any given user's system.

There are many additional options and declarations available but not shown in this example, for constraining which versions of which backends are permitted to provide which backend requirement, under what model-dependent conditions and so on.  Two additional features of note are not shown in Fig.\ \ref{fig::design}: \doublecross{backend initialisation functions}{backend initialisation function}, which always run before any functions or variables in a backend are used, and \doublecross{backend convenience functions}{backend convenience function}, which are agglomerations of functions and variables from one backend, presented to the rest of \GB as if they are single backend functions.

Declaration of backend requirements is covered in detail in Section \ref{declaration_bereq}, and declaration of actual interfaces to backends is covered in Section \ref{backends}.

\begin{figure*}[tp]
\centering
\includegraphics[width=\textwidth]{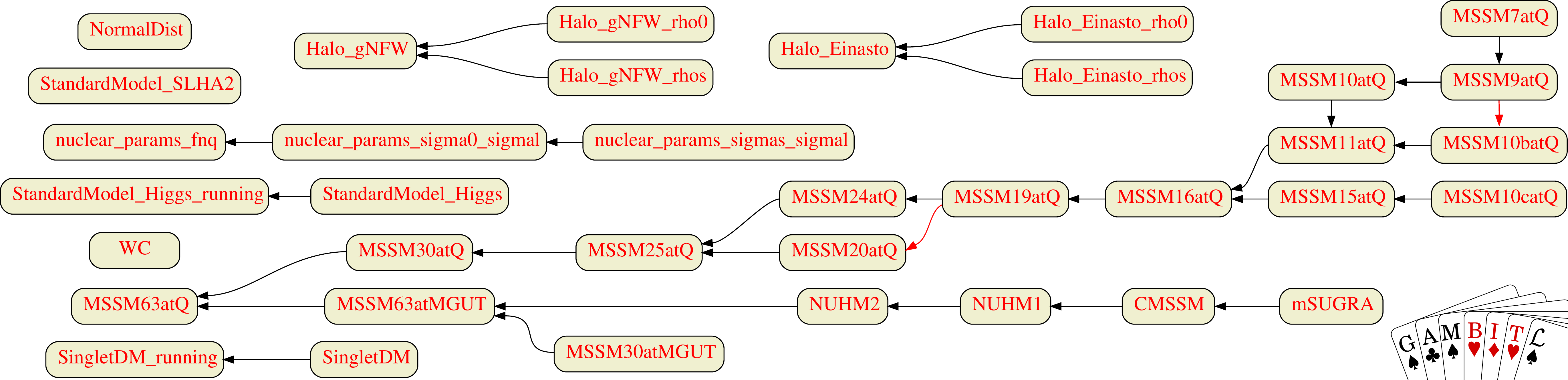}
\caption{The model hierarchy graph of the pre-defined models that ship with \GB \textsf{1.0.0}. The graph forms a set of disconnected directed trees, potentially linked by \doublecross{friend}{friend model} translation pathways. Nodes are individual \doublecross{models}{model}. Black arrows indicate \doublecross{child}{child model}-to-\doublecross{parent}{parent model} translation pathways. The red arrows from \textsf{MSSM9atQ} to \textsf{MSSM10batQ}, and from \textsf{MSSM19atQ} to \textsf{MSSM20atQ}, indicate translations to \doublecross{friend models}{friend model}.  Friend translations can cross between otherwise disconnected family trees, or, as in these two examples, between different branches of the same tree. Graphs like this (including any additional user-specified models) can be generated by running \protect\term{gambit models} from the command line, and following the instructions provided.}
\label{fig::model_tree}
\end{figure*}

\subsubsection{Models}

The models already implemented in \GB \textsf{1.0.0} are shown in Fig.\ \ref{fig::model_tree}, and described in detail in Sec.\ \ref{inventory}.  Instructions for adding new models are given in Sections \ref{model_declaration} and \ref{adding_components}.

\GB automatically activates or disables module and backend functions\footnote{and backend variables --- but from here we will stop explicitly referring to backend functions and backend variables as different things except where it actually matters.} according to their compatibility with the BSM model under investigation.  It does this using a hierarchical model database, where each model is defined as a set of free parameters and a series of relations to other models.  Models can be declared as children of existing models, which implies that there exists a mapping from the child parameter space to some subspace of the \doublecross{parent}{parent model} space.  Each \cross{child model} comes with a function that defines the transformation required to take a parameter point in its space to a corresponding point in the parent parameter space.  \GB uses these transformations at runtime to deliver the same parameter point in different parameterisations to different module functions, according to their declared needs.  Models can also have translations pathways defined to other so-called \doublecross{friend models}{friend model} outside their own family tree.

One important aspect of this arrangement is that models can be arbitrarily `bolted together' for any given scan, so that multiple models can be scanned over simultaneously, and their parameter values delivered together to any module functions that need them.  This allows for the SM parameters to be varied as nuisance parameters when doing an MSSM or other BSM scan, for example.  It also means that in such a joint SM-MSSM scan, the \textit{same} underlying SM model (and therefore the same SM calculations wherever possible) will be used as in any other joint SM-BSM scan.

When a user requests a scan of a particular BSM model, that model and its entire model ancestry are activated.  This makes all module and backend functions that are compatible with any model in the activated ancestry available as valid building blocks of the scan.  This provides maximum safety by forbidding any calculations that are not valid for the model under consideration, and maximum re-usability of modules, backends and their functions with new models, by providing certainty about which existing functions are `safe' to use with new additions to the model hierarchy.

A basic example of model and backend function activation/deactivation can be seen in Fig.\ \ref{fig::design}. Functions A1 and C2 have been specifically declared as model-dependent and therefore require activation or deactivation by the model database.  Only functions that have been declared as model-dependent in this way are allowed to access the values of the underlying parameters in a scan.  No other functions have any such declarations, so they are therefore valid for all models.  Such functions \textit{must} always work for any model, as all they need to do their job is to be confident that \GB will deliver their declared dependencies and backend requirements in the form that they request -- and \GB guarantees precisely this for \textit{all} module functions.

The two examples given in the previous subsections, of the $W$ mass likelihood and the capture rate of DM by the Sun, are both examples of essentially model-independent calculations, where the module function does not need direct access to any of the underlying model parameters.  These functions care only that their dependencies and backend requirements are available; if this is the case, they can do their jobs, irrespective of the underlying model actually being scanned.\footnote{Note the distinction between model-independent \textit{functions} and model-independent \textit{results}.  Model-independent numerical results have the same values regardless of the physics model assumed.  Model-independent functions act on input data according to the values of the data only, not according to the physics model that gave rise to the data.  In general, the input data to model-independent functions are model-dependent quantities, leading to different results for different models. The $W$ mass likelihood is a case in point: the predicted value of $m_W$ and its likelihood are necessarily model-dependent quantities -- but the function that computes the likelihood from a given value of $m_W$ is not dependent on the model for which $m_W$ has been computed.}

An example of an explicitly model-dependent module function is the \darkbit likelihood associated with the nuclear matrix elements relevant for spin-independent DM-nucleon scattering:
\begin{lstcpp}
// Likelihoods for nuclear parameters.
START_CAPABILITY
  #define FUNCTION lnL_sigmas_sigmal
    START_FUNCTION(double)
    ALLOW_MODEL(nuclear_params_sigmas_sigmal)
  #undef FUNCTION
\end{lstcpp}
Here the \cpp{ALLOW_MODEL} declaration is used to indicate that the module function can \textit{only} be used when scanning the \cpp{nuclear_params_sigmas_sigmal} model (or one of its descendants -- but it has none in this version of \GB).  This particular module function directly accesses the values of the model parameters, uses them to compute the joint likelihood and returns the result.  In contrast, when the nuclear matrix elements are needed for calculating the physical DM-nucleon couplings in e.g., the scalar singlet Higgs portal model, they are instead upcast to the \cpp{nuclear_params_fnq} model (an ancestor of \cpp{nuclear_params_sigmas_sigmal}, cf. Fig.\ \ref{fig::model_tree}), and presented as such within the relevant module function:
\begin{lstcpp}
#define FUNCTION DD_couplings_SingletDM
  START_FUNCTION(DM_nucleon_couplings)
  DEPENDENCY(SingletDM_spectrum, Spectrum)
  ALLOW_JOINT_MODEL(nuclear_params_fnq, SingletDM)
#undef FUNCTION
\end{lstcpp}
Here the \cpp{ALLOW_JOINT_MODEL} declaration explicitly forbids \GB from using this function except when scanning \textit{both} the \cpp{nuclear_params_fnq} and \cpp{SingletDM} models, or some pairwise combination of their respective descendants.

The \GB model database, its declarations and features are discussed in much more detail in Sec.\ \ref{models}.

\subsection{Adaptability and flexibility}

After filtering out invalid module and backend functions by checking their compatibility with the model under investigation, \GB works through the remaining functions to properly connect module functions to dependencies and backend functions to backend requirements.  It starts with the quantities requested by the user (observables, likelihood components or other derived quantities), and then progressively resolves dependencies and backend requirements until it either reaches an impasse due to a mutual dependency between groups of module functions, or no outstanding needs remain.  If all needs have been fulfilled, the result is a directed graph of dependencies, with no internal closed loops -- a so-called directed acyclic graph.  Directed acyclic graphs have the mathematical property that they possess an implied topological order.  \GB computes this ordering, and uses it to determine the optimal order in which to evaluate the module functions, such that each module function is guaranteed to run before any other function that depends on its results.  \GB further optimises the ordering beyond the constraint implied by this condition, considering the typical evaluation time of each function as a scan progresses, and its role in ruling out previous parameter combinations.  We explain this overall \cross{dependency resolution} process in detail in Sec.\ \ref{depresolver}.

With a specific module function evaluation order in hand, \GB passes the problem of actually sampling the parameter space to \scannerbit (Sec.\ \ref{stats}).  \scannerbit engages whichever statistical scanning algorithm and output method a user has selected in their input file (see Sec.\ \ref{interface}), choosing parameter combinations, calling the module functions in order, and sending the results to the \GB printer system (Sec.\ \ref{printers}).  Functions log their activity via an extensive internal logging system (Sec.\ \ref{logs}), and invalid parameter combinations, warnings and errors are identified using a dedicated exceptions system (Sec.\ \ref{exceptions}).

This rather abstract formulation of the global fit problem enables a very high degree of automation, in turn providing flexibility and extendibility.  By deferring the actual choice of the function that will provide the requisite physical inputs to each step of a calculation, \GB makes it easy to confidently swap or add functions to existing scans.  It also makes such scans efficient, as only the calculations needed for a given scan are actually activated, and each calculation is guaranteed to run only once for each parameter combination.  Linking this to a hierarchical model database then provides the means for \GB to automatically adapt existing likelihood and observable calculations to new models, to the largest extent theoretically possible.  New components of course need to be added when different physics is to be considered for the first time, but the level of automation allows the user to immediately identify the precise gaps in the theoretical chain in need of new work, rather then wasting time by coding almost identical functions for every new model.  This is facilitated by extensive and informative error messages presented when a scan is attempted but some link in the dependency chain cannot be fulfilled.  These messages explain, for example, when a given dependency cannot be filled by any known function, if a requisite backend appears to be missing, if appropriate functions seem to exist but are not compatible with the model being scanned, if multiple permitted options exist for resolving a given dependency or backend requirement, and so on.

\GB takes this flexibility and automatic adaptation even further by having the backend (Sec.\ \ref{backends}) and build (Sec.\ \ref{cmake}) systems automatically add or disable modules, backends, models, printers and other components when new ones are defined, or when existing ones happen to be missing from a user's system.  \GB also includes extensive command-line diagnostics, which the user can employ to obtain reports on the status and contents of its components at many different levels (Sec.\ \ref{diagnostics}).

\subsection{Performance and parallelisation}

Parallelisation in \GB happens at two levels: at the scanner level via \mpi \cite{MPI}, and at the level of module functions with \omp \cite{openmp}.  This allows \GB to easily scale to many thousands of cores, as most major external sampling packages employ \mpi, and a number of external physics codes make use of \omp (e.g. \textsf{nulike} \cite{IC79_SUSY} and forthcoming versions of \textsf{DarkSUSY} \cite{darksusy}).  Users also have the option of implementing their own module functions using \omp natively in \GB.  In fact, \GB can even automatically connect \omp-aware module functions and have other module functions run them in parallel using \omp.  Sec.\ \ref{declaration_loops} explains how to achieve this.  With this method, the total runtime for a single MSSM parameter combination, even including explicit LHC Monte Carlo simulation, can be reduced to a matter of a few seconds \cite{ColliderBit}.

The performance of \GB is explored in detail in the \scannerbit paper \cite{ScannerBit}.

\subsection{Available examples}

In Sec.\ \ref{examples} we provide a series of examples showing how to run the full \GB code.  Any \GB module can also be compiled with a basic driver into a standalone program.  We also give a number of examples of module standalone drivers in Sec.\ \ref{examples}, as well as dedicated examples for different modules included in first release \cite{DarkBit,ColliderBit,ScannerBit,FlavBit,SDPBit}.

A standalone driver program that calls a \GB module needs to do a number of specific things:
\begin{itemize}
\item specify which model to work with,
\item choose what the parameter values should be,
\item indicate which module functions to run and what to do with the results,
\item indicate which module functions to use to fulfil which dependencies,
\item indicate which backend functions or variables to use to fulfil which backend requirements, and
\item set input options that different module functions should run with.
\end{itemize}
These are all functions that are normally done automatically by \GB.  We provide a series of simple utility functions specifically designed for use in standalone driver programs though, allowing most of these operations to be completed in a single line each.

\section{Modules}
\label{modules}

Other than the six physics and one scanning module in \GB \textsf{1.0.0}, behind the scenes \GB also arranges backend initialisation functions into a virtual module known as \textsf{BackendIniBit}, and puts model parameter translation functions into effective modules of their own.  These are discussed in detail in Secs.\ \ref{backends} and \ref{models}, respectively.

\subsection{Module function declaration}
\label{module_declaration}
\GB modules and their functions are declared in a module's so-called \cross{rollcall header}, using a series of convenient macros.

A module called \metavar{MyBit} would be created simply by creating a header \term{MyBit\_rollcall.hpp} containing
\begin{lstcpp}
#define MODULE @\metavar{MyBit}@
START_MODULE
#undef MODULE
\end{lstcpp}
and then rerunning the build configuration step in order to make \GB locate the new file.\footnote{Re-running the configuration step is a generic requirement whenever adding new source or header files to \GB.  See Sec.\ \ref{cmake} for details.}

Creating a module function requires a user to write it as a standard \Cpp function in a source file, and add a corresponding declaration to the rollcall header.  The function should have return type void, and take exactly one argument by reference: the result of the calculation that the function is supposed to perform.  This result can be of any type.\footnote{At least, any type with a default constructor.  Dealing in types without default constructors requires declaring objects internally in the module and returning pointers to them.}  Taking a double-precision floating point number as an example, the definition of a function \metavar{function\_name} in module \metavar{MyBit} would look like
\begin{lstcpp}
namespace @\metavar{MyBit}@
{
  void @\metavar{function\_name}@(double& result)
  {
    result = ... // something useful
  }
}
\end{lstcpp}
This would traditionally be placed in a file called \term{MyBit.cpp} or similar.

The declaration must state the name of the function, the type of its result, and the capability to assign to it.  Such a declaration would look like
\begin{lstcpp}
#define MODULE @\metavar{MyBit}@
START_MODULE
  #define CAPABILITY @\metavar{example\_capability}@
  START_CAPABILITY
    #define FUNCTION @\metavar{function\_name}@
    START_FUNCTION(double)
    #undef FUNCTION
  #undef CAPABILITY
#undef MODULE
\end{lstcpp}
where \metavar{example\_capability} is the name of the capability assigned to the function \metavar{MyBit::function\_name} in this example.

The following examples in Secs.\ \ref{declaration_allow_model}--\ref{declaration_bereq} will show other specific declarations that may be given between\\\lstinline{START_FUNCTION} and \lstinline{#undef FUNCTION}.

\subsubsection{Model compatibility}
\label{declaration_allow_model}

In the absence of any specific declarations as to the model-dependency of the calculations in a module function, \GB assumes that the function is completely model-independent.  To instead declare that a module function may only be used with a single specific model \metavar{model\_a}, one adds a declaration
\begin{lstcpp}
ALLOW_MODEL(@\metavar{model\_a}@)
\end{lstcpp}
after calling \lstinline|START_FUNCTION|.  To declare that the function may be used with one or more models from a particular set, one instead writes
\begin{lstcpp}
ALLOW_MODEL(@\metavar{model\_a}@)
ALLOW_MODEL(@\metavar{model\_b}@)
...
\end{lstcpp}
or just
\begin{lstcpp}
ALLOW_MODELS(@\metavar{model\_a}@, @\metavar{model\_b}@, ...)
\end{lstcpp}
where the ellipses \cpp{...} indicate that the \lstinline|ALLOW_MODELS| macro is variadic, and can take up to 10 specific models.  Alternatively, to declare that \textit{all} models from a given set must be in use, one declares
\begin{lstcpp}
ALLOW_JOINT_MODEL(@\metavar{model\_$\gamma$}@, @\metavar{model\_$\delta$}@, ...)
\end{lstcpp}
Declaring \lstinline|ALLOW_MODEL|, \lstinline|ALLOW_MODELS| or \lstinline|ALLOW_JOINT_MODEL| also grants the module function access to the values of the parameters of the appropriate model(s) at runtime.  Section \ref{param_pipe} below deals with how to retrieve these parameter values.

\GB is expressly designed for simultaneous scanning of multiple models, where the parameters of each model are varied independently.  This allows for arbitrary combinations of different models, e.g. from including SM parameters as nuisances in a BSM scan, to varying cosmological and BSM parameters simultaneously in some early-Universe cosmological scenario.  In these cases, module functions can be granted access to the parameters of multiple models at the same time, as long as the function is declared from the outset to need all of those parameters in order to operate correctly.

To set rules that constrain module functions' validities to scans of specific combinations of models, rather than simply declaring valid combinations one by one with \lstinline|ALLOW_JOINT_MODEL|, a more involved syntax is required.  Here, the possible individual models involved in the combinations are first listed with \lstinline|ALLOW_MODEL_DEPENDENCE|.  They are then placed into one or more specific \doublecross{model groups}{model group}.  Each allowed model combination is then specified by setting allowed combinations of model groups.  If a given scan includes one model from each group listed in an allowed combination, then the module function is deemed to be compatible with the given scan.

For example, to specify that a function may be used when either \metavar{model\_a} or \metavar{model\_b} is being scanned, \textit{but only if \metavar{model\_c} is also being scanned at the same time}, one must write
\begin{lstcpp}
ALLOW_MODEL_DEPENDENCE(@\metavar{model\_a}@, @\metavar{model\_b}@, @\metavar{model\_c}@)
MODEL_GROUP(group1, (@\metavar{model\_a}@, @\metavar{model\_b}@))
MODEL_GROUP(group2, (@\metavar{model\_c}@))
ALLOW_MODEL_COMBINATION(group1, group2)
\end{lstcpp}
This reveals that \cpp{ALLOW_JOINT_MODEL(}\metavar{model\_$\gamma$}\cpp{,} \metavar{model\_$\delta$}\cpp{,} \cpp{...)} is simply a special case of this extended syntax, precisely equivalent to
\begin{lstcpp}
ALLOW_MODEL_DEPENDENCE(@\metavar{model\_$\gamma$}@, @\metavar{model\_$\delta$}@, ...)
MODEL_GROUP(group1, (@\metavar{model\_$\gamma$}@))
MODEL_GROUP(group2, (@\metavar{model\_$\delta$}@))
...
ALLOW_MODEL_COMBINATION(group1, group2, ...)
\end{lstcpp}

Note that \GB still deems a model to be in use even if its parameters are fixed to constant values during a scan.  Declaring that a module function requires some model or model combination to be in use therefore merely demands that the model parameters have definite values during a scan, not that they are necessarily varied.

An explicit example of the syntax described in this section can be found in the declaration of the function \cpp{DarkBit::DD_couplings_MicrOmegas} in \term{DarkBit/include/gambit/DarkBit/DarkBit_rollcall.hpp}:
\begin{lstcpp}
ALLOW_MODEL_DEPENDENCE(nuclear_params_fnq,
 MSSM63atQ, SingletDM)
MODEL_GROUP(group1, (nuclear_params_fnq))
MODEL_GROUP(group2, (MSSM63atQ, SingletDM))
ALLOW_MODEL_COMBINATION(group1, group2)
\end{lstcpp}
This function computes couplings relevant for direct detection, using \micromegas \cite{micromegas}.  To do this, it needs the parameters of the nuclear matrix element model \doublecrosssf{nuclear\_params\_fnq}{nuclear_params_fnq}, plus the parameters of a dark matter model, which in \GB\ \textsf{1.0.0} may be either the MSSM or the scalar singlet model.

\subsubsection{Dependencies}
\label{declaration_deps}

To indicate that a module function requires some specific quantity as input in order to carry out its own calculation, one must declare that it has a dependency upon the capability, and the corresponding type, of some other module function.  Dependencies are explicitly defined in terms of capabilities, not specific functions: \textit{from the} \GB \textit{perspective functions do not depend on each other, they depend on each others' capabilities}.  This is specifically designed to make module functions genuinely modular, by keeping the use of a module function's result completely independent of its identity.  This has the (entirely intentional) consequence of making it practically impossible to safely use global states for passing information between module functions.

The syntax for declaring that a module function \metavar{function\_name} has a dependency on some capability \metavar{capability} is simply to add a line
\begin{lstcpp}
DEPENDENCY(@\metavar{capability}@, @\metavar{type}@)
\end{lstcpp}
to the module function declaration.  Here \metavar{type} is the actual \Cpp type of the capability that needs to be available for \metavar{function\_name} to use in its function body.

Such a declaration ensures that at runtime, \GB will arrange its \cross{dependency tree} such that it
\begin{itemize}
\item[a)] only runs \metavar{function\_name} after some other module function with capability \metavar{capability} and return type \metavar{type} has already run for the same parameter combination,
\item[b)] delivers the result of the other module function to \metavar{function\_name}, so that the latter can use it in its own calculation.
\end{itemize}

It is also possible to arrange \doublecross{conditional dependencies}{conditional dependency} that only apply when specific conditions are met.  The simplest form is a purely model-dependent conditional dependency,
\begin{lstcpp}
MODEL_CONDITIONAL_DEPENDENCY(@\metavar{capability}@, @\metavar{type}@,
 @\metavar{model\_$\alpha$}@, @\metavar{model\_$\beta$}@, ...)
\end{lstcpp}
which would cause a function to depend on \metavar{capability} only when \metavar{model\_$\alpha$} and/or \metavar{model\_$\beta$} is being scanned.  Here the ellipses again indicate that up to 10 models can be specified.

A concrete example of this is the declaration of the function \cpp{FlavBit::SuperIso_modelinfo} in \term{FlavBit/include/gambit/FlavBit/FlavBit_rollcall.hpp}.  This function is responsible for constructing the data object that will be sent to \superiso \cite{Mahmoudi:2007vz,Mahmoudi:2008tp} to tell it the values of the relevant Lagrangian parameters.  Its declaration includes the lines:
\begin{lstcpp}
MODEL_CONDITIONAL_DEPENDENCY(MSSM_spectrum,
 Spectrum, MSSM63atQ, MSSM63atMGUT)
MODEL_CONDITIONAL_DEPENDENCY(SM_spectrum,
 Spectrum, WC)
\end{lstcpp}
These statements cause the function to have a dependency on an \cpp{MSSM_spectrum} when scanning the MSSM, but a dependency on an \cpp{SM_spectrum} when scanning a low-energy effective theory of flavour (\doublecrosssf{WC}{WC}; see Sec.\ \ref{flavEFT}).

An alternative formulation allows both model conditions and backend conditions to be specified:
\begin{lstcpp}
#define CONDITIONAL_DEPENDENCY @\metavar{capability}@
START_CONDITIONAL_DEPENDENCY(@\metavar{type}@)
ACTIVATE_FOR_MODELS(@\metavar{model\_$\alpha$}@, @\metavar{model\_$\beta$}@, ...)
ACTIVATE_FOR_BACKEND(@\metavar{requirement}@, @\metavar{be\_name1}@)
ACTIVATE_FOR_BACKEND(@\metavar{requirement}@, @\metavar{be\_name2}@)
#undef CONDITIONAL_DEPENDENCY
\end{lstcpp}
In this example, the dependency on \metavar{capability} would not only be activated if \metavar{model\_$\alpha$} or \metavar{model\_$\beta$} were in use, but also if either backend \metavar{be\_name1} or backend \metavar{be\_name2} were used to resolve the backend requirement \metavar{requirement}.  In this case, the \lstinline{CONDITIONAL_DEPENDENCY} declaration must appear \textit{after} the corresponding backend requirement is declared.  Declaration of backend requirements is covered in Sec.\ \ref{declaration_bereq}.

There is currently no way to specify more complicated arrangements like `dependency is activated only if scanning \metavar{model\_$\alpha$} and using \metavar{backend\_name}' or `only if scanning both \metavar{model\_$\alpha$} and \metavar{model\_$\beta$}'.  Wanting to use such complicated scenarios is usually a sign that the intended design of the module function is unnecessarily complicated, and the function would be better just split into multiple functions with different properties.

\subsubsection{Backend requirements}
\label{declaration_bereq}

Backend requirements are declarations that a module function intends to use either a function or a global variable from a backend (external) code.  Backend requirements are specified in a similar way to dependencies: by declaring the type and the capability of the required backend function or variable (\textit{not} the name of a specific backend function).  In contrast to dependencies, however, the type of a backend requirement may be an entire function signature, describing not just the return type, but also the types of an arbitrary number of arguments.  Designating the capability of the backend variable required as \metavar{var\_requirement} and its required type \metavar{var\_type}, the declaration of a backend variable requirement is
\begin{lstcpp}
BACKEND_REQ(@\metavar{var\_requirement}@, (@\metavar{tags}@), @\metavar{var\_type}@)
\end{lstcpp}
If a backend function is required, with capability \metavar{fn\_requirement}, return type \metavar{fn\_return\_type} and function argument types \metavar{arg1\_type, arg2\_type} and so on, the declaration is instead
\begin{lstcpp}
BACKEND_REQ(@\metavar{fn\_requirement}@, (@\metavar{tags}@), @\metavar{fn\_return\_type}@,
 (@\metavar{arg1\_type}@, @\metavar{arg2\_type}@, ...))
\end{lstcpp}
Note that the final argument of \lstinline{BACKEND_REQ} should be absent for backend variable requirements, but should be explicitly specified as \lstinline{()} for backend functions with no arguments --- as is standard \plainC/\Cpp syntax.  The ellipses in the backend function example again indicate that the entry is variadic, so as many function arguments can be specified as required.  If the backend function is \textit{itself} required to be variadic (in the \plainC-style sense that the function required must be able to take a variable number of arguments), then instead of the traditional ellipses used to declare such a function, one must use the keyword \lstinline{etc}, as in
\begin{lstcpp}
BACKEND_REQ(@\metavar{fn\_requirement}@, (@\metavar{tags}@), @\metavar{fn\_return\_type}@,
 (@\metavar{arg1\_type}@, etc))
\end{lstcpp}

The \metavar{tags} entry in the declarations above allows one to specify a set of zero or more comma-separated tags, which can then be used to impose various conditions on how backend requirements can be filled.  Consider the following example:
\begin{lstcpp}
BACKEND_REQ(@\metavar{req\_A}@, (tag1), float, (int, int))
BACKEND_REQ(@\metavar{req\_B}@, (tag1, tag2), int, ())
BACKEND_REQ(@\metavar{req\_C}@, (tag3), int)
ACTIVATE_BACKEND_REQ_FOR_MODELS( (@\metavar{model\_$\alpha$}@,
 @\metavar{model\_$\beta$}@), (tag1) )
BACKEND_OPTION( (@\metavar{be\_name1}@), (tag1) )
BACKEND_OPTION( (@\metavar{be\_name2}@, 1.2, 1.3, 1.5),
 (tag2, tag3) )
FORCE_SAME_BACKEND(tag1)
\end{lstcpp}
In this example, the \lstinline{ACTIVATE_BACKEND_REQ_FOR_MODELS} directive ensures that \metavar{req\_A} and \metavar{req\_B} only exist as backend requirements when \metavar{model\_$\alpha$} and/or \metavar{model\_$\beta$} are in use.  \lstinline{FORCE_SAME_BACKEND} creates a rule that at runtime, both \metavar{req\_A} and \metavar{req\_B} must be filled using functions from the same version of the same backend.

Further rules are given by the \lstinline{BACKEND_OPTION} declarations.  The first of these indicates that \metavar{be\_name1} is a valid backend from which to fill one or both of \metavar{req\_A} and \metavar{req\_B}.  The second \lstinline{BACKEND_OPTION} declaration indicates that \metavar{req\_B} and \metavar{req\_C} may each be filled from versions 1.2, 1.3 or 1.5 only of \metavar{be\_name2}.  Version numbers here are both optional and variadic. Failure to list any version is taken to imply that any version of the backend is permitted.  Presently there is no mechanism for indicating that only specific ranges of version numbers are permitted, short of listing each one explicitly.  Version numbers can actually be specified in the same way when \lstinline{ACTIVATE_FOR_BACKEND} is specified within a \lstinline{CONDITIONAL_DEPENDENCY} declaration.

When \metavar{model\_$\alpha$} or \metavar{model\_$\beta$} is being scanned, the rules in this particular snippet have the effect of forcing \metavar{req\_A} to be filled from some version of \metavar{be\_name1} (due to the first \lstinline{BACKEND_OPTION} declaration), which in turn forces \metavar{req\_B} to be filled from the same version of \metavar{be\_name1} (due to the \lstinline{FORCE_SAME_BACKEND} directive).  If other models are scanned, \metavar{req\_A} and \metavar{req\_B} are simply ignored, and go unfilled. \metavar{Req\_C} is forced to be filled from either version 1.2, 1.3 or 1.5 of \metavar{be\_name2}, regardless of which models are scanned.

As with other \GB rollcall header commands, the lists of models and tags in all backend requirement declarations are variadic.  In this case there is practically no limit to the number of entries that a tag or model list may contain.  Empty lists \lstinline{()} are also permitted.

When a backend requirement has a rule imposed on it by one or more \lstinline{BACKEND_OPTION} declarations, one of the stated options \textit{must} be used.  When none of the tags of a given backend requirement is mentioned in a \lstinline{BACKEND_OPTION} command, any version of any backend is permitted as long as the capability and type match.  Simply omitting \lstinline{BACKEND_OPTION}{\xspace} altogether means that any matching function can be used, from any backend.

\subsubsection{Parallel module functions}
\label{declaration_loops}

\GB can make effective use of \omp parallelistaion either at the backend level, or natively within its own module functions.  The simplest way to use \omp at the module function level is to place \omp directives inside a single module function, keeping the \omp block(s) wholly contained within the module function.  In this case no special declarations are needed at the level of the module's rollcall header.

An alternative method is to have a single module function open and close an \omp block, and to call other module functions (indirectly) within that block, potentially very many times over for a single parameter combination.  In this case we refer to the managing module function as a \cross{loop manager} and the functions it calls \doublecross{nested module functions}{nested module function}.  Loop managers are declared using the \lstinline{CAN_MANAGE_LOOPS} switch
\begin{lstcpp}
START_FUNCTION(@\metavar{type}@, CAN_MANAGE_LOOPS)
\end{lstcpp}
Unlike regular module functions, loop managers may have \metavar{type} = \lstinline{void}.  Nested functions need to declare the capability of the loop manager that they require with
\begin{lstcpp}
NEEDS_MANAGER_WITH_CAPABILITY(@\metavar{management\_cap}@)
\end{lstcpp}
This declaration endows the function with a special dependency on \metavar{management\_cap} that can only be fulfilled by a function that has been declared as a loop manager.  The result type of the loop manager is ignored, i.e.\ loop managers of any return type are equally valid sources of this dependency.

This arrangement allows \GB's \cross{dependency resolver} to dynamically string together various nested module functions and instruct loop managers to run them in parallel.  At runtime, nested functions are arranged into their own mini dependency trees, and pointers to ordered lists of them are handed out to the designated loop managers.

Other functions can depend on nested functions in the regular way.  In this case they receive the final result of the function, the last time it is called by its loop manager for a given parameter combination.  Loop managers are assigned hidden dependencies at runtime by the dependency resolver, on all quantities on which their designated nested functions depend.  This ensures that a loop is not invoked until the dependencies of all functions in the loop have been satisfied.

The \GB \textsf{Core} does not invoke any nested functions itself; this is the express responsibility of loop managers.  The only exception to this rule occurs when for whatever reason a nested function's loop manager executes zero iterations of the loop it manages, but some other module function outside the loop depends on one of the nested functions that never ran; in this case the nested function is run the first time the dependent function tries to retrieve its value (as are any other nested functions that the first nested function depends on).

\subsubsection{One-line module function declaration}
\label{declaration_quick_function}

It is also possible to declare a module function with its allowed models and even dependencies, in a single line:
\begin{lstcpp}
QUICK_FUNCTION( @\metavar{module\_name}@,
    @\metavar{example\_capability}@,
    @\metavar{capability\_vintage}@,
    @\metavar{function\_name}@,
    @\metavar{function\_type}@,
    (@\metavar{model\_$\alpha$}@, @\metavar{model\_$\beta$}@, ...),
    (@\metavar{dep\_cap1}@, @\metavar{dep\_type1}@),
    (@\metavar{dep\_cap2}@, @\metavar{dep\_type2}@),
    ... )
\end{lstcpp}
Here one gives the module name explicitly, meaning that the declaration can even be used after \lstinline{MODULE} has been \mbox{\lstinline{#undef}-ed}.  The argument \metavar{capability\_vintage} tells \GB whether or not \metavar{example\_capability} has been declared previously; this can be set to either \lstinline{NEW_CAPABILITY} or \lstinline{OLD_CAPABILITY}.  As usual, the variadic allowed model list (\metavar{model\_$\alpha$}, \metavar{model\_$\beta$}, ...) can take up to 10 entries.  This can be followed by up to 10 dependencies, given as capability-type pairs.  The model list and dependency entries are optional arguments; specifying dependencies but leaving the allowed models free requires giving () for the allowed model list.

\subsection{Pipes}

Module functions must be entirely self-contained for \GB to safely place them in a dependency tree. They must not call each other directly, nor call functions from specific backends directly.  They should also strongly avoid setting or reading any global variables, especially those where the order of read or write operations might matter at all.  The only safe way for code inside module functions to communicate with the outside world is via the function's own personal set of \doublecross{pipes}{pipe}.

At runtime, \GB's \cross{dependency resolver} (Sec.\ \ref{depresolver}) connects each pipe to the relevant data stream that a module function is permitted to interact with.  This might be the result of a module function deemed appropriate for fulfilling a dependency, a backend function fitting a backend requirement, or some other more specific utility.

Pipes are safe pointers, automatically declared when module functions themselves are declared.  They and the data they point to can be set by the dependency resolver, but not by code inside module functions (except for the special case of data pointed to by a backend variable requirement pipe).  They reside in the namespace \cpp{Pipes::}\metavar{function\_name}.

Here we give a complete list of all the pipes that can be available to a module function, along with information on their usage and the circumstances under which they should be expected to exist.

\subsubsection{Accessing dependencies}
\label{accessing_dep}

A dependency on a \metavar{capability} of \metavar{dep\_type} can be accessed at runtime through the safe pointer
\begin{lstcpp}
Pipes::@\metavar{function\_name}@::Dep::@\metavar{capability}@
\end{lstcpp}
by simply dereferencing it, or calling \metavar{some\_member\_function} of class \metavar{dep\_type}
\begin{lstcpp}
using namespace Pipes::@\metavar{function\_name}@;
@\metavar{dep\_type}@ my_variable = *Dep::@\metavar{capability}@;
Dep::@\metavar{capability}@->@\metavar{some\_member\_function}@();
\end{lstcpp}
e.g. if the function \lstinline{decay_width} had a double-precision dependency on \lstinline{mass}, one would simply type
\begin{lstcpp}
double m = *Pipes::decay_width::Dep::mass;
\end{lstcpp}
The actual host module, name, capability and type of the function providing a dependency can be ascertained from its pipe, e.g.
\begin{lstcpp}
using namespace Pipes::decay_width;
std::string m_module    = Dep::mass->origin();
std::string m_function  = Dep::mass->name();
std::string m_capability= Dep::mass->capability();
std::string m_type      = Dep::mass->type();
\end{lstcpp}

\subsubsection{Accessing backend requirements}

Backend requirements can be used or retrieved by way of the safe pointer
\begin{lstcpp}
Pipes::@\metavar{function\_name}@::BEreq::@\metavar{requirement}@
\end{lstcpp}

Take the example of a double-precision backend variable with capability \lstinline{my_var_req}, declared in function \lstinline{my_func} with
\begin{lstcpp}
BACKEND_REQUIREMENT(my_var_req, (), double)
\end{lstcpp}
This variable is accessed directly as
\begin{lstcpp}
using namespace Pipes::my_func;
double y = 2.5 + *BEreq::my_var_req;
*BEreq::my_var_req = y*y;
\end{lstcpp}

In the case of a backend function, e.g.\ declared as
\begin{lstcpp}
BACKEND_REQUIREMENT(my_fn_req1, (), double,
 (double))
\end{lstcpp}
one can call the corresponding backend function by writing
\begin{lstcpp}
using namespace Pipes::my_func;
double f_of_pi = BEreq::my_fn_req1(3.14159);
\end{lstcpp}

If necessary, the actual underlying function or variable pointer can be retrieved from a backend requirement pipe, by calling its \lstinline{pointer()} method.  This can be useful if a module or backend function requires a pointer to some function in order to perform its duties, as in the following example from \lstinline{DarkBit::nuyield_from_DS}
\begin{lstcpp}
// Hand back the pointer to the DarkSUSY
// neutrino yield function
result.pointer = BEreq::nuyield.pointer();
\end{lstcpp}

There is an important final subtlety to note here: because the arguments are forwarded through a number of different layers of indirection, in order to support the direct use of literals in calls to backend functions it is necessary to indicate explicitly if any non-literal parameters must be passed by value.  The way to do this is to wrap such arguments in the helper function \lstinline{byVal()}.  For example, take a backend requirement of a function \lstinline{my_func} declared as
\begin{lstcpp}
BACKEND_REQUIREMENT(my_fn_req2, (), double,
 (double, double&))
\end{lstcpp}
This can be safely called as
\begin{lstcpp}
using namespace Pipes::my_func;
double x = 2.0;
double y = BEreq::my_fn_req2(3.0, x);
\end{lstcpp}
or
\begin{lstcpp}
using namespace Pipes::my_func;
double x = 2.0;
double y = BEreq::my_fn_req2(byVal(x), x);
\end{lstcpp}
but will fail to compile if
\begin{lstcpp}
using namespace Pipes::my_func;
double x = 2.0;
double y = BEreq::my_fn_req2(x, x);
\end{lstcpp}
is attempted.  The backend requirement system in \GB is entirely typesafe, so if the code compiles one can at least be confident that the types in calls to backend functions correctly match their declarations.

As with dependencies, the name, capability and type of the backend function fulfilling a backend requirement can be extracted from its pipe, along with the host backend and its version, e.g.
\begin{lstcpp}
using namespace Pipes::my_func;
std::string r3_function    = BEreq::r3->name();
std::string r3_capability  =
 BEreq::r3->capability();
std::string r3_type        = BEreq::r3->type();
std::string r3_backend     = BEreq::r3->origin();
std::string r3_bkend_versn = BEreq::r3->version();
\end{lstcpp}

\subsubsection{Accessing model parameters}
\label{param_pipe}

Model parameters are only provided to module functions that have been explicitly declared as model-dependent, and then only for the models actually being used in a particular scan.  A module function is model dependent if it features an \lstinline{ALLOWED_MODELS} or \lstinline{ALLOW_MODEL_DEPENDENCE} declaration, a model-conditional dependency, or a backend requirement activation rule that is conditional on some model.  Once again, this is to enforce modularity; functions that claim to be model-independent through their (lack of) model declarations must operate only on dependencies and backend requirements, i.e. without using the values of the scanned parameters.

For module functions that are permitted access to the parameter values, all parameters of all models are delivered in a simple map of parameter names to their values.  For such a function \metavar{function\_name}, the value of a parameter \metavar{parameter\_name} can then be retrieved with
\begin{lstcpp}
double p = Pipes::@\metavar{function\_name}@::Param["@\metavar{parameter\_}@
 @\metavar{name}@"];
\end{lstcpp}

Whether or not the \lstinline{Param} map contains a given parameter depends on whether or not its model is actually being scanned.  This can be checked with the funtion
\begin{lstcpp}
bool Pipes::@\metavar{function\_name}@::ModelInUse(str);
\end{lstcpp}
which takes as input a string containing the model in question (\lstinline{str} is just a typedef of \lstinline{std::string}).  Note that the models in use in different functions may not always be what one expects --- the nature of the \GB model hierarchy is such that if a module function declares that it can work with a model that is an \textit{ancestor} of the actual model being scanned, the function will be permitted to run but will receive each parameter point delivered \textit{in terms of the parameters of the ancestor model}, rather than directly in the parameters of the model actually being scanned.\footnote{Note that if a module function is explicitly declared to work with \textit{multiple} ancestors of the model being scanned, then only the parameters of the least-distant ancestor will be delivered.  These rules also apply for activation of model-dependent depedencies and backend requirements (cf.\ Secs.\ \ref{declaration_deps} and \ref{declaration_bereq}).}  This is an important feature, as it allows module functions to be re-used unaltered with models that may not have even been invented when the original module function was written.

Although it is possible to scan two models containing a parameter with a common name, it is not possible to retrieve both parameters from the \lstinline{Param} map in the same module function.  By default, \GB raises a runtime error if models with common parameters are declared as allowed (by \lstinline{ALLOWED_MODELS} or \lstinline{ALLOW_MODEL_DEPENDENCE}) in a single module function, and then activated together in a scan.  More adventurous users may wish to deactivate this error and allow such parameter clashes in some very specific circumstances (see Sec.\ \ref{Parameters}).

\subsubsection{Accessing options from the input file}
\label{module_options}

\GB features an extensive system for specifying run options for module functions, discussed in detail in Sec.\ \ref{interface}.  Module functions access these options via a dedicated pipe, which connects to a miniature \YAML structure generated by the dependency resolver from all the entries in the original input \YAML file that actually apply to the module function in question.

The pipe is \lstinline{runOptions}.  It can be queried for the presence of a given option \cpp{"my_option"}
\begin{lstcpp}
using namespace Pipes::@\metavar{function\_name}@;
if (runOptions->hasKey("my_option"))
{
  // Do something exciting
}
\end{lstcpp}
or used to retrieve the value as a variable of type \mbox{\metavar{opt\_type},} either directly
\begin{lstcpp}
using namespace Pipes::@\metavar{function\_name}@;
@\metavar{opt\_type}@ x = runOptions->getValue<@\metavar{opt\_type}@>
 ("my_option");
\end{lstcpp}
or with a default value \metavar{default}
\begin{lstcpp}
using namespace Pipes::@\metavar{function\_name}@;
@\metavar{opt\_type}@ x = runOptions->getValueOrDef<@\metavar{opt\_type}@>
 (@\metavar{default}@, "my_option");
\end{lstcpp}

\subsubsection{Managing parallel module functions}
\label{loopmanagement}

Running \omp loops containing \GB module functions takes a little care, but it is ultimately one of the most efficient ways to speed up computationally challenging likelihood calculations.

A loop manager \metavar{lpman} is responsible for opening and closing the multi-threaded \omp block.  Inside the block, it needs to use the void function
\begin{lstcpp}
Pipes::@\metavar{lpman}@::Loop::executeIteration(long long)
\end{lstcpp}
to execute a single call to the chain of nested functions that it manages.  Here the integer argument of the function is the iteration number, which is passed directly on to each nested function running inside the loop. A nested function \metavar{nested\_fn} can access the iteration using the pipe \lstinline{iteration} as
\begin{lstcpp}
long long it = *Pipes::@\metavar{nested\_fn}@::Loop::iteration;
\end{lstcpp}
Internally, \GB holds the results of each module function in memory, for efficiently handing over results as dependencies and so on. For nested functions, it holds the results in an array of size equal to the number of threads.  Serial module functions access the first element of this array when retrieving dependencies, whereas nested module functions run by the same loop manager access the element corresponding to the thread number.  This is what allows the nested module functions to run safely in parallel, in arbitrary dependency trees arranged by the dependency resolver at runtime.

A consequence of this setup is that any serial module function that depends on a nested module function will only read the result obtained in the last iteration of the first thread (i.e. of index 0).  For this reason, it is generally advisable to run the final iteration of a \GB \omp loop in serial, so as to properly sync the results for use further `downstream'.  Likewise, it is desirable to run the first iteration in serial as well, to allow any nested module functions to initialise any local static variables and other data elements that they might share across threads.  With this consideration in mind, a minimal example of an \omp loop implemented in a loop manager is
\begin{lstcpp}
using namespace Pipes::@\metavar{lpman}@;
Loop::executeIteration(0);
#pragma omp for
for (int i = 1; i < 9; i++)
{
  Loop::executeIteration(i);
}
Loop::executeIteration(9);
\end{lstcpp}
In this example, the first iteration of ten is run serially, the next 8 are done in parallel using however many threads are available, and the tenth and final iteration is again done serially.

The above example assumes that the number of required iterations is known at compile time.  If this is not the case, one may call the void function pipe \lstinline{wrapup()} from within a nested function, in order to signal to the loop manager that the loop can be terminated.  When one of the nested module functions in one of the threads calls \mbox{\lstinline{wrapup()},} the boolean pipe
\begin{lstcpp}
Pipes::@\metavar{lpman}@::Loop::done
\end{lstcpp}
in the function managing the loop is set to true, allowing it to cut the loop short.  This allows constructions like
\begin{lstcpp}
using namespace Pipes::@\metavar{lpman}@;
long long it = 0;
Loop::executeIteration(it);
#pragma omp parallel
{
  #pragma omp atomic
  it++;
  while(not *Loop::done)
  {
    Loop::executeIteration(it);
  }
}
Loop::executeIteration(it++);
\end{lstcpp}
to be used in loop managers.  Note that using this pattern requires that it be safe for a few more iterations of the loop to be performed after the \lstinline{done} flag has been raised, because calling \lstinline{wrapup()} in one thread will not affect other threads until they at least complete their own iterations and return to re-evaluate the while condition.  The final serial iteration should generally also still be run as well, after the loop has terminated.

The \lstinline{done} flag is automatically reset to \lstinline{false} in all nested functions for each new parameter point.  If for whatever reason it needs to be reset manually during a calculation, this can be achieved with the void function pipe
\begin{lstcpp}
Pipes::@\metavar{lpman}@::Loop::reset()
\end{lstcpp}
which is available in all loop managers.

\section{Backends}
\label{backends}

\GB interfaces with backends by using them as runtime plug-ins. Backends are compiled into shared libraries, which \GB then dynamically loads with the \posix-standard \term{dl} library. This approach allows for direct access to the functions of the backend library and efficient data communication via memory, while at the same time keeping the build process of \GB separate from of that of the particular backends used.

The locations of backend shared libraries can be specified in a \YAML file \term{config/backend_locations.yaml}, with entries of the form
\begin{lstyaml}
@\metavar{backend\_name}@:
  @\metavar{backend\_version}@:  @\metavar{path\_to\_shared\_library}@
\end{lstyaml}
Library paths can either be given as absolute paths, or relative to the main \GB directory. If \term{backend_}\term{locations.yaml} does not exist, or if it is missing an entry for a given backend, \GB will instead look for a path in the file \term{config/backend_locations.yaml.default}, which contains default library paths for all backends that \GB has interfaces to.

Similar to module functions, functions in backend libraries are tagged with a \cross{capability} describing what they can calculate. The capability tags are used by \GB to match backend functions to the backend requirements declared by module functions. The layer of abstraction introduced by these tags allows appropriately designed module functions to use different backends interchangeably, given that they calculate the same quantity.

\GB can currently communicate with backends written in \plainC, \Cpp and \Fortran. However, we must pay some attention to the differences between these languages. In particular, using a \Fortran backend requires translating between standard \Fortran and \plainC-family types, and using a \Cpp backend typically involves loading entirely new types from the \Cpp library. We return to these topics in Secs.\ \ref{backend_types} and \ref{boss}.

The interface to a backend library is declared in a frontend header file, located in
\begin{lstterm}
Backends/include/gambit/Backends/frontends
\end{lstterm}
and named after the backend. Thus, a backend called \textsf{MyBackend} would be traditionally interfaced with \GB via a frontend header \term{MyBackend.hpp}. To differentiate multiple versions of the same backend, the version number can be appended to the header name, e.g. \term{MyBackend\_1\_2.hpp} for version 1.2 of \textsf{MyBackend}. Applications such as this, where the periods in the version number are replaced with underscores, make use of what we refer to as the \cross{safe version} of a backend, i.e.\ a representation of the version number that is safe to use when periods would be syntactically hazardous. A frontend header starts by defining the name, version and safe version of the backend, then immediately calls the \lstinline|LOAD_LIBRARY| macro, which takes care of loading the backend shared library:
\begin{lstcpp}
#define BACKENDNAME MyBackend
#define VERSION 1.2
#define SAFE_VERSION 1@\_@2
LOAD_LIBRARY
\end{lstcpp}

\subsection{Backend function declaration}
\label{declaration_be_func}

The main pieces of information required to interface a backend function to \GB are its return type, call signature and library symbol.  The name mangling schemes of \textsf{g\xx}/\textsf{gfortran} and \textsf{icpc}/\textsf{ifort} (the two main compiler suites that \GB supports; cf.\ Appendix \ref{dependencies}) are mostly consistent, so a single symbol name can usually be entered here for both compilers.\footnote{The symbols of a shared library, with names prepended by an additional underscore, can be obtained using the \term{nm} command.  Functions within \Fortran modules are an exception to the consistency of name mangling.  The best way to deal with these is often to use the \plainC-interoperability features of \Fortran to explicitly choose a symbol name, taking the choice out of the hands of the compiler.  An example of this can be seen in \ddcalc \cite{DarkBit}.  In future, \GB will automatically determine the appropriate name mangling itself, according to the scheme of the selected compiler.} In addition, the function must be assigned a name and a capability. This is all specified via the \lstinline|BE_FUNCTION| macro. For instance, a \plainC/\Cpp backend function with the declaration
\begin{lstcpp}
double getMatrixElement(int, int);
\end{lstcpp}
could be registered in the frontend header as
\begin{lstcpp}
BE_FUNCTION(getMatrixElement, double, (int, int),
 "_Z13getMatrixElementii",
 "rotation_matrix_element")
\end{lstcpp}
where \lstinline|"_Z13getMatrixElementii"| is the library symbol and \lstinline|"rotation_matrix_element"| is the capability we assign to this function.

The macro \lstinline|BE_VARIABLE| used to interface global variables in a backend library follows a similar syntax. If the backend contains a global variable
\begin{lstcpp}
double epsilon;
\end{lstcpp}
controlling the tolerance of some calculation, it can be registered as
\begin{lstcpp}
BE_VARIABLE(epsilon, double, "_epsilon",
 "tolerance")
\end{lstcpp}
with \lstinline|"_epsilon"| the library symbol and \lstinline|"tolerance"| the capability assigned to the variable.

Backend functions and variables can be declared as either model-independent or valid for use only with certain models, just like module functions can.  The default is to treat everything as model-independent.  To declare an alternative default that applies to an entire backend, one places
\begin{lstcpp}
BE_ALLOW_MODELS(@\metavar{model\_$\alpha$}@, @\metavar{model\_$\beta$}@, ...)
\end{lstcpp}
directly after \cpp{LOAD_LIBRARY}.  This has the effect of allowing the entire backend to be used only if one or more of the listed models is involved in a scan.  This default can be further overridden at the level of individual backend variables and backend functions, by adding additional model arguments to their declarations:
\begin{lstcpp}
BE_FUNCTION(getMatrixElement, double, (int,int),
            "_Z13getMatrixElementii",
            "rotation_matrix_element",
            (@\metavar{model\_$\alpha$}@, @\metavar{model\_$\beta$}@, ...) )
BE_VARIABLE(epsilon, double,
            "_epsilon",
            "tolerance",
            (@\metavar{model\_$\alpha$}@, @\metavar{model\_$\beta$}@, ...) )
\end{lstcpp}

\subsection{Convenience functions}
\label{be_conv_func}

If several backend function calls or variable manipulations are commonly performed together, it may be useful to combine these into a single backend convenience function. Althought technically defined inside \GB, this function will appear to the rest of \GB as if it were simply another function in the backend library. Convenience functions are registered in the frontend header with the \lstinline|BE_CONV_FUNCTION| macro. The syntax is identical to that of \lstinline|BE_FUNCTION| except that there is no need to specify a library symbol, as the convenience function is not actually part of the backend library.
\begin{lstcpp}
BE_CONV_FUNCTION(getMatrix, Matrix<double,2,2>,
 (), "full_rotation_matrix")
\end{lstcpp}
The definition of the convenience function can then either be given directly in the frontend header, or in a separate source file named after the backend, e.g., \mbox{\term{MyBackend.cpp},} and placed in the directory
\begin{lstterm}
Backends/src/frontends
\end{lstterm}
In either case, the function definition must be placed inside a designated namespace \cpp{Gambit::Backends::}\metavar{backend\_name}\cpp{_}\metavar{safe\_version}, automatically generated with the \lstinline|BE_NAMESPACE| and \lstinline|END_BE_NAMESPACE| macros.
\begin{lstcpp}
BE_NAMESPACE
{
  Matrix<double,2,2> getMatrix()
  {
    // Call getMatrixElement four times
    // and return a complete matrix.
  }
}
END_BE_NAMESPACE
\end{lstcpp}
All backend functions and variables registered with the\\ \lstinline|BE_FUNCTION| and \lstinline|BE_VARIABLE| macros (in the same frontend) can be accessed directly in convenience functions, as long as the body of the convenience function appears after their declarations.  This also applies to calling convenience functions from each other.

Just like backend variables and regular backend functions, backend convenience functions can be declared as model-dependent, e.g.
\begin{lstcpp}
BE_CONV_FUNCTION(getMatrix, Matrix<double,2,2>,
 (), "full_rotation_matrix", (@\metavar{model\_$\alpha$}@,
 @\metavar{model\_$\beta$}@, ...) )
\end{lstcpp}

\subsection{Backend initialisation functions}
\label{be_ini_func}

A backend library will usually have to be initialised in some way before any calculations can be performed. For instance, variables storing masses and couplings may have to be reset for every new parameter point. For this purpose, the user can define a backend initialisation function. This is a special kind of convenience function that automatically runs prior to any other backend operations.  An initialisation function is registered by enclosing it within \lstinline|BE_INI_FUNCTION| and \lstinline|END_BE_INI_FUNCTION|. These macros automatically set up a void function taking no input arguments, so the user only has to supply the function body. As for backend convenience functions, this function definition can be placed either in the frontend header file or in the corresponding source file.
\begin{lstcpp}
BE_INI_FUNCTION
{
  // Point-level initialisation.
}
END_BE_INI_FUNCTION
\end{lstcpp}
If some part of the initialisation only has to happen once for an entire scan, this can be accomplished by using a static flag:
\begin{lstcpp}
BE_INI_FUNCTION
{
  static bool scan_level = true;
  if(scan_level)
  {
    // Scan-level initialisation.
  }
  scan_level = false;

  // Point-level initialisation.
}
END_BE_INI_FUNCTION
\end{lstcpp}
As with convenience functions, all registered backend functions and variables from the same backend are directly accessible from within the body of initialisation functions, so long as the body appears after the functions and variables have been declared.

To help with scan-level initialisation, \GB provides a flag for every registered backend function, variable and convenience function, indicating whether or not it will be used in the upcoming scan.  These flags are accessible only from a backend's initialisation function. The flags consist of pointers to boolean variables placed in the \cpp{InUse} namespace, i.e.
\begin{lstcpp}
bool *InUse::@\metavar{name}@
\end{lstcpp}
where \metavar{name} is the name of the backend function, variable or convenience function as declared in the frontend header. Some example usage of the backend function \cpp{InUse} flags can be found in the fronted source files for \nulike \cite{IC79_SUSY,IC22Methods} and \ddcalc \cite{DarkBit}.

Some backends write temporary files to disk during scan-level initialisation, which means that they cannot be safely initialised simultaneously in different \mpi processes.\footnote{This is also to be discouraged on basic  efficiency grounds.}  For such cases we provide a simple locking utility (\cpp{Utils::FileLock}) that can be employed to force serial execution of any block of code; example usage can be seen in the frontends to \textsf{HiggsBounds} and \textsf{HiggsSignals}~\cite{Bechtle:2008jh,Bechtle:2011sb,Bechtle:2013wla}.

In fact, backend initialisation functions are actually promoted to module function status, and associated with a special \GB-internal module called \textsf{BackendIniBit}.  This is how \GB ensures that they always run before any other functions from their backend are used.  This also allows backend initialisation functions to depend on input from other \GB module functions. This is declared using the \lstinline|BE_INI_DEPENDENCY| and \lstinline|BE_INI_CONDITIONAL_DEPENDENCY| macros.  These follow exactly the same syntax as the \lstinline|DEPENDENCY| and \lstinline|MODEL_CONDITIONAL_DEPENDENCY| macros for module functions (Sec.\ \ref{declaration_deps}):
\begin{lstcpp}
BE_INI_DEPENDENCY(@\metavar{capability}@, @\metavar{type}@)
BE_INI_CONDITIONAL_DEPENDENCY(@\metavar{capability}@, @\metavar{type}@, @\metavar{model\_$\alpha$}@, @\metavar{model\_$\beta$}@, ...)
\end{lstcpp}
Thus, a backend initialisation function that needs to know the particle spectrum for the given parameter point could declare a dependency similar to
\begin{lstcpp}
BE_INI_DEPENDENCY(particle_spectrum, Spectrum)
\end{lstcpp}
This will be fulfilled if some module function can provide the capability \lstinline|particle_spectrum| of type \mbox{\lstinline|Spectrum|}. The dependency can then be accessed from within the function body of the initialisation function,
\begin{lstcpp}
const Spectrum& my_spec = *Dep::particle_spectrum;
\end{lstcpp}
This is similar to the way module functions access their dependencies (Sec.\ \ref{accessing_dep}), except that for backend initialisation functions there is no need to specify the namespace \cpp{Pipes::}\metavar{function\_name}.

\subsection{Backend types}
\label{backend_types}

Backend functions and variables will often require types that are not known to \GB, and which therefore need to be defined. For \plainC and \Fortran backends, these types are typically structs or typedefs involving only built-in \plainC types. In this case the required definitions can be placed directly in a designated backend types header, named after the backend and placed in
\begin{lstterm}
Backends/include/gambit/Backends/backend_types
\end{lstterm}
The types must live within the \lstinline|Gambit| namespace, e.g.,
\begin{lstcpp}
namespace Gambit
{
  struct Triplet
  {
    double x, y, z;
  };
}
\end{lstcpp}
but additional sub-namespaces can of course be used.

To ease the process of generating these type declarations and the \cpp{BE_FUNCTION} and \cpp{BE_VARIABLE} declarations that use them, \GB ships with a simple utility for parsing \Fortran backend code: \textsf{CBGB}, the Common Block harvester for \GB Backends.  \textsf{CBGB} automatically generates \GB code that declares the necessary backend types, functions and variables, according to the list of functions and common blocks that a user chooses to interface with \GB. \textsf{CBGB} is written in \Python and can be found in \term{Backends/scripts/CBGB}.

\textsf{CBGB} takes a single configuration file as input. This file is written in \python syntax and must be placed in \term{Backends/scripts/CBGB/configs}. An annotated example detailing all options and variables can be found in
\begin{lstterm}
Backends/scripts/CBGB/configs/example.py.
\end{lstterm}
The most important variables to set in the configuration file are the three lists \py{input_files}, \py{load_functions} and \py{load_common_blocks}. We illustrate their use with a simple example, assuming a Fortran backend \textsf{FortranBE} v1.1:
\begin{lstpy}
input_files =
  ["../../installed/FortranBE/1.1/src/main.f90"]
load_functions = ["f1", "f2"]
load_common_blocks = ["cb"]
\end{lstpy}
Here \textsf{CBGB} would parse the Fortran file \term{main.f90} and generate the \cpp{BE_FUNCTION} declarations needed to load the functions/subroutines \fortran{f1} and \fortran{f2}, as well as the type and \cpp{BE_VARIABLE} declarations required to load the common block \fortran{cb}. The file paths in \py{input_files} must either be absolute paths or relative to the \term{Backends/scripts/CBGB} directory. To ensure that the library symbol names used in \cpp{BE_FUNCTION} and \cpp{BE_VARIABLE} match those in the backend shared library, \textsf{CBGB} must also know which name mangling scheme to use. This is specified via the variable \py{name_mangling}, which can be set to either \py{"gfortran"}, \py{"ifort"} or \py{"g77"}.

Once the configuration file is ready, \textsf{CBGB} can be run by passing in this file as a command line argument, e.g.\
\begin{lstterm}
python cbgb.py configs/FortranBE.py
\end{lstterm}
The generated \GB code is stored in the output files \term{backend_types_code.hpp} and \term{frontend_code.hpp}. In this example, the code in \term{backend_types_code.hpp} should be used in the backend types header \term{Backends/include/} \term{gambit/Backends/backend_types/FortranBE_1_1.hpp}, while the code in \term{frontend_code.hpp} should go in the frontend header \term{Backends/include/gambit/Backends/} \term{frontends/FortranBE_1_1.hpp}.

As \GB itself is written in \Cpp, interfacing with a \Fortran backend requires translation between the \Fortran types used in the backend and the corresponding \plainC-family types. Therefore, \GB provides several \Fortran-equivalent types and typedefs for use in communicating with \Fortran backends, with names indicating which \Fortran type they correspond to:
\begin{lstcpp}
Flogical
Flogical1
Finteger
Finteger2
Finteger4
Finteger8
Freal
Freal4
Freal8
Freal16
Fdouble
Fdoubleprecision
Fcomplex
Fcomplex8
Fcomplex16
Fdouble_complex
Flongdouble_complex
Fcharacter
\end{lstcpp}
These are the types that \textsf{CBGB} makes use of in the generated \GB code. In cases where \textsf{CBGB} fails to correctly parse the Fortran code, the user must manually specify type, \cpp{BE_VARIABLE} and \cpp{BE_FUNCTION} declarations using the above Fortran-equivalent types.

There are important differences in how arrays are treated in \Fortran compared to \plainC/\Cpp. First, the lower array index in \Fortran is by default 1, in contrast to \plainC/\Cpp arrays, which count from 0. More generally, \Fortran allows the user to specify arbitrary index ranges, something that is not allowed in \plainC/\Cpp. In the case of multidimensional arrays, \plainC/\Cpp arrays are stored in memory in row-major order, whereas \Fortran arrays use column-major ordering, and the two types of arrays are therefore effectively transposed with respect to each other. To save the user from having to deal with these complexities, \GB provides an \lstinline|Farray| class for working with \Fortran arrays.  This class provides basic \Fortran array semantics directly in \Cpp code. The class is templated on the array type and index ranges. Thus, a two-dimensional integer array with index ranges 1--3 and 1--4 can be declared as
\begin{lstcpp}
Farray<Finteger,1,3,1,4> my_f_array;
\end{lstcpp}

We also provide a special \lstinline|Fstring| class for working with \Fortran strings. It takes the string length as a template argument
\begin{lstcpp}
Fstring<4> my_f_string;
\end{lstcpp}
Similar to regular \Fortran strings, any string longer than the specified length will be truncated, and shorter strings will be padded with trailing spaces.

More information about the \GB \Fortran compatibility types can be found in the in-code \GB documentation (cf.\ Sec.\ \ref{component_databases}), and in \term{Utils/include/gambit/Utils/util_types.hpp}.

\subsection{Loading \textsf{C{\smaller ++}} classes at runtime with BOSS}
\label{boss}

Most physics tools written in \plainC or \Fortran are fundamentally just collections of functions and variables of standard types. In contrast, \Cpp tools typically define a number of new classes for the user to work with. Unfortunately, there exists no standard way of loading an {\it arbitrary} \Cpp class from a shared library at runtime. The \term{dl} library, itself written in \plainC, only provides access to functions and global variables. This limitation can be overcome if the main application has a predefined class interface that classes in the shared library are forced to adhere to; this is the so-called `factory' pattern. This is unproblematic as long as all plug-ins are developed after the main application, which is normally the case. In \GB, however, we face the reverse problem of turning already existing \Cpp physics tools into plug-ins for \GB. To solve this problem we have developed the \python-based Backend-On-a-Stick Script (\doublecrosssf{BOSS}{BOSS}), which we describe here.

Strategies for runtime loading of classes are essentially always based on the \Cpp concept of {\it polymorphism}. One constructs a class interface from a base class containing a set of {\it virtual} member functions. These are functions for which the signature is defined, but where the actual implementation is expected to be overridden by classes that inherit from the base class. The idea can be illustrated by considering a base class \cpp{Polygon} containing a virtual member function \cpp{calculateArea}. From this base class two derived classes \cpp{Triangle} and \cpp{Rectangle} can be defined. Both classes should contain a \cpp{calculateArea} member function, but their implementations of this function would differ.

In plug-in, i.e. factory-based, systems, the main application defines the base class, while the plug-ins provide the specialized classes deriving from the base class. The main application can then be designed with the assumption that any future class passed in from a plug-in will have the predefined set of member functions, whose implementations live in the shared library that is loaded at runtime. The shared library also contains {\it factory functions}, one for each class it provides. These are functions that return a pointer to a newly created instance of a plug-in class. When a new class instance is required, the main application calls the correct factory function and interprets the pointer it receives as pointing to an instance of the known base class.

\begin{figure}
\centering
\includegraphics[width=\columnwidth]{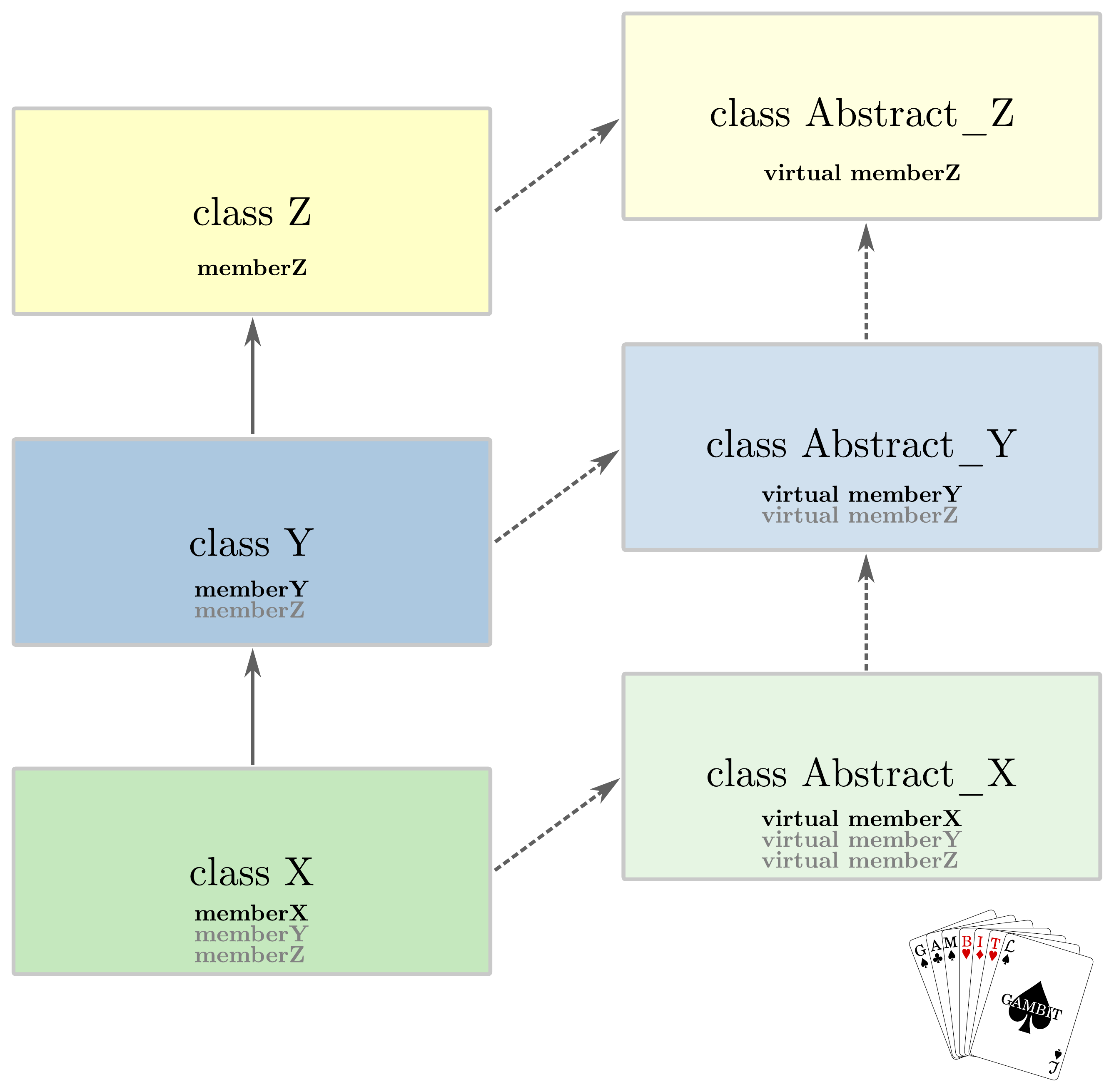}
\caption{The basic class `ladder' pattern generated by \BOSS in order to allow runtime loading of classes \cpp{X}, \cpp{Y} and \cpp{Z}, where \cpp{Z} is the parent of \cpp{Y}, which is in turn the parent of \cpp{X}. The original class hierarchy is mirrored by the abstract parent classes generated by \BOSS. Virtual inheritance, illustrated here by dashed arrows, is used to avoid ambiguities. Member functions in the original classes are matched by pure virtual member functions in the abstract classes.}
\label{fig::boss_class_pattern}
\end{figure}

The purpose of \BOSS is to reverse-engineer such a plug-in system for every backend class that is to be used from \GB. Starting from a class \cpp{X} defined in the backend library, \BOSS must generate source code for a base class with matching pure virtual member functions, as well as code for factory functions corresponding to the constructors \cpp{X(...)}. The generated base class is called \mbox{\cpp{Abstract\_X},} as classes containing pure virtual member functions are generally referred to as {\it abstract classes}. The source code for \cpp{Abstract\_X} is added to both the backend source code and to \GB. On the backend side, some additional source code is also inserted in the original class \mbox{\cpp{X},} most importantly adding \cpp{Abstract\_X} to the inheritance list of \cpp{X}. If class \cpp{X} originally inherits from a parent class \cpp{Y}, the abstract classes generated by \BOSS mirror this structure.  The resulting `ladder pattern' is illustrated in Fig.\ \ref{fig::boss_class_pattern}.

When the ladder structure is complete, the basic ingredients for a plug-in system are in place. However, from the user perspective there are several limitations and inconveniences inherrent in such a minimal system. For example, the factory functions must be called to create class instances, and class member variables cannot be accessed directly.  To overcome such limitations, \BOSS generates an additional layer in the form of an \textit{interface class}, which mimics the user interface of the original class. It is this interface class that a user interacts with from within \GB. The generated class is placed in a namespace constructed from the backend name and version, so if our example class \cpp{X} is part of \textsf{MyBackend} v1.2 the full name of the interface class will be \cpp{MyBackend_1_2::X}. However, from within a \GB module function using this class, the shorter name \cpp{X} can be used.

\begin{figure}
\centering
\includegraphics[width=\columnwidth]{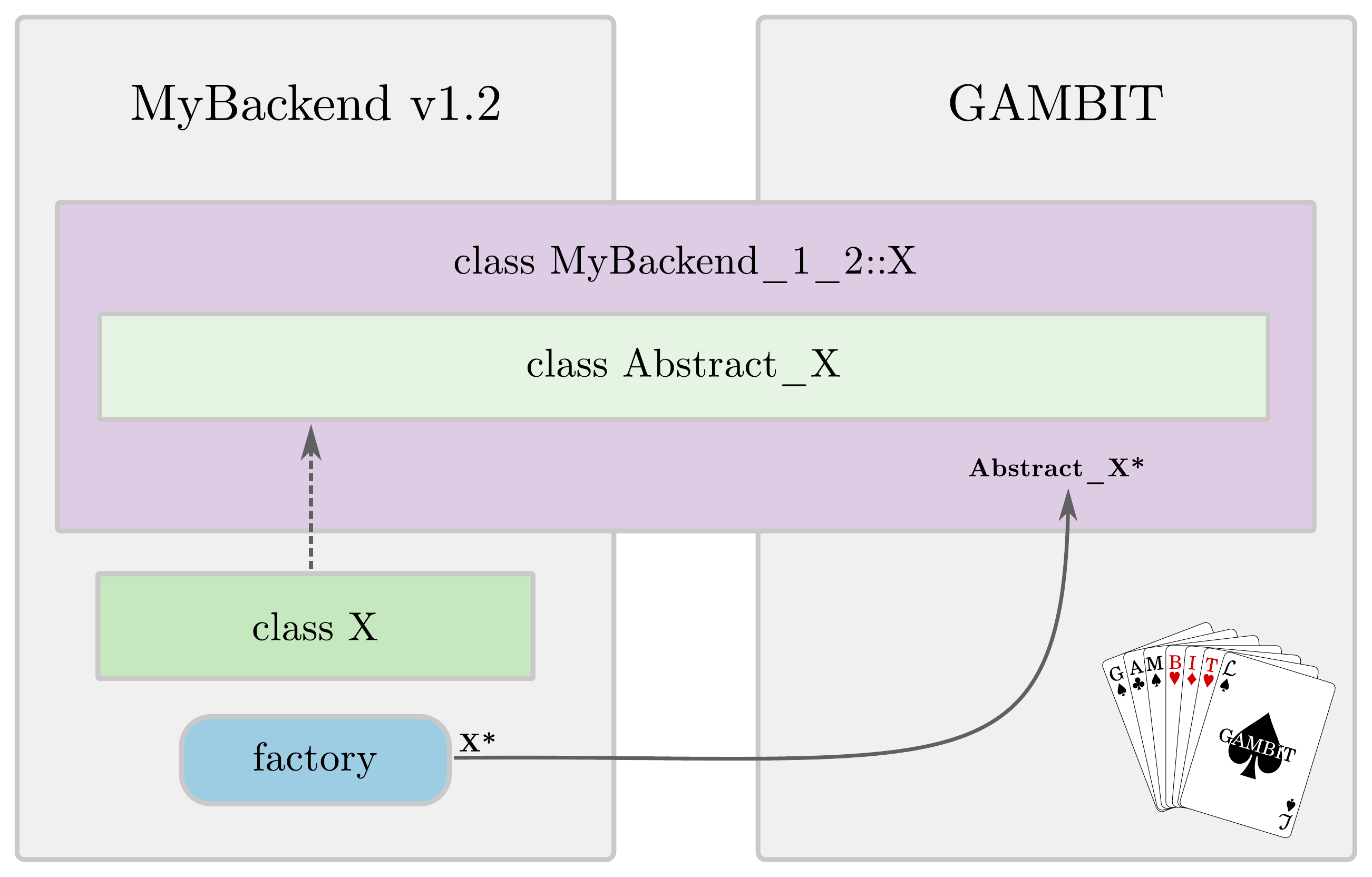}
\caption{Schematic diagram of the plug-in system generated by \BOSS for the case where a backend library \textsf{MyBackend 1.2} contains a single class \cpp{X}. For every constructor in \cpp{X}, a factory function returning a pointer to a new \cpp{X} instance is added to the library. An abstract base class \cpp{Abstract\_X} and an interface class \cpp{MyBackend_1_2::X} are generated and added to both the backend library and \GB. The interface class \cpp{MyBackend_1_2::X} wraps a pointer to \cpp{Abstract\_X}. The factory function initialises this pointer with an instance of \cpp{X}, allowing \GB to communicate with the original library class.}
\label{fig::boss}
\end{figure}

Fundamentally, the interface class is just a wrapper for a pointer to the abstract class. Through a factory function, this pointer is initialised to point to an instance of the orginal class, thus establishing the connection between \GB and the original class living in the backend library. In the example considered above, the class \cpp{MyBackend_1_2::X} would hold a pointer of type \mbox{\cpp{Abstract\_X},} pointing to an instance of \cpp{X}. This system is illustrated in Fig.\ \ref{fig::boss}. Note that the source code for the interface class is also inserted into the backend library. This allows \BOSS to generate wrapper functions for any global library functions where the original class appears in the declaration.

When a \GB module function requires classes from a backend library, this must be specified in the function's rollcall header entry by adding the macro
\begin{lstcpp}
NEEDS_CLASSES_FROM(@\metavar{backend\_name}@, @\metavar{versions}@)
\end{lstcpp}
Here \metavar{versions} is an optional comma-separated list of permitted backend version numbers. If \metavar{versions} is left out or set to \mbox{\cpp{default},} \GB will use the default backend version specified in the header file \term{Backends/include/} \mbox{\term{gambit/Backends/default_bossed_versions.hpp}}. Here a default version can be chosen by setting a precompiler variable \cpp{Default\_}\metavar{backend\_name} to the desired safe version number, e.g.\
\begin{lstcpp}
#define Default_MyBackend 1@\_@2
\end{lstcpp}

\BOSS itself is the stand-alone \python program
\begin{lstterm}
Backends/scripts/BOSS/boss.py
\end{lstterm}
For parsing the backend library source code \BOSS employs the open-source tool \textsf{CastXML}.\footnote{\url{http://github.com/CastXML/CastXML}} The basic input to \BOSS is a configuration file, written in \python, containing information about the backend library code that is to be `BOSSed'. The configuration file should be placed in the \term{configs} subdirectory of the main \BOSS directory. Here we will briefly go through the most important parts of the configuration file. For a complete list of options and variables we refer the reader to the example
\begin{lstterm}
Backends/scripts/BOSS/configs/Example_1_234.py
\end{lstterm}

First the name and version number that \GB should associate with the BOSSed library is set via the two variables \yaml{gambit_backend_name} and \yaml{gambit_backend_version}.
\begin{lstpy}
gambit_backend_name    = "MyBackend"
gambit_backend_version = "1.2"
\end{lstpy}
Then follows a set of path variables. All paths must be given either as absolute paths or relative to the main \BOSS directory. Consider the following example:
\begin{lstpy}
input_files =
    ["../../installed/MyBackend/1.2/include/X.hpp"]
include_paths =
    ["../../installed/MyBackend/1.2/include"]
base_paths = ["../../installed/MyBackend/1.2"]
\end{lstpy}
Here we assume that our example backend \textsf{MyBackend 1.2} is located in
\begin{lstterm}
Backends/installed/MyBackend/1.2
\end{lstterm}
The \py{input_files} variable is a list of the header files that contain the declarations for the classes and functions that are to be used from \GB. Next, \py{include_paths} lists the paths where \textsf{CastXML} should search for any header files that are included from one of the input files. Finally, \py{base_paths} is a list of the base directories of the backend library. This is used by \BOSS to differentiate between classes that are native to the backend and classes that are pulled in from external libraries.

\BOSS generates a number of header and source files that must be included when the BOSSed backend is compiled into a shared library. The output paths for these files are set with the variables \py{header_files_to} and \py{src_files_to}, for instance
\begin{lstpy}
header_files_to =
    "../../installed/MyBackend/1.2/include"
src_files_to = "../../installed/MyBackend/1.2/src"
\end{lstpy}

The next two variables, \py{load_classes} and \py{load_functions}, are lists containing the fully qualified names of the classes and functions to be loaded for use in \GB. If we assume that in addition to the class \cpp{X}, \textsf{MyBackend} also contains a global function \cpp{addX} for adding two instances of \mbox{\cpp{X},} we may have
\begin{lstpy}
load_classes = ["X"]
load_functions = ["addX(X, X)"]
\end{lstpy}

Typically users will only need access to a subset of all the classes defined in the library, so only a subset of the available classes will be listed in \py{load_classes}. \BOSS will then automatically limit the user interface of the BOSSed library to make sure that only functions and variables that make use of the loaded library classes and standard \Cpp types are accessible from \GB. However, if the backend library includes some classes that are also independently included in \GB, functions and variables relying on these classes should also be allowed as part of the BOSSed library interface. Such classes can be listed in the dictionary \py{known_classes}. Here the dictionary key is the class name and the corresponding value is the header file where the class is declared.

\BOSS is run by passing in the configuration file as a command line argument. For instance, with a configuration file \term{configs/MyBackend_1_2.py}, the command is simply
\begin{lstterm}
python boss.py configs/MyBackend_1_2.py
\end{lstterm}
When \BOSS finishes, a short set of instructions on how to connect the BOSSed library with \GB is printed to \term{stdout}. Several of the variables in the configuration file can also be set directly as command line arguments to \BOSS. For a complete list of arguments with explanations, see the output of the command
\begin{lstterm}
python boss.py --help
\end{lstterm}

Although \BOSS is able to provide runtime loading for most \Cpp classes and functions, there are some cases that the plug-in system generated by \BOSS cannot handle yet. Most importantly, \BOSS currently does not work with backend template classes, nor specialisations of \Cpp standard template library (STL) classes where the template parameter is a backend class. Further, the use of function pointers as function arguments or return types, and the use of \Cppeleven features in function declarations, is only partially supported. When a limitation only affects a class member function or variable, \BOSS will attempt to generate a limited class interface where the problematic element is excluded.  Future versions of \BOSS will improve on these limitations.

\subsection{Backend information utility}
\label{backendinfo}

Although most users will never have need to access it directly, we briefly point out here that a global backend information object exists in \GB.  It can be accessed by reference from any module function using the function \cpp{Backends::backendInfo()}.  It provides a plethora of runtime information about which backends are presently connected, their versions, functions, classloading status and so on.  The mostly likely use cases from within module functions for this object are to determine the folder in which a loaded backend resides:
\begin{lstcpp}
std::string path_to_MyBackend_1_2 = Backends::
 backendInfo().path_dir("MyBackend", "1.2");
\end{lstcpp}
or to get the default version of a BOSSed backend required by an unversioned \cpp{NEEDS_CLASSES_FROM} declaration:
\begin{lstcpp}
std::string default_MyBackend = Backends::
 backendInfo().default_version("MyBackend");
\end{lstcpp}
The full interface to this object can be found in \term{Backends/}\term{include/gambit/}\term{Backends/}\term{backend_info.hpp}.

\section{Hierarchical model database}
\label{models}
In \GB, a \cross{model} is defined to be a collection of named parameters.  These parameters are intended to be sampled by some scanning algorithm, according to some chosen prior probability distribution.\footnote{For frequentist sampling, the prior simply defines the distance measure on the parameter space to be used internally by the scanning algorithm when choosing new points.} The physical meaning of these parameters is defined entirely by how they are interpreted by \doublecross{module functions}{module function}. It is up to the writers of modules to ensure that parameters are used in a consistent manner. Consistent usage is facilitated by the \GB model database that the \cross{dependency resolver} (Sec.\ \ref{depresolver}) employs in order to automatically determine which module functions are compatible with which models. Module functions that are incompatible with the model(s) selected for scanning are disabled at runtime, and not considered during dependency resolution.
\subsection{Model declaration} \label{model_declaration}
\GB ships with a pre-defined selection of common models (Sec.\ \ref{inventory}). New models can be defined easily by adding an appropriate declaration in a new \Cpp header file located in the folder
\begin{lstterm}
Models/include/gambit/Models/models
\end{lstterm}
When the \GB build configuration is next re-run (see Sec.\ \ref{cmake}), the new model will be automatically detected and registered.  The declarations of all the pre-defined models can also be found in this folder.

The syntax for declaring a simple two parameter model \metavar{my\_model} with parameters \metavar{my\_par1} and \metavar{my\_par2} is:
\begin{lstcpp}
#define MODEL @\metavar{my\_model}@
START_MODEL
DEFINEPARS(@\metavar{my\_par1}@, @\metavar{my\_par2}@)
#undef MODEL
\end{lstcpp}
The \cpp{START_MODEL} command creates a \cpp{ModelParameters} object for the given model, which will hold the values of the parameters chosen at runtime by \scannerbit, and communicate them to relevant module functions during a scan.  The macro \cpp{DEFINEPARS} is variadic, and can take up to 64 parameters (or more, depending on the user's version of \textsf{Boost}).  If one prefers to break a long list of parameters into several pieces, this macro can be reused as many times as desired.

It is often the case that models will be subsets of a more general model, in the sense that a mapping from the general model to the more constrained model can be constructed. This hierarchical relationship between models is handled in \GB by defining the general model to be a \doublecross{parent}{parent model} of the constrained model, with the constrained model being reciprocally defined as a \doublecross{child}{child model} of that parent. The mapping from the child parameters to the parent parameters is encoded in a translation function, which \GB will call automatically when needed. Each parent model may have multiple children, however, a child model has only one parent. The ``family tree'' of any given model is thus a directed rooted tree graph, with the root of the tree being the common ancestor of all other models in the graph. The complete model database consists of a disconnected set of such family trees, see Fig.\ \ref{fig::model_tree}. When assessing the compatibility of module function with the model(s) being scanned, the \GB dependency resolver automatically treats all module functions that declare compatibility with a given model as \textit{also} compatible with all descendents of that model.

To declare that a model has a parent model \mbox{\metavar{parent},} and assign a function \metavar{to\_parent} capable of performing the translation from the child to the parent parameter set, the model declaration can be expanded to the following:
\begin{lstcpp}
#define PARENT @\metavar{parent}@
  #define MODEL @\metavar{my\_model}@
  START_MODEL
  DEFINEPARS(@\metavar{my\_par1}@, @\metavar{my\_par2}@)
  INTERPRET_AS_PARENT_FUNCTION(@\metavar{to\_parent}@)
  #undef MODEL
#undef PARENT
\end{lstcpp}
If a model is declared to have a parent but no translation function, any attempt to use another function that depends on the translation will trigger a runtime error from the dependency resolver. Further details on declaring and defining translation functions can be found in Sec.~\ref{model_translation}.  Note that we are only dealing with the abstract concept of translation functions between different model parameter spaces at this stage, not the actual physics of any translations in any given class of models. The actual translations between the models implemented in \GB \textsf{1.0} are implied by the relations between parent and child models described in Sec.\ \ref{inventory}.

Putting these aspects together, complete model declarations can be very simple, as in the \doublecrosssf{CMSSM}{CMSSM}:
\begin{lstcpp}
// Must include models that are targets of
// translation functions
#include "gambit/Models/models/NUHM1.hpp"

#define MODEL CMSSM
#define PARENT NUHM1
  START_MODEL
  DEFINEPARS(M0,M12,A0,TanBeta,SignMu)
  INTERPRET_AS_PARENT_FUNCTION(CMSSM_to_NUHM1)
  // Translation functions defined in CMSSM.cpp
#undef PARENT
#undef MODEL
\end{lstcpp}
This declaration can be found in the model header \term{Models/include/gambit/Models/models/CMSSM.hpp}.

Directed cross-links between branches of a family tree, or even between trees, are also possible. Models related in this way are denoted as \doublecross{friend models}{friend model}, though the relationship is not automatically mutual. If a model \metavar{my\_model} has a friend model \metavar{friend}, then a function \metavar{to\_friend} must also exist that can translate the parameters of \metavar{my\_model} into those of \metavar{friend}. To declare such a relationship, one inserts the following into the model declaration for \metavar{my\_model}:
\begin{lstcpp}
INTERPRET_AS_X_FUNCTION(@\metavar{friend}@, @\metavar{to\_friend}@)
\end{lstcpp}

With the addition of friend translations, the model hierarchy graph can become arbitrarily complicated.  To avoid painful manual resolution of ambiguous translation pathways between models, it can be advisable to limit the number of friend links. Nevertheless, a large number of the possible ambiguities are automatically resolved by the default behaviour of the dependency resolver to prefer child-to-parent links over child-to-friend links. This behaviour can be disabled by switching the \yaml{prefer\_model\_specific\_functions} option in the \yaml{KeyValues} section of the initialisation file to \yaml{false}. Manual resolution of all translation pathway ambiguities will then be required.  Alternatively, one can simply add a \yaml{Rule} that overrides the default in a specific case.  See Sec.\ \ref{interface} for details.

In some cases the translation from child to parent model, or to a friend model, may require the result of a calculation from a module function. It is therefore possible to declare \doublecross{dependencies}{dependency} for translation functions, which are directly analogous to the dependencies declared by module functions.\footnote{Internally, the translation functions actually {\it are} module functions, each belonging to a virtual module named after their source model.}

In general, translation functions can depend on any other capability, which may be filled by functions from any module.  The dependency resolution system ensures consistency of the requested and provided dependencies of all translation functions in such cases.  For example, the translation functions might depend on some aspect of the particle spectrum, and may involve translation of parameters from one renormalisation scheme to another, from a UV-complete theory to an EFT, or from one renormalisation scale to another.  In these examples, the dependencies would be most naturally resolved from \specbit, if it possesses the relevant capability for the model in question; we refer readers to Ref.\ \cite{SDPBit} for details of the functionalities available from this module.

To declare a dependency on some \metavar{capability} with \Cpp \metavar{type} for a child-to-parent translation function, one adds the following to the model declaration for the child:
\begin{lstcpp}
INTERPRET_AS_PARENT_DEPENDENCY(@\metavar{capability}@, @\metavar{type}@)
\end{lstcpp}
To declare such a dependency for a translate-to-friend function, one instead adds:
\begin{lstcpp}
INTERPRET_AS_X_DEPENDENCY(@\metavar{friend}@, @\metavar{capability}@, @\metavar{type}@)
\end{lstcpp}
where \metavar{friend} is the name of the target friend model.  These declarations must appear after the declaration of the corresponding translation function.

The full machinery for declaring dependencies with complicated conditions on groups of models and backend choices --- which is available for standard module functions --- is not available for the dependencies of model translation functions. If this machinery is required, one should write a module function that uses it and returns a result associated with a new capability, which can then be made accessible in a translation function using the above declarations. In the most extreme case, this module function may perform the complete parameter translation and then simply store the results in a temporary \cpp{ModelParameters} object, which can then be retrieved by the ``true'' translation function via the above mechanism and simply copied into the target model \cpp{ModelParameters} object (see Sec. \ref{model_translation}).
\subsection{Model capabilities}
In some cases a parameter in a \cross{model} may directly correspond to a physically meaningful quantity.  This quantity may be available already, computed in an alternate way, as the \cross{capability} of some existing \cross{module function}.  One may wish to have the alternative of simply using the value of the parameter to satisfy the dependencies of other module functions on this quantity, rather than the module function calculation. It can therefore be convenient to directly inform \GB of this correspondence when declaring a model. To declare this kind of relationship between a parameter \metavar{my\_par} and a capability \metavar{capability}, one adds the following to the declaration of the model containing \metavar{my\_par}:
\begin{lstcpp}
MAP_TO_CAPABILITY(@\metavar{my\_par}@, @\metavar{capability}@)
\end{lstcpp}
Of course the same could be achieved by manually creating a trivial module function that takes the model parameters as input, and then directly outputs one of them as its capability. Internally, \cpp{MAP_TO_CAPABILITY} causes \GB to create a virtual module function of precisely this kind, but it is convenient to have this task performed automatically.

The module function so created has the same name as the parameter being mapped, and lives in the module corresponding to the model to which it belongs.  Take the example of the top mass, a parameter of the \doublecrosssf{demo\_A}{demos} model found (commented out) in \term{Models/include/gambit/Models/models/demo.hpp}:
\begin{lstcpp}
MAP_TO_CAPABILITY(Mstop, Mstop_obs)
\end{lstcpp}
This declaration creates a new module function called \cpp{Mstop}, with capability \cpp{Mstop_obs} and return type \cpp{double}, and places it within the module named after \textsf{demo\_A}.  The function \cpp{demo_A::Mstop} simply returns the value of the \cpp{Mstop} parameter as it varies during a scan of \textsf{demo\_A}.

This convenience facility exists for the simplest case only. In the case where the correspondence is not direct --- for example, if a factor of two or a change of units is required, or if a dependency on some other calculation exists --- then manually adding an additional module function to do the transformation is the only option.
\subsection{Defining translation functions}
\label{model_translation}
In Sec.\ \ref{model_declaration} we discussed how to declare a translation pathway between two models; we now turn to how to define the functions that actually perform the translation.  These may or may not involve calculations relating to
the spectrum (as in the example the referee is thinking of). , so in this
case, they would depend on functions from SpecBit. The full details of how
those functions work is provided in the SpecBit, DecayBit and PrecisionBit
paper (1705.07936). In particular, this includes translations between pole and
running masses in different schemes and EFTs.

The function definition can either be placed directly into the header file in which the source model is declared, or into a separate source file that includes the header.  In the former case, the function body must appear \textit{after} the \cpp{INTERPRET\_AS} macro that declares it.  In the latter case, the source file should be placed in
\begin{lstterm}
Models/src/models
\end{lstterm}
to be properly auto-detected by the \GB build system. Some extra headers providing additional helper macros should be included, and the names of the model and its parent redefined in order for the helpers to work properly. A basic template for such a file is:
\begin{lstcpp}
#include "gambit/Models/model_macros.hpp"
#include "gambit/Models/model_helpers.hpp"
#include "gambit/Models/models/@\metavar{my\_model}@.hpp"

#define MODEL  @\metavar{my\_model}@
#define PARENT @\metavar{parent}@

// function definition

#undef PARENT
#undef MODEL
\end{lstcpp}

Consider the following example function definition:
\begin{lstcpp}s
void MODEL_NAMESPACE::@\metavar{to\_parent}@(const
 ModelParameters& myparams, ModelParameters&
 parentparams)
{
  double x = myparams["@\metavar{my\_par}@"];
  parentparams.setValue("@\metavar{parent\_par}@", 2*x);
}
\end{lstcpp}
Although this example is a child-to-parent translation function, the syntax is the same for child-to-friend functions.  The translation function must return \mbox{\cpp{void},} and take two arguments by reference: the source model parameters (which are \cpp{const}), and the target model parameters (of either the parent or friend model). The helper macro \cpp{MODEL\_NAMESPACE} places the function in the correct namespace (\cpp{Gambit::Models::MODEL}), and relies on \cpp{MODEL} having been defined appropriately.  On the first line of the function body, a parameter \metavar{my\_par} is retrieved from the \cpp{ModelParameters} object, which contains the parameters of the source model.  The value of this parameter is stored in the variable \cpp{x}. This is then multiplied by two, and used to set the value of the target model parameter \metavar{parent\_par}, completing the parameter translation.

This example assumes that the target model has only one parameter, \metavar{parent\_par}.  Often a source and target model will have many overlapping parameters, and it is convenient to have a mechanism for copying all of these automatically, without modification. This can be done using the \cpp{setValues} member function of the target \cpp{ModelParameters} object:
\begin{lstcpp}
parentparams.setValues(myparams, true);
\end{lstcpp}
The second parameter is optional, and \cpp{true} by default.  This triggers an error if any of the parameters in \cpp{myparams} (from \metavar{my\_model}) are missing from \cpp{parentparams} (from \metavar{parent}), i.e. if the source model parameter names are not a subset of the target model parameter names. Setting this \cpp{false} causes matching parameters to be copied but unmatched parameters to be ignored.

A real-world example that make use of \cpp{setValues} is the the \doublecrosssf{CMSSM}{CMSSM}-to-\doublecrosssf{NUHM1}{NUHM1} translation function (which was declared in Sec.\ \ref{model_declaration}):
\begin{lstcpp}
#define MODEL CMSSM

  void MODEL_NAMESPACE::CMSSM_to_NUHM1
   (const ModelParameters &myP,
    ModelParameters &targetP)
  {
     logger()<<"Running interpret_as_parent "
             <<"calculations for CMSSM --> NUHM1."
             <<LogTags::info<<EOM;

     // Send all parameter values upstream
     // to matching parameters in parent.
     targetP.setValues(myP);

     // Set NUHM1 parameter @\cpppragma{$m_H$}@ equal to @\cpppragma{$m_0$}@.
     targetP.setValue("mH", myP["M0"]);
  }

#undef MODEL
\end{lstcpp}
This function can be found in \term{Models/src/models/} \term{CMSSM.cpp}.

To retrieve dependencies on externally-calculated quantities, one uses regular module function syntax
\begin{lstcpp}
USE_MODEL_PIPE(@\metavar{target}@)
const @\metavar{type}@* my_variable = *Dep::@\metavar{capability}@;
\end{lstcpp}
where the \cpp{USE\_MODEL\_PIPE} macro simply expands to a \cpp{using} statement that brings the \doublecross{pipes}{pipe} for the translation function into the current namespace, making the \cpp{Dep::}\metavar{capability} pointer easily accessible. The argument should be the name of the target (parent or friend) model, i.e. \cpp{USE_MODEL_PIPE(}\metavar{friend}\cpp{)} or \cpp{USE_MODEL_PIPE(PARENT)} (remembering that \cpp{PARENT} is a macro holding the actual parent model, defined in the model header).

\subsection{Models defined in \GB \textsf{1.0.0}}
\label{inventory}

Here we list the models already defined in the first release of \GB, along with their parameters.  The relationships between these models can be seen in Fig.\ \ref{fig::model_tree}.

\subsubsection{Standard Model}

The SM exists in two parts within \GB.  The Higgs mass must be specified separately from the rest of the SM parameters, as it is often contained within the definition of BSM theories featuring BSM Higgs sectors.  For those theories that do not include their own Higgs sector, e.g. \doublecrosssf{SingletDM}{SingletDM}, we therefore provide additional models containing the Higgs mass as a parameter: \doublecrosssf{StandardModel\_Higgs}{SM_h} and \doublecrosssf{StandardModel\_Higgs\_running}{SM_h_run}. Typically, one of these models should be scanned over in tandem with the rest of the SM (\doublecrosssf{StandardModel\_SLHA2}{SM_SLHA2}) and the BSM theory in question.  To investigate just the SM itself, one should perform a simultaneous scan of \doublecrosssf{StandardModel\_SLHA2}{SM_SLHA2} and either \doublecrosssf{StandardModel\_Higgs}{SM_h} or \doublecrosssf{StandardModel\_Higgs\_running}{SM_h_run}.
\begin{description}

\item[\textbf{\textsf{StandardModel\_SLHA2}}\label{SM_SLHA2}:] \term{CKM_A, CKM_etabar,} \term{CKM_lambda, CKM_rhobar, GF, alpha1, alpha2,} \term{alphaS, alphainv, delta13, mBmB, mCmC, mD, mE,} \term{mMu, mNu1, mNu2, mNu3, mS, mT, mTau, mU, mZ,} \term{theta12, theta13, theta23}.\\
This model contains the SM parameters defined in the \texttt{SMINPUTS}, \texttt{VCKMIN} and \texttt{UPMNSIN} blocks of the second SUSY Les Houches Accord (SLHA2; \cite{Allanach:2008qq}).  This includes the $Z$ pole mass, the Fermi (weak) coupling $G_\mathrm{F}$, the strong and electromagnetic couplings at scale $m_Z$ in the \MSbar renormalisation scheme, pole masses for leptons, neutrinos and the top quark, running masses for other quarks in the \MSbar scheme (at scale $m_b$ for $b$, $m_c$ for $c$ and 2\,GeV for $u, d$ and $s$), the CKM mixing matrix in Wolfenstein parameterisation, and the PMNS matrix, characterised by three mixing angles and three $CP$-violating phases.  To convert the Wolfenstein parameters into $V_{\rm CKM}$ entries internally, we use the 9th-order expansions of Ref.\ \cite{CKMFitter}.  More detailed definitions of these parameters can be found in Appendix \ref{SMdefs}. \vspace{2mm}

\item[\textbf{\textsf{StandardModel\_Higgs\_running}}\label{SM_h_run}:] \term{QEWSB, mH}.\\
This model provides a description of the SM Higgs sector in terms of $m^2_H$, the bare Higgs mass parameter in the SM Lagrangian at scale $m_Z$.  The vacuum expectation value of the SM Higgs field at $m_Z$ can be obtained from the \doublecrosssf{StandardModel\_SLHA2}{SM_SLHA2} as $v_0 = (\sqrt{2}G_\mathrm{F})^{-1/2}$.  This model is intended for use in situations where the Higgs potential is run to different scales, e.g. for calculating pole masses or investigating vacuum stability.  It therefore also contains one additional parameter: $Q_{\rm EWSB}$, the scale at which the electroweak symmetry-breaking (EWSB) consistency condition that the Higgs potential possess a tree-level minimum is imposed.  Although in principle physical properties should not depend on its value, typically one prefers to take $Q_{\rm EWSB}\sim m_t$ in order to minimise errors from neglecting higher-order loops.  It is common to vary this parameter by a factor of a few in order to try to quantify the uncertainty in resulting pole masses from missing loop terms. \vspace{2mm}

\item[\textbf{\textsf{StandardModel\_Higgs}}\label{SM_h}:] \term{mH}.\\
Unlike the \doublecrosssf{StandardModel\_Higgs\_running}{SM_h_run} model, the tree-level Higgs mass $m_h$ is taken as the free parameter of \doublecrosssf{StandardModel\_Higgs}{SM_h}, but interpreted directly as the pole mass for most calculations.  This generally removes the need to calculate it via renormalisation group running in any higher-energy theory.  For simple calculations, this allows a cut-down \GB\ \cpp{Spectrum} object to be produced, with no ability to run, and the Higgs `pole' mass extracted from it by simply accessing the value of the tree-level parameter for the given point in parameter space.  When observables are to be calculated that genuinely need to use running parameters, the model point is up-translated to a parameter point in the \doublecrosssf{StandardModel\_Higgs\_running}{SM_h_run} (the parent model), where $m^2_H$ at scale $m_Z$ is set equal to the square of the tree-level mass, and $Q_{\rm EWSB}$ is set to $m_t$.  This is useful for more detailed calculations involving module functions that explicitly require the running mass parameter of \doublecrosssf{StandardModel\_Higgs\_running}{SM_h_run}, and/or functions that need accurate pole masses calculated by including the Higgs sector in renormalisation calculations.

\end{description}

\subsubsection{Scalar singlet}
The scalar singlet is the simplest possible model for DM, consisting of a single additional scalar field $S$ uncharged under the gauge symmetries of the SM, and stabilised by a $\mathbb{Z}_2$ symmetry.  The additional renormalisable Lagrangian terms permitted by general symmetry arguments are
\begin{equation}
\mathcal{L}_\mathrm{SS} = \frac12 \mu_S^2 S^2 + \frac12\lhs S^2|H|^2 + \frac14\ls S^4 + \frac12\partial_\mu S \partial^\mu S.
\end{equation}
From left to right, these are: the bare $S$ mass, the dimension-4 Higgs-portal coupling, the $S$ quartic self-coupling, and the kinetic term.  The latter plays no role in phenomenology, leaving three free parameters of the theory: $\mu_S^2$, $\lhs$ and $\ls$.  After EWSB, the singlet mass receives an additional contribution from the Higgs-portal term, leading to a tree-level mass of
\begin{equation}
\label{m_S_tree}
m_S = \sqrt{\mu_S^2 + \frac12{\lhs v_0^2}}.
\end{equation}
This model has been subjected to global fits in Refs.\ \cite{Cheung:2012xb,Cuoco:2016jqt,SSDM}.

\begin{description}

\item[\textbf{\textsf{SingletDM\_running}}\label{SS_run}:] \term{lambda_S, lambda_hS, mS}.\\
This model has the \MSbar couplings $\lhs$ and $\ls$ at scale $m_Z$ as free parameters, as well as the tree-level mass $m_S$, which is matched internally to the \MSbar value of $\mu_S$ at scale $m_Z$ using $\lhs(m_Z)$ and Eq.\ \ref{m_S_tree}.  This allows full calculation of pole masses, renormalisation group running and vacuum stability.\vspace{2mm}

\gsfitemc{SingletDM} \term{lambda_hS, mS}.\\
The relationship between \doublecrosssf{SingletDM}{SingletDM} and \doublecrosssf{SingletDM\_running}{SS_run} is analogous to the one between \doublecrosssf{StandardModel\_Higgs}{SM_h} and \doublecrosssf{StandardModel\_Higgs\_running}{SM_h_run}.  \textsf{SingletDM} has $m_S$ as a free parameter, leading to two use cases.  The first is to interpret the model parameter directly as the pole mass for $S$ and do phenomenology without any spectrum calculation; the second is to take the parameter $m_S$ as the tree-level estimate of the mass, use Eq.\ \ref{m_S_tree} to recover $\mu_S^2$ matched to the tree-level mass at scale $m_Z$, and calculate the pole mass accordingly in the parent model \doublecrosssf{SingletDM\_running}{SS_run}.  One chooses between these two options by selecting which function from \specbit to obtain a \cpp{Spectrum} object from.  \textsf{SingletDM} includes the Higgs-portal coupling $\lhs$ identically to the parent model, which we also set the running coupling at $m_Z$ to when translating a \doublecrosssf{SingletDM}{SingletDM} model point to a \doublecrosssf{SingletDM\_running}{SS_run} point.  It however does not include any description of the quartic coupling.  This is because the $S$ self-coupling term only plays a role in observables via RGE running, such as in the calculation of pole masses and analysis of Higgs vacuum stability.  When translating to the parent model, we therefore explicitly choose the quartic coupling to be absent at scale $m_Z$ (even though it will be regenerated at other scales under RGE flow):
\begin{align}
&\ls(m_Z) = 0.
\end{align}

\end{description}

\subsubsection{Weak-scale MSSM}

These models feature MSSM soft SUSY-breaking Lagrangian parameters defined at a chosen scale $Q$, typically set to something near the weak scale.

The MSSM is the version of SUSY containing the least additional field content beyond the SM.  Its Lagrangian (see e.g. \cite{BaerTata})
\begin{equation}
\mathcal{L}_\mathrm{MSSM} = \mathcal{L}_\textrm{SUSY-SM} + \mathcal{L}_\mathrm{soft},
\label{lmssm}
\end{equation}
is obtained by supersymmetrising the (pre-EWSB) SM Lagrangian to find $\mathcal{L}_\textrm{SUSY-SM}$, and augmenting it with all possible renormalisable soft SUSY-breaking terms that conserve both baryon ($B$) and lepton number ($L$).  These soft terms are
\begin{subequations}
\label{lsoft}
\begin{align}
\mathcal{L}_\mathrm{soft} =& -\frac12\left[M_1\bar{\tilde{B}}^0\tilde{B}^0 + M_2\bar{\tilde{W}}_A\tilde{W}_A + M_3\bar{\tilde{g}}_B\tilde{g}_B\right] \label{gauginos} \\
&  -\frac{i}{2}\left[M_1'\bar{\tilde{B}}^0\gamma_5\tilde{B}^0 + M_2'\bar{\tilde{W}}_A\gamma_5\tilde{W}_A + M_3'\bar{\tilde{g}}_B\gamma_5\tilde{g}_B\right] \label{CP-violating gauginos} \\
&  - \epsilon_{ab}\Big[b{H}^a_u{H}^b_d + \mathrm{h.c.}\Big] - m^2_{H_u}|H_u|^2 - m^2_{H_d}|H_d|^2 \label{Higgses} \\
&  +\textstyle\sum_{i,j=1,3} \left\{ -\left[\tilde{Q}_i^\dagger(\mathbf{m}^\mathbf{2}_Q)_{ij}\tilde{Q}_j +
\tilde{d}_{\mathrm{R}i}^\dagger(\mathbf{m}^\mathbf{2}_d)_{ij}\tilde{d}_{\mathrm{R}j} \right. \right. \nonumber \\
&  + \tilde{u}_{\mathrm{R}i}^\dagger(\mathbf{m}^\mathbf{2}_u)_{ij}\tilde{u}_{\mathrm{R}j}
   + \tilde{L}_{i}^\dagger(\mathbf{m}^\mathbf{2}_L)_{ij}\tilde{L}_{j}
   + \tilde{e}_{\mathrm{R}i}^\dagger(\mathbf{m}^\mathbf{2}_e)_{ij}\tilde{e}_{\mathrm{R}j} \Big] \label{sfermion masses} \\
 & - \epsilon_{ab}\left[(\mathbf{T}_u)_{ij}\tilde{Q}^a_iH^b_u\tilde{u}^\dagger_{\mathrm{R}j}
   - (\mathbf{T}_d)_{ij}\tilde{Q}^a_iH^b_d\tilde{d}^\dagger_{\mathrm{R}j} \right. \nonumber\\
&  \hspace{1cm} - \left. (\mathbf{T}_e)_{ij}\tilde{L}^a_iH^b_d\tilde{e}^\dagger_{\mathrm{R}j} + \mathrm{h.c.}\right] \label{trilinear couplings} \\
&  - \epsilon_{ab}\left[(\mathbf{C}_u)_{ij}\tilde{Q}^a_iH^{*b}_d\tilde{u}^\dagger_{\mathrm{R}j}
   - (\mathbf{C}_d)_{ij}\tilde{Q}^a_iH^{*b}_u\tilde{d}^\dagger_{\mathrm{R}j} \right. \nonumber \\
&  \hspace{1cm} - \left.\left. (\mathbf{C}_e)_{ij}\tilde{L}^a_iH^{*b}_u\tilde{e}^\dagger_{\mathrm{R}j} + \mathrm{h.c.}\right] \right\}. \label{trilinear C couplings}
\end{align}
\end{subequations}
Here we explicitly sum over the generation indices $i$ and $j$, and imply summation over the gauge generator indices $A=1\,..\,3$ and $B=1\,..\,8$, as well as the $SU(2)_\mathrm{L}$ indices $a,b=1,2$.  Here $\epsilon_{ab}$ is the two-dimensional completely antisymmetric tensor, defined such that $\epsilon_{12} = -\epsilon_{21} = 1$.  Superparticle fields are denoted by tilded operators ($\tilde{Q}_j$, $\tilde{u}^\dagger_{\mathrm{R}j}$, etc.), where $\tilde{B}^0$, $\tilde{W}_A$ and $\tilde{g}_B$ are the superpartners of the SM gauge bosons.  Fields denoted with capital letters are $SU(2)_\mathrm{L}$ doublets ($Q_i\equiv(u_{\mathrm{L}i},d_{\mathrm{L}i})^\mathrm{T}$, etc), whereas lowercase fields are $SU(2)_\mathrm{L}$ singlets. The subscripts $u$ and $d$ refer to the two Higgs doublets, which give masses separately to up- and down-type quarks.

The first two terms in Eq.~\ref{lsoft} (\ref{gauginos} and \ref{CP-violating gauginos}) are explicit gaugino masses associated with the real parameters $M_1,M_2,M_3$ and $M'_1,M'_2,M'_3$.  The second set of these violates $CP$, so $M'_1,M'_2$ and $M'_3$ should be very small to agree with experiment.  The Higgs sector (\ref{Higgses}) includes explicit mass terms with real parameters $m^2_{H_u}$ and $m^2_{H_d}$, as well as a bilinear coupling with complex parameter $b$.  Explicit sfermion masses (\ref{sfermion masses}) come from the five $3\times3$ Hermitian mass-squared matrices $\mathbf{m}^\mathbf{2}_Q$, $\mathbf{m}^\mathbf{2}_u$, $\mathbf{m}^\mathbf{2}_d$, $\mathbf{m}^\mathbf{2}_L$ and $\mathbf{m}^\mathbf{2}_e$.  The final two terms (\ref{trilinear couplings} and \ref{trilinear C couplings}) denote trilinear couplings between the Higgs and squarks or sleptons, with general Yukawa-type complex $3\times3$ matrices $\mathbf{T}_u, \mathbf{T}_d, \mathbf{T}_e$ and $\mathbf{C}_u, \mathbf{C}_d, \mathbf{C}_e$.  The $\mathbf{C}$ terms are often omitted from the definition of the MSSM, as they end up strongly suppressed in many SUSY-breaking schemes.  SUSY-breaking scenarios often imply universality relations between the Yukawa-scaled soft-breaking trilinear couplings $\mathbf{A}_u, \mathbf{A}_d$ and $\mathbf{A}_e$,  which are defined as
\begin{equation}
(\mathbf{A}_f)_{ij} \equiv (\mathbf{T}_f)_{ij}/(\mathbf{Y}_f)_{ij} \quad f \in \{u,d,e\},
\end{equation}
where $i$ and $j$ run over all three generations, but summation is not implied.

$\mathcal{L}_\textrm{SUSY-SM}$ contains derivatives of the superpotential
\begin{eqnarray}
\hat{W} =& \epsilon_{ab} \left\{ \textstyle\sum_{i,j=1,3} \left[(\mathbf{Y}_u)_{ij}\hat{Q}^a_i\hat{H}^b_u\hat{U}^c_j - (\mathbf{Y}_d)_{ij}\hat{Q}^a_i\hat{H}^b_d\hat{D}^c_j \right.\right. \nonumber\\
         & - \left.\left. (\mathbf{Y}_e)_{ij}\hat{L}^a_i\hat{H}^b_d\hat{E}^c_j\right] - \mu\hat{H}^a_u\hat{H}^b_d \right\}.
\label{superpot}
\end{eqnarray}
Here the indices $i$ and $j$ are again generation number, $a$ and $b$ are $SU(2)_\mathrm{L}$ indices and $\epsilon_{ab}$ is the two-dimensional antisymmetric tensor.  Carets indicate superfields.  The terms $\hat{U}^c_j$, $\hat{D}^c_j$ and $\hat{E}^c_j$ are the left chiral superfields containing the charge conjugates of the right-handed $SU(2)_\mathrm{L}$ singlets: up-type (s)quarks, down-type (s)quarks and (s)electrons, respectively.  Derivatives of $\hat{W}$ with respect to its scalar fields give rise to all non-gauge interaction terms in $\mathcal{L}_\textrm{SUSY-SM}$.  It plays a similar role to the non-gauge part of the scalar potential in non-supersymmetric theories, specifying the Higgs potential via the complex parameter $\mu$ and the Higgs-fermion interactions and fermion masses via the complex $3\times3$ Yukawa coupling matrices $\mathbf{Y}_u$, $\mathbf{Y}_d$ and $\mathbf{Y}_e$.

$R$ parity is conserved in Eqs.\ \ref{lsoft} and \ref{superpot} by construction, by virtue of $B$ and $L$ being conserved individually.  This makes the lightest SUSY particle (LSP) absolutely stable; in general we impose the condition that this must be the lightest neutralino.  Unless otherwise noted, we neglect the phenomenology of the gravitino, assuming it to be sufficiently heavy that it is not the LSP and its decays at early times are irrelevant.

\begin{description}
\gsfitemc{MSSM63atQ} \term{Ad_11, Ad_12, Ad_13, Ad_21, Ad_22,}
                     \term{Ad_23, Ad_31, Ad_32, Ad_33, Ae_11, Ae_12, Ae_13,}
                     \term{Ae_21, Ae_22, Ae_23, Ae_31, Ae_32, Ae_33, Au_11,}
                     \term{Au_12, Au_13, Au_21, Au_22, Au_23, Au_31, Au_32,}
                     \term{Au_33, M1, M2, M3, Qin, SignMu, TanBeta, mHd2,}
                     \term{mHu2, md2_11, md2_12, md2_13, md2_22, md2_23,}
                     \term{md2_33, me2_11, me2_12, me2_13, me2_22, me2_23,}
                     \term{me2_33, ml2_11, ml2_12, ml2_13, ml2_22, ml2_23,}
                     \term{ml2_33, mq2_11, mq2_12, mq2_13, mq2_22, mq2_23,}
                     \term{mq2_33, mu2_11, mu2_12, mu2_13, mu2_22, mu2_23,}
                     \term{mu2_33}.\\
This model contains 65 free parameters: the scale $Q$, the sign of the $\mu$ parameter, and 63 parameters of the MSSM Lagrangian.   Because of their usual irrelevance in (known) SUSY-breaking schemes, here we set the $\mathbf{C}$ terms to zero.  Apart from this omission, the MSSM63 is the most general formulation of the $CP$-conserving MSSM: here the $CP$-violating gaugino masses $M'$ and complex phases are all also explicitly set to zero.  This leaves 3 free gaugino masses $M_1$, $M_2$ and $M_3$, 6 real parameters each from the mass-squared matrices $\mathbf{m}^\mathbf{2}_Q$, $\mathbf{m}^\mathbf{2}_u$, $\mathbf{m}^\mathbf{2}_d$, $\mathbf{m}^\mathbf{2}_L$ and $\mathbf{m}^\mathbf{2}_e$ and a further 9 each from the trilinear couplings $\mathbf{A}_u, \mathbf{A}_d$ and $\mathbf{A}_e$.  The final three parameters come from the Higgs sector, where we have $m^2_{H_u}$ and $m^2_{H_d}$, and trade $b$ and $\mu$ for the sign of $\mu$ and the ratio of the up-type to down-type Higgs vacuum expectation values $\tan\beta\equiv v_\mathrm{u}/v_\mathrm{d}$.  All parameters are defined in the \DRbar scheme at the scale $Q$, except for $\tan\beta$, which is defined at $m_Z$.

\quad Relative to the general MSSM, the additional constraints applied in this model are:
\begin{align}
&M'_1 = M'_2 = M'_3 = 0,\\
&\mathbf{C}_u = \mathbf{C}_d = \mathbf{C}_e = \mathbf{0},\\
&\mathbf{m}^\mathbf{2}_Q, \mathbf{m}^\mathbf{2}_u, \mathbf{m}^\mathbf{2}_d, \mathbf{m}^\mathbf{2}_L, \mathbf{m}^\mathbf{2}_e, \mathbf{A}_u, \mathbf{A}_d, \mathbf{A}_e\ \mathrm{all~real}.
\end{align}

\gsfitemc{MSSM30atQ} \term{Ad_1, Ad_2, Ad_3, Ae_1, Ae_2, Ae_3,}
                     \term{Au_1, Au_2, Au_3, M1, M2, M3, Qin, SignMu,}
                     \term{TanBeta, mHd2, mHu2, md2_1, md2_2, md2_3, me2_1,}
                     \term{me2_2, me2_3, ml2_1, ml2_2, ml2_3, mq2_1, mq2_2,}
                     \term{mq2_3, mu2_1, mu2_2, mu2_3}.\\
As per the \doublecrosssf{MSSM63atQ}{MSSM63atQ}, but with all off-diagonal elements in $\mathbf{m}^\mathbf{2}_Q$, $\mathbf{m}^\mathbf{2}_u$, $\mathbf{m}^\mathbf{2}_d$, $\mathbf{m}^\mathbf{2}_L$, $\mathbf{m}^\mathbf{2}_e$, $\mathbf{A}_u, \mathbf{A}_d$ and $\mathbf{A}_e$ set to zero, in order to suppress flavour-changing neutral currents:
\begin{align}
&\mathbf{m}^\mathbf{2}_Q, \mathbf{m}^\mathbf{2}_u, \mathbf{m}^\mathbf{2}_d, \mathbf{m}^\mathbf{2}_L, \mathbf{m}^\mathbf{2}_e, \mathbf{A}_u, \mathbf{A}_d, \mathbf{A}_e\ \mathrm{diagonal}.
\end{align}

\gsfitemc{MSSM25atQ} \term{Ad_3, Ae_12, Ae_3, Au_3, M1, M2, M3,}
                     \term{Qin, SignMu, TanBeta, mHd2, mHu2, md2_1, md2_2,}
                     \term{md2_3, me2_1, me2_2, me2_3, ml2_1, ml2_2, ml2_3,}
                     \term{mq2_1, mq2_2, mq2_3, mu2_1, mu2_2, mu2_3}.\\
This was the model investigated in Ref.\ \cite{Silverwood12}.  As per the \doublecrosssf{MSSM30atQ}{MSSM30atQ}, but with first and second-generation trilinear couplings degenerate in the slepton sector, and set to zero for squarks:
\begin{align}
&(\mathbf{A}_e)_{11} = (\mathbf{A}_e)_{22}, \\
&(\mathbf{A}_u)_{11} = (\mathbf{A}_u)_{22} = (\mathbf{A}_d)_{11} = (\mathbf{A}_d)_{22} = 0.
\end{align}

\gsfitemc{MSSM24atQ} \term{Ad_3, Ae_3, Au_3, M1, M2, M3, Qin,}
                     \term{SignMu, TanBeta, mHd2, mHu2, md2_1, md2_2, md2_3,}
                     \term{me2_1, me2_2, me2_3, ml2_1, ml2_2, ml2_3, mq2_1,}
                     \term{mq2_2, mq2_3, mu2_1, mu2_2, mu2_3}.\\
As per the \doublecrosssf{MSSM25atQ}{MSSM25atQ}, but with first and second-generation trilinear couplings in the slepton sector also set to zero:
\begin{align}
(\mathbf{A}_u)_{ii} = (\mathbf{A}_d)_{ii} = (\mathbf{A}_e)_{ii} = 0 \hspace{2mm} \forall\ i\in\{1,2\}.
\end{align}

\gsfitemc{MSSM20atQ} \mbox{\term{Ad_3, Ae_12, Ae_3, Au_3, M1, M2, M3,}}
                     \term{Qin, SignMu, TanBeta, mHd2, mHu2, md2_12, md2_3,}
                     \term{me2_12, me2_3, ml2_12, ml2_3, mq2_12, mq2_3,}
                     \term{mu2_12, mu2_3}.\\
As per the \doublecrosssf{MSSM25atQ}{MSSM25atQ}, but with degenerate first and second-generation sfermion mass parameters:
\begin{align}
&(\mathbf{A}_e)_{11} = (\mathbf{A}_e)_{22},\\
&(\mathbf{A}_u)_{11} = (\mathbf{A}_u)_{22} = (\mathbf{A}_d)_{11} = (\mathbf{A}_d)_{22} = 0, \\
&(\mathbf{m}^\mathbf{2}_X)_{11} = (\mathbf{m}^\mathbf{2}_X)_{22}\hspace{2mm} \forall\ X\in\{Q,u,d,L,e\}.
\end{align}

\gsfitemc{MSSM19atQ} \term{Ad_3, Ae_3, Au_3, M1, M2, M3, Qin,}
                     \term{SignMu, TanBeta, mHd2, mHu2, md2_12, md2_3,}
                     \term{me2_12, me2_3, ml2_12, ml2_3, mq2_12, mq2_3,}
                     \term{mu2_12, mu2_3}.\\
This is the model that is sometimes referred to as the ``phenomenological'' MSSM (pMSSM).  It has been the focus of many non-statistical random parameter scans, e.g. \cite{Berger09,LATpMSSM,Conley11,Arbey12,ATLAS15}. As per the \doublecrosssf{MSSM20atQ}{MSSM20atQ}, but with first and second-generation trilinear couplings in the slepton sector also set to zero:
\begin{align}
&(\mathbf{A}_u)_{ii} = (\mathbf{A}_d)_{ii} = (\mathbf{A}_e)_{ii} = 0 \hspace{2mm} \forall\ i\in\{1,2\},\\
&(\mathbf{m}^\mathbf{2}_X)_{11} = (\mathbf{m}^\mathbf{2}_X)_{22}\hspace{2mm} \forall\ X\in\{Q,u,d,L,e\}.
\end{align}

\gsfitemc{MSSM16atQ} \term{Ad_3, Ae_3, Au_3, M1, M2, M3, Qin,}
                     \term{SignMu, TanBeta, mHd2, mHu2, md2_3, me2_3, ml2_12,}
                     \term{ml2_3, mq2_12, mq2_3, mu2_3}.\\
As per the \doublecrosssf{MSSM19atQ}{MSSM19atQ}, but with all first and second generation squark mass parameters degenerate, and all first and second generation slepton mass parameters degenerate:
\begin{align}
&(\mathbf{m}^\mathbf{2}_Q)_{ii} = (\mathbf{m}^\mathbf{2}_u)_{jj} = (\mathbf{m}^\mathbf{2}_d)_{kk} \hspace{2mm} \forall\ i,j,k\in\{1,2\},\\
&(\mathbf{m}^\mathbf{2}_L)_{ii} = (\mathbf{m}^\mathbf{2}_e)_{jj} \hspace{2mm} \forall\ i,j\in\{1,2\}.
\end{align}

\gsfitemc{MSSM15atQ} \term{A0, At, M1, M2, M3, Qin, SignMu,}
                     \term{TanBeta, mHd2, mHu2, md2_3, me2_3, ml2_12, ml2_3}
                     \term{mq2_12, mq2_3, mu2_3}.\\
This is the model explored in Ref.\ \cite{Strege15}, up to reparameterisation of the Higgs sector.  As per the \doublecrosssf{MSSM16atQ}{MSSM16atQ}, but with down-type and sleptonic trilinear couplings degenerate:
\begin{align}
(\mathbf{A}_d)_{33} = (\mathbf{A}_e)_{33}.
\end{align}

\gsfitemc{MSSM11atQ} \term{Ad_3, Ae_3, Au_3, M1, M2, M3, Qin,}
                     \term{SignMu, TanBeta, mHd2, mHu2, ml2, mq2}.\\
As per the \doublecrosssf{MSSM16atQ}{MSSM16atQ}/\doublecrosssf{MSSM19atQ}{MSSM19atQ}, but with universal squark ($m_{\tilde q}^2$) and slepton ($m_{\tilde l}^2$) mass parameters:
\begin{align}
&(\mathbf{m}^\mathbf{2}_X)_{ii} \equiv m_{\tilde q}^2 \hspace{2mm}\forall\ i\in\{1..3\}, X\in\{Q,u,d\},\\
&(\mathbf{m}^\mathbf{2}_Y)_{ii} \equiv m_{\tilde l}^2 \hspace{2mm}\forall\ i\in\{1..3\}, Y\in\{L,e\}.
\end{align}

\gsfitemc{MSSM10atQ} \term{Ad_3, Au_3, M1, M2, M3, Qin, SignMu,}
                     \term{TanBeta, mHd2, mHu2, ml2, mq2}.\\
As per the \doublecrosssf{MSSM11atQ}{MSSM11atQ}, but with no sleptonic trilinear coupings:
\begin{align}
(\mathbf{A}_e)_{33} = 0.
\end{align}

\gsfitemc{MSSM10batQ} \term{Ad_3, Ae_3, Au_3, M1, M2, M3, Qin,}
                      \term{SignMu, TanBeta, mHd2, mHu2, mf2}.\\
As per the \doublecrosssf{MSSM11atQ}{MSSM11atQ}, but with a universal sfermion mass parameter $m_{\tilde f}^2$:
\begin{align}
m_{\tilde q}^2 = m_{\tilde l}^2 \equiv m_{\tilde f}^2.
\end{align}

\gsfitemc{MSSM10catQ} \term{A0, M1, M2, M3, Qin, SignMu, TanBeta,}
                      \term{mHd2, mHu2, ml2, mq2_12, mq2_3}.\\
This is the model explored in Ref.\ \cite{MasterCodeMSSM10}, up to reparameterisation of the Higgs sector.  As per the \doublecrosssf{MSSM15atQ}{MSSM15atQ}, but with a universal trilinear coupling $A_0$, 3rd generation squark mass ($m_{\tilde q3}^2$) and slepton mass ($m_{\tilde l}^2$) parameters:
\begin{align}
&(\mathbf{A}_u)_{33} = (\mathbf{A}_d)_{33} = (\mathbf{A}_e)_{33} \equiv A_0,\\
&(\mathbf{m}^\mathbf{2}_Q)_{33} = (\mathbf{m}^\mathbf{2}_u)_{33} = (\mathbf{m}^\mathbf{2}_d)_{33} \equiv m_{\tilde q3}^2,\\
&(\mathbf{m}^\mathbf{2}_L)_{ii} = (\mathbf{m}^\mathbf{2}_e)_{jj} \equiv m_{\tilde l}^2 \hspace{2mm} \forall\ i,j\in\{1..3\}.
\end{align}

\gsfitemc{MSSM9atQ} \term{Ad_3, Au_3, M1, M2, M3, Qin, SignMu,}
                    \term{TanBeta, mHd2, mHu2, mf2}
As per the \doublecrosssf{MSSM11atQ}{MSSM11atQ}, but with both the approximations introduced in the \doublecrosssf{MSSM10atQ}{MSSM10atQ} and \doublecrosssf{MSSM10batQ}{MSSM10batQ}, i.e. universal sfermion masses and no sleptonic trilinear couplings:
\begin{align}
&(\mathbf{A}_e)_{33} = 0,\\
&m_{\tilde q}^2 = m_{\tilde l}^2 \equiv m_{\tilde f}^2.
\end{align}

\gsfitemc{MSSM7atQ} \term{Ad_3, Au_3, M2, Qin, SignMu, TanBeta,}
                    \term{mHd2, mHu2, mf2}.\\
This model has been used extensively in \ds papers, e.g. \cite{BergstromGondolo96,darksusy,BMSSM}.  As per the \doublecrosssf{MSSM9atQ}{MSSM9atQ}, but assuming a Grand Unified Theory (GUT)-inspired relationship between the gaugino masses.
\begin{align}
\frac{3}{5}\cos^2\theta_\mathrm{W}M_1 = \sin^2\theta_\mathrm{W}M_2 = \frac{\alpha}{\alpha_\mathrm{s}}M_3.
\end{align}
When implementing this relationship, we use $\sin^2\theta_\mathrm{W}$ at the $Z$ pole mass scale, which we calculate directly from the \doublecrosssf{StandardModel\_SLHA2}{SM_SLHA2} parameters $G_\mathrm{F}$, $m_Z$ (pole) and $\alpha^{-1}_{\overline{\rm MS}}(m_Z)$.

\end{description}

\subsubsection{GUT-scale MSSM}

These models feature MSSM soft SUSY-breaking Lagrangian parameters defined at the scale of gauge coupling unification, typically referred to as the GUT scale.

\begin{description}

\gsfitemc{MSSM63atMGUT} \mbox{\term{Ad_11, Ad_12, Ad_13, Ad_21, Ad_22,}}
                        \term{Ad_23, Ad_31, Ad_32, Ad_33, Ae_11, Ae_12, Ae_13,}
                        \term{Ae_21, Ae_22, Ae_23, Ae_31, Ae_32, Ae_33, Au_11,}
                        \term{Au_12, Au_13, Au_21, Au_22, Au_23, Au_31, Au_32,}
                        \term{Au_33, M1, M2, M3, SignMu, TanBeta, mHd2, mHu2,}
                        \term{md2_11, md2_12, md2_13, md2_22, md2_23, md2_33,}
                        \term{me2_11, me2_12, me2_13, me2_22, me2_23, me2_33,}
                        \term{ml2_11, ml2_12, ml2_13, ml2_22, ml2_23, ml2_33,}
                        \term{mq2_11, mq2_12, mq2_13, mq2_22, mq2_23, mq2_33,}
                        \term{mu2_11, mu2_12, mu2_13, mu2_22, mu2_23, mu2_33}.\\
As per the \doublecrosssf{MSSM63atQ}{MSSM63atQ}, but with $Q$ set to the GUT scale.  Translation to \doublecrosssf{MSSM63atQ}{MSSM63atQ} requires having already solved the renormalisation group equations (RGEs) for the model, in order to determine the value of the GUT scale.\vspace{2mm}

\gsfitemc{MSSM30atMGUT} \term{Ad_1, Ad_2, Ad_3, Ae_1, Ae_2,}
                        \term{Ae_3, Au_1, Au_2, Au_3, M1, M2, M3, SignMu,}
                        \term{TanBeta, mHd2, mHu2, md2_1, md2_2, md2_3, me2_1,}
                        \term{me2_2, me2_3, ml2_1, ml2_2, ml2_3, mq2_1, mq2_2,}
                        \term{mq2_3, mu2_1, mu2_2, mu2_3}.\\
This is the \doublecrosssf{MSSM30atQ}{MSSM30atQ} with $Q=M_\mathrm{GUT}$; as per the \doublecrosssf{MSSM63atMGUT}{MSSM63atMGUT}, but with all off-diagonal elements in $\mathbf{m}^\mathbf{2}_Q$, $\mathbf{m}^\mathbf{2}_u$, $\mathbf{m}^\mathbf{2}_d$, $\mathbf{m}^\mathbf{2}_L$, $\mathbf{m}^\mathbf{2}_e$, $\mathbf{A}_u, \mathbf{A}_d$ and $\mathbf{A}_e$ set to zero, in order to suppress flavour-changing neutral currents:
\begin{align}
\mathbf{m}^\mathbf{2}_Q, \mathbf{m}^\mathbf{2}_u, \mathbf{m}^\mathbf{2}_d, \mathbf{m}^\mathbf{2}_L, \mathbf{m}^\mathbf{2}_e, \mathbf{A}_u, \mathbf{A}_d, \mathbf{A}_e\ \mathrm{diagonal}.
\end{align}

\gsfitemc{NUHM2} \term{A0, M0, M12, SignMu, TanBeta, mHd, mHu}.\\
The second Non-Universal Higgs Mass model. Descended from the \doublecrosssf{MSSM63atMGUT}{MSSM63atMGUT}.  All off-diagonal elements in $\mathbf{m}^\mathbf{2}_Q$, $\mathbf{m}^\mathbf{2}_u$, $\mathbf{m}^\mathbf{2}_d$, $\mathbf{m}^\mathbf{2}_L$ and $\mathbf{m}^\mathbf{2}_e$ are set to zero, and all diagonal elements are set equal to a universal sfermion mass $m_0$.  All gaugino masses are set to the universal mass $m_{1/2}$, and all entries in $\mathbf{A}_u$, $\mathbf{A}_d$ and $\mathbf{A}_e$ are set to a universal trilinear coupling $A_0$.  Global fits of this model have been performed in Refs.\ \cite{arXiv:1405.4289,Buchmueller:2014yva,CMSSM}.
\begin{align}
&\mathbf{m}^\mathbf{2}_Q, \mathbf{m}^\mathbf{2}_u, \mathbf{m}^\mathbf{2}_d, \mathbf{m}^\mathbf{2}_L, \mathbf{m}^\mathbf{2}_e\ \mathrm{diagonal},\\
&M_1 = M_2 = M_3 \equiv m_{1/2},\\
&(\mathbf{m}^\mathbf{2}_X)_{ii} \equiv m_0^2\hspace{1.5mm} \forall\ i\in\{1..3\}, X\in\{Q,u,d,L,e\},\\
&(\mathbf{A}_Y)_{ij} \equiv A_0\hspace{2mm} \forall\ i,j\in\{1..3\}, Y\in\{u,d,e\}.
\end{align}

\gsfitemc{NUHM1} \term{A0, M0, M12, SignMu, TanBeta, mH}.\\
The first Non-Universal Higgs Mass model, fitted in Refs.\ \cite{Buchmueller09,Mastercode12b,Strege13,MastercodeCMSSM,CMSSM}. As per the \doublecrosssf{NUHM2}{NUHM2}, but with a single Higgs mass parameter $m_H$:
\begin{align}
&m^2_{H_u} = m^2_{H_d} \equiv (m_H)^2.
\end{align}

\gsfitemc{CMSSM} \term{A0, M0, M12, SignMu, TanBeta}.\\
The Constrained MSSM, most notably fitted in recent years in Refs.\ \cite{Fittinocoverage,MastercodeCMSSM,Han:2016gvr,CMSSM}. As per the \doublecrosssf{NUHM1}{NUHM1}, but with $m_0$ playing the role of a fully universal scalar mass parameter:
\begin{align}
&m_H = m_0.
\end{align}

\gsfitemc{mSUGRA} \term{A0, M0, M12, SignMu, TanBeta}.\\
The most common definition of the minimal supergravity model; just a pseudonym for the \doublecrosssf{CMSSM}{CMSSM}.\footnote{Other authors define mSUGRA as a smaller subspace of the CMSSM; see Ref.\ \cite{Dudas:2012hx} for discussion and further references.}\vspace{2mm}

\end{description}

\subsubsection{Flavour EFT}
\label{flavEFT}

The study of rare meson decays is typically done within the framework of effective field theory (EFT), where squared matrix elements for decays from initial states $i$ to final states $f$ are calculated from $|\langle f |{\cal H}_{\rm eff}|i\rangle|^2$, using an interaction Hamiltonian
\begin{equation}
\mathcal{H}_{\rm eff}  =  -\frac{4G_{F}}{\sqrt{2}} V_{tb} V_{ts}^{*} \sum_{x} C_{x}(\mu) \mathcal{O}_x(\mu)\;.
\end{equation}
Here $\mu$ specifies the scale of the process, $V$ is the CKM matrix and $G_\text{F}$ is the Fermi constant.  $\mathcal{H}_{\rm eff}$ is decomposed into a linear combination of effective interactions $\mathcal{O}_x$ with Wilson coefficients $C_x$.  Some such interactions exist already in the SM, e.g.\
\begin{align}
\label{O7910}
&\mathcal{O}_7 = \frac{e}{(4\pi)^2} m_b (\overline{s} \sigma^{\mu\nu} P_R b) F_{\mu\nu} \;,\nonumber \\
&\mathcal{O}_9 =  \frac{e^2}{(4\pi)^2} (\overline{s} \gamma^\mu P_L b) (\bar{\ell} \gamma_\mu \ell) \;,  \nonumber \\
&\mathcal{O}_{10} =  \frac{e^2}{(4\pi)^2} (\overline{s} \gamma^\mu P_L b) (\bar{\ell} \gamma_\mu \gamma_5 \ell) \;,
\end{align}
whereas others, such as
\begin{align}
\label{OQ}
\mathcal{Q}_{1}& = \frac{e^2}{(4\pi)^2}(\bar{s} P_R b)(\bar{\ell}\,\ell) \;,\nonumber \\
\mathcal{Q}_{2}& = \frac{e^2}{(4\pi)^2}(\bar{s} P_R b)(\bar{\ell}\gamma_5 \ell) \;,
\end{align}
are almost exclusively the purvey of new physics.  In general, the interesting quantities for new physics are therefore the differences between the expected SM and BSM values,
\begin{align}
\Delta C_x \equiv C_{x,\text{BSM}} - C_{x,\text{SM}}.
\end{align}
More details can be found in the \flavbit paper \cite{FlavBit} and Ref.\ \cite{Mahmoudi:2008tp}.

\begin{description}

\gsfitemc{WC} \term{Re_DeltaC7, Im_DeltaC7, Re_DeltaC9,}\\
              \term{Im_DeltaC9, Re_DeltaC10, Im_DeltaC10,}\\
              \term{Re_DeltaCQ1, Im_DeltaCQ1, Re_DeltaCQ2,}\\
              \term{Im_DeltaCQ2}.\\

This model incorporates enhancements and suppressions to the real and imaginary parts of the Wilson coefficients of the effective operators $\mathcal{O}_7$, $\mathcal{O}_9$, $\mathcal{O}_{10}$, $\mathcal{Q}_{1}$ and $\mathcal{Q}_{2}$ (Eqs.\ \ref{O7910} and \ref{OQ}).

\end{description}

\subsubsection{Nuisance parameters}

These models contain values with significant uncertainties that can be essential for calculating signal rates (particularly in DM searches), but which are not part of a BSM model or the Standard Model.

\begin{description}

\item[\textbf{\textsf{Halo\_gNFW}}\label{Halo_gNFW}:] \term{alpha, beta, gamma, r_sun, rho0, rhos,}
                                                      \term{rs, v0, vesc, vrot}.\\
This as well as all other halo models specify the radial dark matter distribution $\rho(r)$ in the Milky Way and the local properties of dark matter relevant for direct detection and capture in the Sun. Specifically, this model corresponds to the generalized NFW profile
\begin{equation}
\rho(r) = \frac{2^{(\beta - \gamma)/\alpha} \rho_s}{(r/r_s)^\gamma \left[1 + (r/r_s)^\alpha \right]^{(\beta - \gamma) / \alpha}} \, ,
\label{eqn:gNFW}
\end{equation}
where $\gamma$ ($\beta$) describes the inner (outer) slope of the profile, $\alpha$ is the shape in the transition region around the scale radius $r=r_s$, and $\rho_s \equiv \rho(r_s)$ is the scale density. Furthermore, the local properties of dark matter are described by means of the local density $\rho_0$ as well as a Maxwell-Boltzmann velocity distribution boosted to the rest frame of the Earth,
\begin{equation}
\label{fv}
f(\mathbf{u}) = \frac{e^{-\left(\frac{\mathbf{u}+\mathbf{v}_{\rm LSR} + v_{\rm \odot, pec} + V_\oplus}{v_0}\right)^2}}{\pi^{3/2} v_0^3\erf \left(\frac{v_{\rm esc}}{v_0} \right) - 2\pi v_0^2 v_{\rm esc} e^{-\left(\frac{v_{\rm esc}}{v_0}\right)^2}}.
\end{equation}
Here, $v_0$ is the most probable speed of a DM particle with respect to the galactic halo, while $v_{\rm esc}$ denotes the local escape velocity~\cite{Akrami:2010dn}. The remaining parameters describe the relative motion of the Earth and the Galactic rest frame: $\mathbf{v}_{\rm LSR} = (0, v_{\rm rot}, 0)$ is the motion of the Local Standard of Rest in Galactic coordinates, with $v_{\rm rot}$ being the local disk circular velocity, $\mathbf{v}_{\rm \odot, pec} = (11, 12, 7)$\,km\,s$^{-1}$ is the peculiar velocity of the Sun \cite{Schoenrich:2009bx}, and $V_\oplus = 29.78$\,km\,$^{-1}$ is the Keplerian velocity of the Earth around the Sun. Notice that in this halo model the scale density $\rho_s$ and the local density $\rho_0$ are treated as independent parameters.

\item[\textbf{\textsf{Halo\_gNFW\_rho0}}\label{Halo_gNFW_rho0}:] \term{alpha, beta, gamma, r_sun, rho0,}
                                                                 \term{rs, v0, vesc, vrot}.\\
Same as \doublecrosssf{Halo\_gNFW}{Halo_gNFW}, but deriving the scale density $\rho_s \equiv \rho(r_s)$ from a given value of the local density $\rho_0 \equiv \rho(r_{\rm sun})$ via Eq.~\ref{eqn:gNFW}. Here, $r_{\rm sun}$ denotes the distance from the solar system to the Galactic center.

\item[\textbf{\textsf{Halo\_gNFW\_rhos}}\label{Halo_gNFW_rhos}:] \term{alpha, beta, gamma, r_sun, rhos,}
                                                                 \term{rs, v0, vesc, vrot}.\\
Same as \doublecrosssf{Halo\_gNFW}{Halo_gNFW}, but deriving the local density $\rho_0 \equiv \rho(r_{\rm sun})$ from a given value of the scale density $\rho_s \equiv \rho(r_s)$ via Eq.~\ref{eqn:gNFW}.

\item[\textbf{\textsf{Halo\_Einasto}}\label{Halo_Einasto}:] \term{alpha, r_sun, rho0, rhos, rs, v0,}
                                                            \term{vesc, vrot}.\\
Same as \doublecrosssf{Halo\_gNFW}{Halo_gNFW}, but assuming instead the Einasto profile for the radial distribution of dark matter in the Milky Way:
\begin{equation}
\rho(r) = \rho_s \exp \left\{-\frac{2}{\alpha} \left[ \left(\frac{r}{r_s}
\right)^\alpha -1 \right] \right\} \, ,
\label{eqn:Einasto}
\end{equation}
with $r_s$ referring to the scale radius, $\rho_s$ to the scale density, and
$\alpha$ describing the shape of the profile.

\item[\textbf{\textsf{Halo\_Einasto\_rho0}}\label{Halo_Einasto_rho0}:] \term{alpha, r_sun, rho0, rs, v0,}
                                                                       \term{vesc, vrot}.\\
Same as \doublecrosssf{Halo\_gNFW\_rho0}{Halo_gNFW_rho0}, but using the Einasto profile given by Eq.~\ref{eqn:Einasto}.

\item[\textbf{\textsf{Halo\_Einasto\_rhos}}\label{Halo_Einasto_rhos}:] \term{alpha, r_sun, rhos, rs, v0,}
                                                                       \term{vesc, vrot}.\\
Same as \doublecrosssf{Halo\_gNFW\_rhos}{Halo_gNFW_rhos}, but using the Einasto profile given by Eq.~\ref{eqn:Einasto}.

\item[\textbf{\textsf{nuclear\_params\_fnq}}\label{nuclear_params_fnq}:] \term{deltad, deltas, deltau, fnd,}
                                                                         \term{fns, fnu, fpd, fps, fpu}.\\
This model contains the nuclear matrix elements that parameterise the quark content of protons and neutrons, $f^{(N)}_{T_q}$, defined by
\begin{equation}
m_N f^{(N)}_{T_q} \equiv \langle N | m_q \bar{q} q | N \rangle \, ,
\end{equation}
where $N \in \{p,n\}$ and $q \in \{u, d, s\}$ \cite{Ellis:2008hf}. The model also contains the parameters $\Delta^{(p)}_q$ that describe the spin content of the proton.\vspace{2mm}

\item[\textbf{\textsf{nuclear\_params\_sigma0\_sigmal}}\label{nuclear_params_sigma0_sigmal}:] \term{deltad, deltas,}
                                                                                              \term{deltau, sigma0, sigmal}.\\
The same as \doublecrosssf{nuclear\_params\_fnq}{nuclear_params_fnq}, but with the 6 $f^{(N)}_{T_q}$ parameters replaced by the light quark content of the nucleon $\sigma_l$ and the quantity $\sigma_0$, defined as
\begin{align}
\sigma_l &\equiv m_l \langle N | \bar{u}u + \bar{d}d | N \rangle \label{eq:sigma_l}\,,\\
\sigma_0 &\equiv m_l \langle N | \bar{u}u + \bar{d}d - 2 \bar{s}s | N \rangle \label{eq:sigma_0}\, ,
\end{align}
where $m_l \equiv (1/2) (m_u + m_d)$. We take $\sigma_l$ and $\sigma_0$ to be the same for protons and neutrons \cite{Young:2013nn}.\vspace{2mm}

\item[\textbf{\textsf{nuclear\_params\_sigmas\_sigmal}}\label{nuclear_params_sigmas_sigmal}:] \term{deltad, deltas,}
                                                                                              \term{deltau, sigmal, sigmas}.\\
The same as \doublecrosssf{nuclear\_params\_fnq}{nuclear_params_fnq}, but with the 6 $f^{(N)}_{T_q}$ parameters replaced by $\sigma_0$ from Eq.~\ref{eq:sigma_0} and the strange quark content of the nucleon $\sigma_s$, which is defined as
\begin{equation}
\sigma_s \equiv m_s \langle N | \bar{s}s | N \rangle \, .
\end{equation}
Again, $\sigma_0$ and $\sigma_s$ are assumed to be the same for protons and neutrons \cite{Young:2013nn}.

\end{description}

\subsubsection{Toys}

\begin{description}

\gsfitemc{NormalDist} \term{mu, sigma}.\\
A simple test model consisting of two parameters: the width and central value of a Gaussian distribution.  This model is used in most of the toy examples discussed in this paper.

\gsfitemc{TestModel1D} \term{x}.\\
A one-dimensional test model, typically used for debugging simple prior transformations, or when a dummy model is required (as in the external model example of the \colliderbit paper \cite{ColliderBit}).

\item[\textbf{\textsf{demo\_A}, \textsf{demo\_B}, etc\label{demos}}:] \hspace{2mm}\\
These are additional example models available in the same header as \textsf{NormalDist} and \textsf{TestModel1D}, but commented out in order to keep from cluttering up the model hierarchy with fake models.

\end{description}

\section{User interface and input file}
\label{interface}

In this section we describe the general user interface of \GB.  This includes a
description of the available command line switches as well as a detailed
walk-through of the structure and content of the main initialisation file.
Further details about the functionality of the
dependency resolver, printers and scanners are given in the following
sections.

\subsection{Command line switches and general usage}
\label{switches}

\GB is run by executing the \term{gambit} executable.  The canonical way to launch a scan is to specify an initialisation file \metavar{myfile.yaml} with the \term{-f} switch, as in
\begin{lstterm}
gambit -f @\metavar{myfile.yaml}@
\end{lstterm}
The full set of command-line switches available is:
\begin{description}
\item\term{--version}\\ Print the \GB version number and exit.\vspace{2mm}
\item\term{-h/--help}\\ Display usage information and exit.\vspace{2mm}
\item\term{-f} \metavar{file}\\ Use instructions in \metavar{file} to start a scan.\vspace{2mm}
\item\term{-v/--verbose}\\ Run with full verbose output.\vspace{2mm}
\item\term{-d/--dryrun}\\ Perform a dry run of a scan.  \GB will resolve all dependencies and backend requirements, then list the function evaluation order to \term{stdout}, but won't actually start the scan. It will also produce necessary files and instructions for plotting the \cross{dependency tree} (see Sec.\ \ref{depresolver}).  Requires \term{-f}.\vspace{2mm}
\item\term{-r/--restart}\\ Restart a scan, overwriting any existing output. Requires \term{-f}. If \term{-r} is not specified and previous output exists matching the instructions in \metavar{file}, \GB will attempt to resume scanning based on that output.
\end{description}

\GB also has various diagnostic modes that provide information about its current configuration from the command line.  See Sec.\ \ref{diagnostics} for further information.

\subsection{The master initialisation file}

The master initialisation file of \GB is written in the \YAML
format.\footnote{See \url{http://www.yaml.org} for a definition of the
standard.  A compact introduction can be found at
\url{http://en.wikipedia.org/wiki/YAML}.  Note that \GB is also fully compatible at the module level with the SLHA1 \cite{Skands:2003cj} and SLHA2 \cite{Allanach:2008qq} formats for SUSY models; see Refs.\ \cite{SDPBit,DarkBit,ColliderBit,FlavBit} for details.}  \YAML is a `human-friendly, cross-language, Unicode-based data serialization language' that provides a general
framework for setting up nested structures of common native data types.  The
format is reminiscent of \python.  As
such, leading whitespace (\ie the indentation level) matters, and is part of the
syntax.

The top node of the master initialisation file is a dictionary that contains
eight entries.
\begin{description}
\item \yaml{Parameters} describes the scan
parameters for different models.
\item \yaml{Priors} describes the priors
to be placed on the scan parameters.
\item \yaml{ObsLikes} describes
observables and likelihoods that the user would like to be calculated in a scan.
\item \yaml{Rules} specifies additional rules to guide the resolution
of dependencies and backend requirements.
\item \yaml{Printer} provides details about how and where to
store the results of the scan.
\item \yaml{Scanner} provides information about the scanning algorithm to be adopted in a scan.
\item \yaml{Logger} chooses options for logging \GB messages during the scan.
\item \yaml{KeyValues} is an additional global option section.
\end{description}

Any number of other \YAML files can be imported to any section of the master initialisation file, using the {\footnotesize\color{black}\ttfamily{!}}\yaml{import} \metavar{other\_file.yaml} directive.  Imported files may import files of their own, and so on.

\subsection{Model and parameters}

\subsubsection{General setup and fast priors}
\label{Parameters}

Selecting models to scan and setting options for their parameters is done in the \yaml{Parameters} section of the master \YAML file, using the syntax:
\begin{lstyaml}
Parameters:
  @\metavar{model\_1}@:
     @\metavar{parameter\_1}@:
        # optional fast prior statements
     @\metavar{parameter\_2}@:
        # optional fast prior statements
     ...
  @\metavar{model\_2}@:
     # content as above
  @\metavar{model\_3}@:
  ...
\end{lstyaml}

For example, in the scalar singlet \YAML file that ships with \GB, \term{yaml_files/SingletDM.yaml}, this looks like:
\begin{lstyaml}
Parameters:

  # SM non-Higgs parameters.
  StandardModel_SLHA2: !import
   @\yamlvalue{include/StandardModel\_SLHA2\_scan.yaml}@

  # Nuclear matrix parameters.
  nuclear_params_sigmas_sigmal:
    sigmas:
      range: [19, @\yamlvalue{67}@]
    sigmal:
      range: [31, @\yamlvalue{85}@]
    deltau:  0.842
    deltad:  -0.427
    deltas:  -0.085

  # SM Higgs-sector parameters
  StandardModel_Higgs:
    mH:
      range: [124.1, @\yamlvalue{127.3}@]

  # Scalar singlet dark matter parameters
  SingletDM:
    mS:
      range: [45., @\yamlvalue{10000.}@]
      prior_type: log
    lambda_hS:
      range: [0.0001, @\yamlvalue{10.00}@]
      prior_type: log

  # Dark matter halo parameters
  Halo_gNFW_rho0:
    rho0:
      range: [0.2, @\yamlvalue{0.8}@]
    v0: 235.0
    vesc: 550.0
    vrot: 235.0
    rs: 20.0
    r_sun: 8.5
    alpha: 1
    beta: 3
    gamma: 1
\end{lstyaml}
Here we see that the SM parameters are imported from the \YAML fragment \term{yaml_files/include/} \term{StandardModel_SLHA2_scan.yaml}.

As this layout suggests, multiple models can be scanned simultaneously; for example a particle physics model, plus a DM halo model, plus a set of nuclear physics parameters. This allows for arbitrary physics models to be combined easily and fluidly.  This makes it simple to add new observables to existing scans even if they bring `baggage' in the form of additional free parameters.  The typical example is that of nuisance parameters.  Adding an observable that depends not only on the particle physics scenario, but also the assumed value of the top mass, for example, is easy: one adds the new observable to the \yaml{ObsLikes} section, and adds the value or range of top masses to consider when calculating that observable to the \yaml{Parameters} section.  Broader examples of the utility of this arrangement include observables with dual implications for both particle physics and cosmology, or for both BSM and neutrino physics.

The subsection following each parameter is an optional `fast prior' definition.  For the purposes of sampling parameter values, a prior is the portion of the probability distribution function for choosing parameter values that is independent of the likelihood, i.e. the sampling distribution determined \textit{prior} to any contact with data.  Many sampling algorithms (indeed, essentially all useful ones) apply \textit{additional} conditions designed to preferentially sample points that constitute better fits to data --- but one must always choose what initial prior to employ, independent of the sampling algorithm to be employed. The simplest example would be assigning independent flat distributions for each parameter, viz. `flat priors'.  When paired with a naive random scanner, this would lead to simple uniform sampling of the parameter values.

Using the \yaml{Prior} section (see Sec. \ref{priors} and Ref.\ \cite{ScannerBit}), \GB makes it possible to use any arbitrary prior in a scan --- but in most cases a very simple prior will suffice. The fast prior subsection provides a streamlined way to directly set such simple priors in the \yaml{Parameters} section, for each parameter.

Note that every parameter of every model mentioned in the \yaml{Parameters} section must be associated with some prior, either in the \yaml{Priors} section or via a fast prior in the \yaml{Parameters} section.  This applies even if the parameter is not actually used in a given scan.  In this case, the parameter should normally simply be set to some arbitrary constant value in the \YAML file.

In it's simplest form, the fast prior section can just specify such a value to fix a parameter to during a scan (a so-called `delta-function prior'):
\begin{lstyaml}
  @\metavar{model\_1}@:
     @\metavar{parameter\_1}@: 125.0
     @\metavar{parameter\_2}@: 750.0
\end{lstyaml}
The same thing can be achieved with
\begin{lstyaml}
  @\metavar{model\_1}@:
     @\metavar{parameter\_1}@:
       fixed_value: 125.0
     @\metavar{parameter\_2}@:
       fixed_value: 750.0
\end{lstyaml}
This syntax naturally extends to specifying an ordered set of points to cycle through, e.g.
\begin{lstyaml}
  @\metavar{model\_1}@:
     @\metavar{parameter\_1}@: [@\yamlvalue{125.0, 142.5, 119.0}@]
     @\metavar{parameter\_2}@:
       fixed_value: [@\yamlvalue{750.0, 2015.0, 38.0}@]
\end{lstyaml}

There may be cases where parameters spanning multiple models are equivalent and should thus be described as a single parameter.  \GB allows model parameters to be combined using the \yaml{same_as} keyword.  Thus, \metavar{parameter\_1} of \metavar{model\_1} can be set to be equal to \metavar{parameter\_2} of \metavar{model\_2} via a fast prior entry such as
\begin{lstyaml}
  @\metavar{model\_1}@:
     @\metavar{parameter\_1}@:
        same_as: @\metavar{model\_2}@::@\metavar{parameter\_2}@
        scale: @\metavar{scale}@
        shift: @\metavar{shift}@
\end{lstyaml}
Here, \metavar{model\_1}::\metavar{parameter\_1} will be automatically set from the value assigned to \metavar{model\_2}::\metavar{parameter\_2} at each point in the scan.  The keywords \yaml{scale} and \yaml{shift} can also be optionally specified; these scale the parameter by an amount \metavar{scale} and shift it by \metavar{shift}.  Thus, in the above example,
\begin{eqnarray}
 \lefteqn{\metavar{model\_1}::\metavar{parameter\_1} =} \nonumber\\
 & & \quad \metavar{shift} + \metavar{scale} * \metavar{model\_2}::\metavar{parameter\_2}.
\end{eqnarray}

When two models being scanned have parameter names in common, extra care needs to be taken.  \scannerbit treats each parameter of each model as fully separate by default, but any module functions that declare both models as allowed (either individually or in combination) will trigger a runtime error when \GB attempts to add the values of all parameters in both models to the \cpp{Params} pipe (cf. Sec. \ref{param_pipe}).  Usually this indicates poor module function design, although there are some use cases, where the \yaml{same_as} directive is in use, when it may be simplest to proceed without worrying which of the two models' common parameters appears in the \cpp{Params} pipe.  Users wishing to hack their way through such a situation can set the \cpp{ALLOW_DUPLICATES_IN_PARAMS_MAP} precompiler variable in \term{Elements/include/gambit/}\term{Elements/} \term{module\_}\term{macros\_}\term{incore.hpp} to \cpp{1}.

Other fast priors can be chosen via the \yaml{prior_type} keyword, which can be set to \yamlvalue{flat}, \yamlvalue{log} (uniform in the log of the parameter value), or various trigonometric functions (\yamlvalue{cos}, \yamlvalue{sin}, \yamlvalue{tan} or \yamlvalue{cot}), as in
\begin{lstyaml}
  @\metavar{model\_1}@:
     @\metavar{parameter\_1}@:
        prior_type: @\metavar{chosen\_prior}@
        range: [@\metavar{low}@@\yamlvalue{,}@ @\metavar{high}@]
     @\metavar{parameter\_2}@:
        prior_type: log
        range: [@\yamlvalue{5, 75}@]
\end{lstyaml}
The allowed values of the parameters are given by setting \yaml{range}.  The \yaml{scale} and \yaml{shift} parameters also work with \yaml{prior_type}, in just that same way as with \yaml{same_as}.

If no fixed value is given for a parameter, and both \yaml{prior_type} and \yaml{same_as} are absent but \yaml{range} is given, a \yamlvalue{flat} prior is assumed.

Additional custom priors can be be written as plugins for \scannerbit, and accessed by setting \yaml{prior_type: plugin}; details can be found in Ref.\ \cite{ScannerBit}.

\subsubsection{More involved priors}
\label{priors}
Certain priors introduce correlations between parameters.  This makes specifying a separate, unique prior for each parameter impossible.  Such multidimensional priors, operating on multiple parameters simultaneously, can only be declared in a separate \yaml{Prior} section of the main \YAML file.
\begin{lstyaml}
Priors:
  @\metavar{prior\_name}@:
    parameters: [@\metavar{model\_1}@::@\metavar{param1}@, @\metavar{model\_1}@::@\metavar{param2}@,
                   ...]
    prior_type: @\metavar{prior\_type\_1}@
    @\metavar{options}@
  @\metavar{other\_prior\_name}@:
    parameters: [@\metavar{model\_2}@::@\metavar{paramA}@, @\metavar{model\_2}@::@\metavar{paramB}@,
                   ...]
    prior_type: @\metavar{prior\_type\_2}@
    @\metavar{options}@
  ...
\end{lstyaml}
A multidimensional prior is defined under a new user-defined key such as \metavar{prior\_name}.  Each prior declared in this way must specify a vector of input parameters, a prior type, and any options required by the prior.  A list of prior types and their options can be obtained with the \GB diagnostic \term{gambit priors} (see Sec.\ \ref{priors diagnostic}).  Available multidimensional priors include Gaussian and Cauchy distributions, as well as the ability to specify any additional \scannerbit\ prior plugin present on a user's system; these are discussed in detail in Ref.\ \cite{ScannerBit}.

\subsection{\cpp{ObsLikes}: Target observables and likelihoods}
\label{ObsLikes}

Entries in this section determine what is calculated during a scan.  Each
entry lists a likelihood contribution or an observable that should be calculated
during the scan. (Likelihood functions and observables are largely the same
within \GB, the main difference being that the former are used to drive the scan,
whereas the latter are simply recorded.)  The minimal allowed entry has the form
\begin{lstyaml}
ObsLikes:
  - capability: @\metavar{example\_capability}@
    purpose: @\metavar{example\_purpose}@
  - ...
\end{lstyaml}
Here, \metavar{example\_capability} is the capability of the likelihood or
observable to be calculated, while \metavar{example\_purpose} is its role in
the scan.  The latter determines its treatment by the scanner and the printer system.  In
the simplest cases, \yaml{purpose} will be set to either
\yamlvalue{LogLike} or \yamlvalue{Observable}.\footnote{Alternative purposes
are relatively easy to arrange, but these are the conventional ones.  See Sec.\
\ref{stats} for further discussion.} In the case of a \yamlvalue{LogLike}, the
calculated quantity will be used as one of the likelihoods in the scan.  As a convention in
\GB, all likelihoods are given in terms of $\log\mathcal{L} = \ln$(likelihood).
In the case of an \mbox{\yamlvalue{Observable},} the calculated quantity will
be simply written as additional output and will be available for
later post-processing.

For example, the following entries from \term{yaml_files/} \term{SingletDM.yaml} ensure that the
likelihood from the dark matter relic density is included in the overall likelihood
function, and that the value of the relic density itself is saved in the output of the scan, for
every valid combination of model parameters:
\begin{lstyaml}
ObsLikes:

  # Relic density likelihood contribution
  - capability: lnL_oh2
    purpose:    LogLike

  # Relic density prediction
  - capability: RD_oh2
    purpose:    Observable

\end{lstyaml}

It will often happen that several module functions can provide the same
capability.  In order to remove such ambiguities, it is possible to specify the
requested quantity further by adding one or more of the following optional
arguments
\begin{lstyaml}
ObsLikes:
  - capability: @\metavar{capability}@
    purpose: @\metavar{purpose}@
    type: @\metavar{type}@
    function: @\metavar{function}@
    module: @\metavar{module}@
  - ...
\end{lstyaml}
Here, \metavar{type} specifies the \Cpp type of the module function that should be used to fulfil the requested capability,
\metavar{function} explicitly gives the name of a module function, and \metavar{module} demands that the function must come from
a specific module.  These additional
specifications in the \yaml{ObsLikes} section  are in fact just a convenient
shortcut for setting up the most common \doublecross{rules}{rule} for dependency resolution.  Dependency resolution rules
can be set up in far more generality in the separate \yaml{Rules} section, which we discuss below.

In the case of the purpose \mbox{\yamlvalue{LogLike},} the
\metavar{type} of the module function selected \emph{must} be \mbox{\cpp{double},} \mbox{\cpp{float},}
\cpp{std::vector<double>} or \cpp{std::vector<float>}, as the result will be
sent to the \cross{likelihood container} to contribute to the total likelihood function.  (This applies regardless of whether the user has specified the \metavar{type} explicitly, or left it to the dependency resolver to work out.) In the case of vectors, the likelihood container automatically sums all entries.

Finally, the additional option \yaml{printme} can be set for each \yaml{ObsLikes} entry, for example
\begin{lstyaml}
ObsLikes:
  - capability: @\metavar{example\_capability}@
    purpose: @\metavar{example\_purpose}@
    printme: true
  - ...
\end{lstyaml}
This option is \yaml{true} by default, meaning that by default \GB will attempt to record to disk (i.e. `print'; see Sec. \ref{printers}) the computed result of the each of the target observables/likelihoods. This is the behaviour that one almost always wants during a production scan, however by setting \yaml{printme} to \yaml{false} the user can tell \GB not to try to output the result of the thusly-flagged computation. It is useful to do this when testing and debugging new module functions, for example, because these often produce results that are not of a printable \Cpp type (and so attempting to print them would cause a runtime error, see Sec \ref{print_overloads}), yet one will often want to set these functions as \yaml{ObsLikes} targets just to ensure that \GB will run them.

\subsection{Rules: Dependency resolution and module options}
\label{Rules}

Entries in the \yaml{Rules} section determine the details of how the
likelihoods and observables listed in the
\yaml{ObsLikes} section are calculated in the scan.
In the rather common case that several different module functions can provide a capability
requested in the \yaml{ObsLikes} section, or several module functions can provide the neccessary
capability-type pair requested in another module function's \cross{dependency}, then
further specifications in the \yaml{Rules} section are required to fully define the scan.
The \yaml{Rules} section can likewise be used to control the
resolution of backend requirements, and to set options for individual module
functions, modules and backend initialisation functions.

\subsubsection{Module function dependencies}
\label{regular_dependencies}
In the rather common case that several different module functions provide the
same requested quantity, further rules are necessary to define the scan.  Note
that with quantity, we refer here specifically to capability/type pairs,
\metavar{quantity} $\equiv$ (\metavar{capability}, \metavar{type}).  These rules
can be specified in the \yamlvalue{Rules} section of the initialisation file.
Furthermore, this section is used to control the resolution of backend
dependencies, and to set options for individual module functions, modules and
backend initialisation functions.  In this sense, the rules determine how an
individual point is calculated during the scan.

In the simplest case, a rule has the form
\begin{lstyaml}
Rules:
  - capability: @\metavar{capability}@
    type: @\metavar{type}@
    function: @\metavar{function}@
    module: @\metavar{module}@
\end{lstyaml}
where \yaml{capability} is required, \yaml{type} is optional, and one
or both of the entries \yaml{function} and \yaml{module} must be
given.  This entry translates into the rule: Any \metavar{capability} with \Cpp
type \metavar{type} should be resolved by module function \metavar{function}
from the module \metavar{module}.  Assigning the empty string \yamlvalue{""} or the wildcard character \yamlvalue{"*"} to an
entry is equivalent to omitting it.  If regex is activated (this is
\textit{not} the default; see Sec.\ \ref{keyvalues}), all
entries are actually treated as regular expressions, allowing rules to be made
arbitrarily complex.\footnote{For details about regular expressions we refer the reader to
\url{https://en.wikipedia.org/wiki/Regular\_expression}.}

A simple example of such a rule is the one in \term{yaml_files/SingletDM.yaml} that specifies that the observed relic density
should be treated as an upper limit only when computing the likelihood.  This allows for the possibility that
some of the dark matter is not in the form of scalar singlet particles.
\begin{lstyaml}
# Choose to implement the relic density likelihood
# as an upper bound, not a detection
- capability: lnL_oh2
  function: lnL_oh2_upperlimit
\end{lstyaml}
This rule says that wherever the capability \cpp{lnL_oh2} is needed in a scan, \GB must use a function with the name \cpp{lnL_oh2_upperlimit}.  As it turns out, there is only one function with such a name in \GB\ \textsf{1.0.0}, and it lives in \darkbit\ -- so this rule forces \cpp{DarkBit::lnL_oh2_upperlimit} to be used.

The simple form shown above applies a rule to the resolution of dependencies of \emph{any} module functions matching the specified \yaml{capability} and \yaml{type}.  In order to set up rules that only affect the
dependency resolution of a specific module function, one can add a dedicated \yaml{dependencies} subsection, and optionally omit any of the top-level keys \yaml{capability}, \yaml{type},
\yaml{function} and \yaml{module} (or equivalently, set them to \yamlvalue{""} or \yamlvalue{"*"}).
\begin{lstyaml}
Rules:
  - capability: @\metavar{capability}@
    type: @\metavar{type}@
    function: @\metavar{function}@
    module: @\metavar{module}@
    dependencies:
    - {capability: @\metavar{cap\_A}@, type: @\metavar{type\_A}@,
       function: @\metavar{func\_A}@, module: @\metavar{mod\_A}@}
    - {capability: @\metavar{cap\_B}@, type: @\metavar{type\_B}@,
       function: @\metavar{func\_B}@, module: @\metavar{mod\_B}@}
    - ...
  - ...
\end{lstyaml}
If regex is activated, the values are treated as regular expressions.
The entry translates into the following rule:  when resolving dependencies
of module function \metavar{function} in module \metavar{module}, which
provides capability \metavar{capability} with \Cpp type \metavar{type}, apply
the rules listed under the keyword \yaml{dependencies}.

If conflicting rules are found during dependency resolution, \GB will throw an error.
This is intended to reduce side effects that changes in some parts of the initialisation file can
have on other parts.  However, rules can be \emph{explicitly} declared as
weak and over-rideable, by using the flag \yaml{!weak}, as per
\begin{lstyaml}
Rules:
  - !weak
    capability: @\metavar{capability}@
    type: @\metavar{type}@
    function: @\metavar{function}@
    module: @\metavar{module}@
  - ...
\end{lstyaml}
Note that the flag affects the \emph{entire} rule for which it is set, not only
specific nearby keywords.

A special case can occur if several module functions depend on
the same quantity as they provide.  In this case these module functions can be
chained, and setting up such chains in the rules section is simplified by using
the keyword \yaml{functionChain}.  This is illustrated in the following
example, where \metavar{func1}, \metavar{func2} and \metavar{func3} are
supposed to provide as well as depend on \metavar{capability} with
\metavar{type}.  These functions will be chained together, with \metavar{func1}
fulfulling the depenencies of \metavar{func2} etc.
\begin{lstyaml}
Rules:
  - capability: @\metavar{capability}@
    type: @\metavar{type}@
    functionChain: [@\metavar{func1}@, @\metavar{func2}@, @\metavar{func3}@]
    module: @\metavar{module}@
\end{lstyaml}

Finally, when performing \yaml{type} matching, the dependency resolver takes all type equivalences defined in \term{config/}\term{resolution\_}\term{type\_equivalency\_}\mbox{\term{classes.yaml}} into account.  We discuss this type equivalency database for dependency resolution in more detail in Sec.\ \ref{types}.

\subsubsection{Backend requirements}
\label{backend_rules}

After a module function has been selected to take part in a scan, its backend requirements are resolved.  This process can be guided and controlled using rules for backend requirements, which have the form:
\begin{lstyaml}
Rules:
  - capability: @\metavar{capability}@
    type: @\metavar{type}@
    function: @\metavar{function}@
    module: @\metavar{module}@
    backends:
    - {capability: @\metavar{cap\_A}@, type: @\metavar{type\_A}@,
       function: @\metavar{func\_A}@, backend: @\metavar{backend\_A}@,
       version: @\metavar{backend\_A\_version\_number}@}
    - {capability: @\metavar{cap\_B}@, type: @\metavar{type\_B}@,
       function: @\metavar{func\_B}@, backend: @\metavar{backend\_B}@,
       version: @\metavar{backend\_B\_version\_number}@}
    - ...
\end{lstyaml}
The usage is essentially identical to the one discussed above for dependencies, except that
\yaml{backend} may be specified rather than \yaml{module}, and a specific version of a backend may be requested, as e.g.
\begin{lstyaml}
  - capability: Higgs_Couplings
    backends:
    - {backend: FeynHiggs, version: 2.11.3}
\end{lstyaml}
There are also a number of other restrictions that can be applied via rules declared in the module function's rollcall header entry (Sec~\ref{declaration_bereq}).  These include backend requirements that are only activated for specific models (analogous to model-conditional dependencies of module functions), restricted lists of permitted backends and versions, and the condition that certain combinations of backend requirements must be resolved by the \emph{same} version of the \emph{same} backend.

\subsubsection{Options for module functions}
\label{rules_options}

Besides setting rules for the resolution of dependencies and backend requirements, the \yaml{Rules} section can also be used to set options for module functions.  This is done with the \yaml{options} keyword, as
\begin{lstyaml}
Rules:
  - capability: @\metavar{capability}@
    type: @\metavar{type}@
    function: @\metavar{function}@
    module: @\metavar{module}@
    options:
      @\metavar{key\_A}@: @\metavar{value\_A}@
      @\metavar{key\_B}@: @\metavar{value\_B}@
      ...
  - ...
\end{lstyaml}
This rule sets the option \metavar{key\_A} to \metavar{value\_A} and option \metavar{key\_B} to \metavar{value\_B}, for any module function
that matches the indicated \metavar{capability}, \metavar{type},
\metavar{function} and \metavar{module}.  Any of these keywords can be omitted; if
regex is activated, they are treated as regular expressions.  This allows, for
instance, module-wide options to be set using just the name of the module,
whilst omitting the other three keywords or setting them to wildcards:
\begin{lstyaml}
Rules:
  - module: DarkBit
    options:
      DM_is_made_of: axions
\end{lstyaml}
Here, the key \yaml{DM_is_made_of} is accessible by all module functions
in the module \yamlvalue{DarkBit}.

This last example is a bit glib, as in reality \yaml{DM_is_made_of} is not a recognised option of any functions in \darkbit, so
setting it doesn't actually have any effect in \GB\ \textsf{1.0.0}.  A more realistic example is:
\begin{lstyaml}
Rules:
  # Use the DarkBit native calculator
  # to compute the relic density
  - capability: RD_oh2
    function: RD_oh2_general
    options:
      fast: 1
\end{lstyaml}
This can be seen in e.g.\ \term{yaml_files/SingletDM.yaml}.  This rule specifically selects the \cpp{RD_oh2_general} function from \darkbit for calculating capability \cpp{RD_oh2} (i.e.\ the relic density), and passes it the option \cpp{fast = 1}, to set the accuracy required when solving the Boltzmann Equation for the thermal relic density of scalar singlet particles.

The key-value pairs specified in this way are
easily accessed by any module function that matches a given rule, using
\lstinline{runOptions->getValue} (cf. Sec.~\ref{module_options}).

In most cases, module functions will interpret option values as simple \Cpp types (commonly
\cpp{float}, \cpp{int}, \cpp{bool} or \cpp{std::string}), but composite
types like \cpp{std::vector<double>} can also be set.  The necessary syntax for doing this is defined by the \YAML standard.  Options can also be easily nested, with the \Cpp type of the top-level option to be retrieved itself a \cpp{YAML::Node}\footnote{This class is defined in the contributed package \textsf{yaml-cpp}, which ships with \GB.  Documentation is available at \href{http://github.com/jbeder/yaml-cpp}{http://github.com/jbeder/yaml-cpp}.}, from which lower-level options can then be retrieved.

Information about what options are available for which module function can be
found in the module function documentation.  Options that are never requested by module functions at runtime are silently ignored.

In case of ambiguity, such as when an option requested by a module function is
listed in several matching rules, \GB throws an error during initialisation.

\subsection{Printer}
\label{printer_setup}
The \GB ``printer'' system handles the output of all scan results, whether to disk, a network resource or any other output stream.  This system allows all \GB output to be handled in an abstract way throughout the code, with the actual format of the output being decided by the choice of an output plugin (a \cross{printer}) at runtime, via the master \YAML file.  Therefore, setting up \GB output consists primarily of choosing a printer and setting options for it.  In this section we describe how to do this; full details of the printer system can be found in Sec.\ \ref{printers}.

Note that output handled by the \GB printer system is essentially independent of other output that might be created by any backend or scanner codes. This allows the output to remain as uniform as possible, regardless of the scanning algorithm and external codes being used.

\GB \textsf{1.0.0} ships with two printers: \textsf{ascii} and \textsf{hdf5}. The \textsf{ascii} printer outputs data as a simple ASCII table, whereas the \textsf{hdf5} printer writes data to a binary file in \textsf{HDF5} format\footnote{\href{https://www.hdfgroup.org/HDF5/}{https://www.hdfgroup.org/HDF5/}}. The former format is useful for its simplicity, however the latter is far superior when dealing with large datasets, particularly in terms of disk usage and read/write speed. We have also upgraded the external analysis tool \pippi \cite{pippi} to accept \GB input in these formats; it can be easily retrieved via the \GB build system (Sec.\ \ref{misc_build}).

Most options that affect the output system are entered in the \yaml{Printer} section of the master \YAML file. The basic layout of this section is:
\begin{lstyaml}
Printer:
  printer: @\metavar{plugin\_name}@
  options:
    @\metavar{option\_1}@: @\metavar{value\_1}@
    @\metavar{option\_2}@: @\metavar{value\_2}@
    ...
\end{lstyaml}
That is, one chooses a plugin \metavar{plugin\_name} and sets its options, which vary with the plugin. In the next sections we describe the options available in each printer.

\subsubsection{Common options}
\label{common_printer_setup}
These options are common to both the \textsf{ascii} and \textsf{hdf5} printers:
\begin{lstyaml}
  options:
    output_path: @\metavar{default\_output\_path}@
    output_file: @\metavar{filename}@
\end{lstyaml}

\begin{description}
\item[\YAMLkeystyle \texttt{output\_path}] specifies the directory in which the printer output will be stored. By default it is set to the value of \yaml{default_output_path} as set in the \yaml{KeyValues} section of the input \YAML file (see Sec. \ref{keyvalues}), however if a value is set here it will override that default.
\item[\YAMLkeystyle \texttt{output\_file}] specifies the name of the file in which to store data generated during the run. If it does not exist then it will be created.
\end{description}

\subsubsection{Specific options: \textsf{ascii} printer}
\label{ascii_printer_setup}
The only specific option for this plugin is \yaml{buffer_length}, which defaults to a value of \yamlvalue{100}:
\begin{lstyaml}
Printer:
  printer: ascii
  options:
    buffer_length: 100
\end{lstyaml}
This specifies the size of the internal buffer used by the printer. A value of \metavar{N} will cause output to be written to disk every \metavar{N} model points.  If model points are slow to evaluate, it can be useful (particularly during testing) to set \yaml{buffer_length} to \yamlvalue{1} so that output is generated frequently. However, if model points are evaluated extremely rapidly then frequent writing of output will create a significant bottleneck, and a high value of \yaml{buffer_length} will be more appropriate.

\subsubsection{Specific options: \textsf{hdf5} printer}
\label{hdf5_printer_setup}
There are three specific options for this plugin:
\begin{lstyaml}
Printer:
  printer: hdf5
  options:
    group: "/"
    delete_file_on_restart: false
\end{lstyaml}
The first is \yaml{group}, which defaults to \yaml{"/"}. This option specifies the name of the group \emph{within} the host \textsf{HDF5} \yaml{output_file} in which data will be stored. \textsf{HDF5} files are structured similarly to a filesystem (i.e.\ hierarchically) and a `group' is analogous to a directory. Various objects (such as datasets, and other groups) are then stored within groups\footnote{See \href{https://www.hdfgroup.org/HDF5/doc/Glossary.html}{https://www.hdfgroup.org/HDF5/doc/Glossary.html} for further description of `groups' and `datasets' in \textsf{HDF5}.} The default value of \yaml{"/"} specifies the root group, and this option should rarely need to be set to anything else. A deeper-layer group can be specified e.g. as \yaml{"/group1/group2/etc/"}. Absent groups at any layer will be automatically created.

The second option is \yaml{delete_file_on_restart}. This option is mainly a convenience for performing repeated test scans, and causes the file specified by \yaml{output_file} to be deleted if it already exists when a run restarts (i.e. if the \term{-r} command line flag is used, see Sec. \ref{switches}). By default this is false, meaning that if a \textsf{HDF5} file already exists matching the name given in \yaml{output_file} then \GB will attempt to \emph{add} the data for the run to this pre-existing file.

Further details of the \textsf{HDF5} objects that \GB writes to disk via this printer can be found in Sec. \ref{printers}. Note that results from several runs can be stored inside the same \textsf{HDF5} file by storing the data in different groups, however it is safer to use separate files because \textsf{HDF5} files are vulnerable to corruption from write errors (which in principle can occur if \GB terminates abnormally; see Sec. \ref{resume} for safe early shutdown methods), and data recovery is difficult. If \yaml{delete_file_on_restart} is \yaml{false} and the chosen \yaml{group} already exists, \GB will throw a runtime error telling you to choose a different group or overwrite the whole file. Groups can be deleted, however the disk space they occupy cannot be reclaimed without copying the entire contents of the \textsf{HDF5} file into a new file, e.g. using the \textsf{h5repack} command line tool\footnote{\href{https://www.hdfgroup.org/HDF5/doc/RM/Tools.html\#Tools-Repack}{https://www.hdfgroup.org/HDF5/doc/RM/Tools.html\#Tools-Repack}}. We leave these kind of file manipulations to the user.

\subsubsection{Output selection}
The outputs handled by the printer system are simply the results of module function evaluations. However, not all module function results are of a \Cpp type that can be `printed' with every printer (see Sec.\ \ref{print_overloads} for the restrictions), so \GB cannot automatically output \emph{all} results. To instruct \GB to write the result of a calculation to an output stream, the module function that computes a result must be selected to fulfil one of the capabilities requested in the \yaml{ObsLikes} section of the master \YAML file. Intermediate results, computed by functions run by the dependency resolver only in order to fulfil dependencies of other functions, are not output.

\subsection{Scanner}
\label{scanner_yaml}
\GB ships with a variety of \doublecross{scanner plugins}{scanner plugin} that can be used in a ``plug and play'' manner.  A full list of scanner plugins can be obtained from the \GB scanners diagnostic (Sec.\ \ref{scanners diagnostic}).  A scanner is selected by specifying one of these plugins and any plugin-specific options in the \yaml{Scanner} section of the \YAML file, e.g.
\begin{lstyaml}
Scanner:
  use_scanner: nested_sampler
  scanners:
    nested_sampler:
      plugin: MultiNest
      like:  LogLike
      nlive: 4000
      tol: 0.5
      mmodal: 1
    other_sampler:
      plugin: ...
      ...
    ...
\end{lstyaml}
The \yaml{Scanner} section can contain multiple scanner definitions with user-defined names, such as \yaml{nested_sampler} and \yaml{other_sampler} in the above example.  The scanner that will actually be used in a given scan is specified with the \yaml{use_scanner} key.  Within the \YAML scanner definitions, the \yaml{plugin} option must be set to a valid scanner plugin known to \GB, and any necessary/desired options for that scanner should also be set.  Note that a typical scanner plugin requires a \cross{purpose} to use for its objective function, such as \yamlvalue{LogLike} or \yamlvalue{Observable}; this is provided by setting the \yaml{like} option in the example of the \yamlvalue{MultiNest} plugin.  Valid and required plugin options, plugin descriptions, and the status of a plugin can be obtained through the \GB free-form diagnostic (see Sec. \ref{free-form diagnostic}),
\begin{lstterm}
gambit @\metavar{plugin\_name}@
\end{lstterm}
where \metavar{plugin\_name} is the name of the scanner plugin.

\GB also ships with a number of simple objective test functions, which can be used as objective functions for a scan in place of the regular \GB \cross{likelihood container} output, for testing scanners and other parts of the code.  These exist as \doublecross{test function plugins}{test function plugin} in \scannerbit, and are accessed from the main \YAML file with similar syntax to scanners, e.g.
\begin{lstyaml}
Scanner:
  use_objectives: my_test_function
  objectives:
    my_test_function:
      plugin: uniform
      parameter_A: 10
      parameter_B: false
    other_test_function:
      plugin: ...
      ...
    ...
\end{lstyaml}
As the \yaml{use_objectives} directive suggests, multiple test functions can be specified with the regular \YAML \yaml{[}\yamlvalue{x},\yamlvalue{y}\yaml{]} syntax if desired, in which case all the listed objectives will be multiplied to form the actual objective function to be used in the scan.  Details of the available test functions and their options can be found in the \scannerbit paper \cite{ScannerBit}.

\subsection{Logger}

The logging output of a scan can be directed to various output files.  This is
done using entries of the form:
\begin{lstyaml}
Logger:
  prefix: @\metavar{output\_path}@
  redirection:
    [Scanner, Warning] : "scanner_warnings.log"
    [ExampleBit_A] : "ExampleBit_A.log"
    ...
\end{lstyaml}
Here \yaml{prefix} specifies the output location for log files (defaulting to \yaml{default_output_path}; cf. Sec.\ \ref{keyvalues}), and the entries in the \yaml{redirection} subsection dictate which logging messages go to which file.  These options are discussed further in Sec.\ \ref{logs}.

\subsection{\cpp{KeyValues}: general purpose options}
\label{keyvalues}

Most of the general behaviour of \GB is controlled by various options in the
\yamlvalue{KeyValues} section.  The syntax is the same as described above in the
context of the module function options.  We provide here a complete list of
available options.  Where we indicate concrete values, these are the default
values that will be used if the option is omitted; where no default is indicated, the option is required.

\begin{lstyaml}
KeyValues:

  likelihood:
    # The value of the log-likelihood to assign to
    # invalid points.  Also the log-likelihood value
    # below which an otherwise valid point is declared
    # invalid.
    model_invalid_for_lnlike_below: @\metavar{lnlike\_min}@
    # Alternative value of the log-likelihood to
    # assign to invalid points later in a scan (e.g.
    # with the MultiNest scanner; see @\cite{ScannerBit}@).
    model_invalid_for_lnlike_below_alt: #defaults to
    # 0.5*@\metavar{lnlike\_min}@.

    # Print likelihood debug information to stdout and
    # logs, including parameter values and
    # contributions of individual likelihood
    # components. Set true automatically if the master
    # debug flag (below) is true.
    debug: false

  exceptions:
    # Set the fatality of different exceptions (see
    # Sec.@\,\ref{exceptions}@).  By default, all
    # errors are fatal and all warnings non-fatal.
    core_warning: non-fatal
    core_error: fatal
    ExampleBit_A_warning: non-fatal
    ExampleBit_A_error: non-fatal
    ...

  dependency_resolution:
    # If multiple module functions can resolve the
    # same dependency, prefer the one that is more
    # tailored for the scanned model.  See Sec.@\,\ref{depres_general}@.
    prefer_model_specific_functions: true
    # Interpret rules in terms of regular expressions
    use_regex: false
    # Print running average runtime for all functions
    # in dependency resolver logs
    log_runtime: false

  # Print timing information into hdf5 output
  print_timing_data: false

  # Root prefix to use in all output paths. The
  # default value is based on the input \YAML file
  # name, with the (final) file extension removed.
  default_output_path: "runs/@\metavar{inifile\_name}@/"

  # Call MPI_ABORT when attempting to shut down. Many
  # implementations of MPI_ABORT are buggy and do not
  # abort other MPI processes properly; in these
  # cases, set this option false to let GAMBIT try to
  # abort things its own way.
  use_mpi_abort: true

  # Pick a random number generator engine.
  # See Sec.@\,\ref{random numbers}@ for details.
  rng: default # default = mt19937_64 in GAMBIT 1.0.0

  # Turn on master debug mode. Implies
  # Logger:debug=true and
  # KeyValues:likelihood:debug=true
  debug: false

\end{lstyaml}

\section{Dependency Resolver}
\label{depresolver}

\begin{figure*}[tp]
\centering
\includegraphics[width=0.85\textwidth]{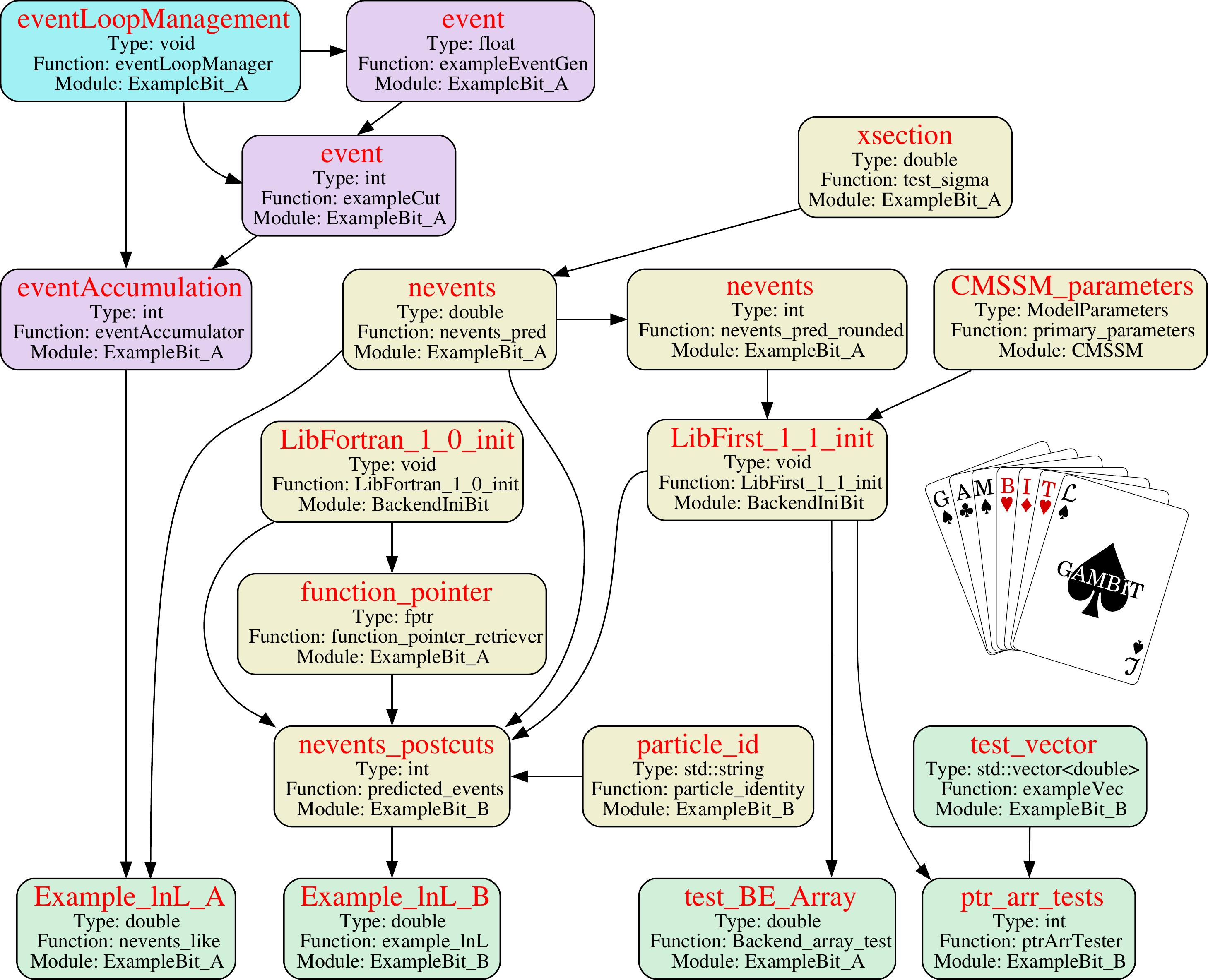}
\caption{An example \protect\cross{dependency tree} generated in the initialisation stage of a \GB scan.  Each block corresponds to a single \protect\cross{module function}, with the red text indicating its \cross{capability}.
Arrows indicate resolution of \doublecross{dependencies}{dependency} of different module functions with the results of others.  The functions selected by the \protect\cross{dependency resolver} to provide the observables and likelihoods requested in the \protect\yaml{ObsLikes} section of the scan's input \YAML file are shaded in green.  Module functions shown shaded in purple are \protect\doublecross{nested module functions}{nested module function}.  These run in an automatically-parallelised loop managed by a \protect\cross{loop manager} function, which is shown shaded in blue.  This example is included in the \GB distribution as \protect\term{spartan.yaml}; see Sec.\ \protect\ref{examples} for more details.  Figures like this can be generated for any scan by following the instructions provided after calling \GB with the \protect\term{-d} switch; see Sec.\ \protect\ref{switches} for details.}
\label{fig::deptree}
\end{figure*}

The \cross{dependency resolver} runs during the initialisation stage of a \GB scan.
It determines which module functions are required for a specific scan, infers
their initial evaluation order, and connects their \doublecross{pipes}{pipe}.  A major part of this plumbing exercise is constructing the \cross{dependency tree} of a scan, a
directed acyclic graph with dependency pipes as the connectors (`edges' in graph language) and module functions as the nodes.
Roughly speaking, the dependency tree starts at its `top' with the scanned
models and their parameters, and terminates at the `bottom' with functions that provide the likelihoods
and observables requested in the \yaml{ObsLikes} section of the scan's initialisation file (Sec.\ \ref{ObsLikes}).  An example can be seen in Fig.\ \ref{fig::deptree}. The construction of a
valid dependency tree will happen mostly automatically, and depends only on the
declarations in the module and backend \doublecross{rollcall headers}{rollcall
header}.  However, it is rather common in \GB that there are several ways to
calculate the same thing, in which case additional rules have to be specified
in the input file (Sec.\ \ref{Rules}).

\subsection{General procedure}
\label{depres_general}

The steps of dependency resolution are:

\begin{enumerate}
  \item Disable all module and backend functions not compatible with the models being scanned.
  \item Based on the entries of the \yaml{ObsLikes}
    section, make a list of initially requested quantities; this is the initial dependency queue.
  \item Pick an unresolved quantity from the dependency queue, along with a designated target.
    Entries in the initial dependency queue can be thought of as having the chosen printer as their target.
  \item Make a list of module functions that can provide the requested quantity.
  \item If the \yaml{KeyValues} entry \yaml{prefer_model_specific_functions} is \yaml{true}:\begin{itemize}
    \item If any module functions on the list are tailor-made for the scanned models, remove all other module functions from the list.
    \item If any module functions on the list are tailor-made for ancestors of the scanned models, keep only the module functions most closely related to the scanned models.
  \end{itemize}
  \item Adopt the \yaml{Rules} specified in the initialisation file (see Sec.\ \ref{Rules}), removing non-matching module functions from the list.
  \item If exactly one module function is left on the list, resolve the
    quantity requested by the target function with the capability provided by
    that module function.  This automatically connects the pipe of the
    target function to the result of the resolving function.
  \item If the resolving function was not already activated for the scan,
    activate it and add its dependencies to the dependency queue (with the
    resolving function as new target function).
  \item Resolve backend requirements, as described below.
  \item Resolve module function options, as described below.
  \item Repeat from step 3 until the dependency queue is empty.
\end{enumerate}

\subsection{Evaluation order}

After building up the dependency tree of module functions, the dependency resolver determines
the initial runtime ordering of its chosen module functions.  An obvious minimal
requirement is that if the output of module function $\mathcal{A}$ is required
by module function $\mathcal{B}$, then $\mathcal{A}$ must be evaluated before
$\mathcal{B}$.  We do this by topologically sorting the directed dependency
tree, using graph-theoretic methods from the \textsf{Boost Graph Library}.\footnote{\href{http://www.boost.org/doc/libs/1_63_0/libs/graph/doc/}{http://www.boost.org/doc/libs/1\_63\_0/libs/graph/doc/}.  The reader may wonder how mutually-dependent quantities should be dealt with, i.e.\ in cases where the values of $\mathcal{A}$ and $\mathcal{B}$ are defined by a set of equations that must be solved simultaneously, by means of iteration.  Take the calculation of precision values of $m_W$ and $m_h$ in the MSSM for example, where each depends on the other.  GAMBIT does not provide any option for doing such iterative calculations directly through the dependency tree.  Generally the way to deal with such a situation is to either\begin{enumerate}
\item write a module function that can compute the two quantities simultaneously and self-consistently (i.e. that does the iteration internally), returning them both as its result, or
\item use function pointers as return values of module functions.
\end{enumerate}
For option 2, take the Higgs mass example. If a module author wishes to permit the user to choose from two possible expressions for $m_h$ that depend on $m_W$, they would first write the two expressions as functions of $m_W$.  Call these expressions $f(m_W)$ and $g(m_W)$.  The author would then write one or more module functions that return a pointer to $f$ or $g$.  The module function that computes $m_W$ should then depend on a pointer to a Higgs mass function, and then just call it (either $f$ or $g$, depending on which one the user chooses) whilst it does its iterative calculation of $m_W$.  It should then return its final value of $m_W$.  Another module function responsible for computing $m_h$ should then depend on both the value of $m_W$, and the pointer to the same Higgs mass function ($f$ or $g$).  This module function then simply takes the previously computed value of $m_W$, passes it to the function pointed to by its dependency on the Higgs mass function pointer, retrieves the final value of the Higgs mass, and returns it as its own result.}

In most cases, the evaluation order of the observables and likelihoods listed in the
\yaml{ObsLikes} section (Sec.~\ref{ObsLikes}) remains unconstrained by the topological
sorting.  The dependency resolver first orders the likelihoods
by estimating the expected evaluation time for each one, including all dependent module functions,
along with the probability that each likelihood will invalidate a point.  (A point may be invalidated if the likelihood is extremely close to zero, the point is unphysical, etc.)  These estimates are based on the
runtime and invalidation frequency of the previously calculated points, and
updated on the fly during the scan.  The dependency resolver then sorts the evaluation order of likelihoods
such that the expected average time until a point is invalidated is
minimised.  In practice this means that, for instance, the relatively fast
checks for consistency of a model with physicality constraints,
such as perturbativity and the absence of tachyons, would be automatically
performed before the often time-consuming evaluation of collider constraints.
This gives a significant efficiency gain in a large scan, because expensive
likelihoods are not even evaluated for points found to be invalid or sufficiently
unlikely on the basis of faster likelihoods.

Observables not associated with likelihoods used to drive a scan (cf.\ \ref{ObsLikes}) are always calculated after the likelihood components, as they do not have the power to completely invalidate a model point.  Invalid observable calculations can still be flagged, but they will not trigger the termination of all remaining calculations for that point in the way that an invalid likelihood component will.

\subsection{Resolution of backend requirements}

Resolving backend requirements is in some sense a lot easier than resolving module function dependencies, in that backend requirements cannot themselves have explicit backend requirements nor dependencies, so there is no equivalent of the dependency tree to build.  However, the ability to specify groups of backend functions from which only one requirement must be resolved, along with rules that apply to them (Sec.\ \ref{declaration_bereq}), especially the declaration that backend requirements that share a certain tag must be resolved from the same backend --- without necessarily specifying \textit{which} backend --- makes backend resolution a uniquely challenging problem.

The dependency resolver employs an iterative approach to backend resolution.  It performs multiple passes over the list of backend requirements, choosing to defer resolution of ambiguous requirements until resolution of other requirements makes it possible to uniquely resolve the initial requirements.  The overall strategy proceeds as follows:
\begin{enumerate}
\item Create a new group of backend requirements, consisting of all requirements that were declared in the rollcall header without a group.  This will be a special group; unlike declared groups, all requirements in this group must be resolved rather just one.
\item Create a queue of all groups of backend requirements.
\item Choose a group from the queue.
\item \begin{enumerate}\item If the group is a regular group, iterate through all available backend functions and retain those that fulfil all rules of the group.  If no backend function exists that satisfies all rules, throw a runtime error.  If only one such function exists, resolve the group backend requirement with it.  If multiple solutions are found, but one or more of them is subject to a rule linking it to another backend requirement, flag the group as one whose resolution should be deferred until other backends have been resolved.
\item If instead the group is the special one, iterate over all requirements in the group, attempting one by one to find a unique backend function that fulfils each requirement.  Fail if no solution exists to any one of these requirements. If just one solution exists to a given requirement, resolve the requirement with it.  If no unique solution is found for some requirement, but one or more candidates is subject to a rule linking it to another requirement, flag the group for deferral and come back to its unresolved members later.\end{enumerate}
\item If it has been flagged for deferral, add the group again to the end of the queue.
\item Repeat from step 3 until either \begin{enumerate} \item all groups have been fully resolved, or \item the queue stagnates, i.e. a full iteration has been carried out through the entire queue of groups without any successful backend resolutions.  In this case, disable the possibility to defer resolution, and try one last iteration through the queue, ultimately failing if any backend groups fail to resolve on the final attempt.
\end{enumerate}\end{enumerate}

\subsection{Resolution of loops and nested functions}

As discussed in Sec.~\ref{declaration_loops}, it is possible to write
special module functions (\doublecross{loop managers}{loop manager}) that control the parallel
execution of other module functions (\doublecross{nested module functions}{nested module function}).  Nested functions
explicitly declare a dependency on a loop manager with a certain capability.
The dependency resolution proceeds then as for non-nested module functions.
The main difference is that loop managers have access to pointers to the nested
module functions that they control.  The dependency resolver generates a miniature dependency
tree for each loop manager, consisting of all nested functions assigned to run inside the loop managed by that manager.
The loop manager is then given responsibility for executing the nested functions, in the order provided to it by the
dependency resolver.  Further details can be found in Sec.~\ref{declaration_loops}.

\subsection{Option resolution}

Each time a module function is activated during the process of dependency
resolution, the dependency resolver searches the \yaml{Rules} section of the initialisation file for relevant option entries (see Sec.~\ref{rules_options} for the format of option entries). All
options matching the characteristics of the activated module function are collected into a new object, which is then connected to the function's \cpp{runOptions} pipe (cf.~Sec.\ \ref{module_options}).  If
the same option is set to conflicting values in multiple entries in the \yaml{Rules} section of the initialisation file, the dependency resolver will throw an error.

\section{Statistics and scanning}
\label{stats}

In this section we explain the statistical strategy employed by \GB (Sec \ref{stats:strategy}), how to obtain final inferences from its outputs (Sec \ref{stats:results}), and the generic likelihood forms available within \GB for use by module functions that do not define their own dedicated likelihoods (Sec \ref{stats:likelihoods}).

\subsection{The role of \scannerbit}
\label{stats:strategy}

To launch a \GB run, a user requests a parameter scan of a certain model, specifying ranges and priors of the model parameters, how to sample them, and the quantities that should be calculated and included in the scan.  The \GB model database activates the relevant model ancestry, which the dependency resolver uses together with the capabilities and types of the user's requested quantities to select and connect appropriate module and backend functions into a dependency tree.  Choosing which values of the model parameters to run through this dependency tree is the job of \scannerbit, the sampling and statistics module \cite{ScannerBit}.  \scannerbit applies any prior transformations requested by the user, and activates the appropriate \cross{scanner plugin} in order to run the requested sampling algorithm.  \scannerbit presently contains plugins for nested sampling (\multinest \cite{MultiNest}), Markov Chain Monte Carlo (\great \cite{GreAT}), a population-based Monte Carlo (\twalk \cite{ScannerBit}), differential evolution (\diver \cite{ScannerBit}), and various grid, random and other toy samplers \cite{ScannerBit}.  It also contains a dedicated \textsf{postprocessor} scanner plugin, which can be used for reprocessing samples obtained in a previous scan, either to recompute some output quantities or add new ones. See Ref.\ \cite{ScannerBit} for details.

When requesting a quantity in a scan, users are required to assign it a \cross{purpose} in the context of that scan.  The purpose may be \yamlvalue{Observable} or \mbox{\yamlvalue{Test},} indicating that the quantity should be computed and output for every parameter combination sampled during a scan.  Alternatively, a user can assign a \yaml{purpose} with a specific statistical meaning, such as \yamlvalue{LogLike} or \yamlvalue{Likelihood}.  Interfaces to parameter sampling algorithms in \scannerbit allow the user to choose which \yaml{purpose} to associate with the objective function for the scanner at runtime. Following dependency resolution, \GB creates a \cross{likelihood container} from the module functions of the dependency tree that have been assigned the purpose(s) associated with the sampling algorithm.  The likelihood container packages the module functions' combined results into a simple objective function for the sampler to call.  The sampler then chooses parameter combinations to sample, sends each to the likelihood container, and receives the final likelihood for the parameter combination in return.

The \GB convention is to assign \yaml{purpose: LogLike} to each component of a fit that is to be associated with the scanner, and for the module functions in question to return the natural log of the likelihood $\ln\mathcal{L}$.  The likelihood container then combines the results of all such module functions by simply summing their return values, returning the result to the scanner as the total log-likelihood.  All sampling algorithms interfaced in \textsf{ScannerBit 1.0.0} allow only a single designated \yaml{purpose} to drive a scan, although other scanners to be connected in future versions will make use of multiple, different purposes within a single scan, for example to split likelihood calculations into `fast' and `slow' subsets \cite{Lewis_slowfast}.

\subsection{Analysing samples}
\label{stats:results}

As it samples different parameter values, \scannerbit ensures that those values are output using whichever generalised print stream the user has selected (see Sec.\ \ref{printers} for details), along with all requested observables and likelihood components.  The final task of statistical interpretation then requires parsing the printed samples and processing them into meaningful statistical quantities, whether Bayesian (posterior probability distribution functions, credible intervals and/or evidence ratios) or frequentist (profile likelihoods, confidence intervals and/or $p$ values).  Depending on the sampler employed, not all of these options may be valid (we return to this discussion in more detail in Ref.\ \cite{ScannerBit}).

Although processing the saved samples into statistical measures and producing corresponding plots are tasks technically outside the scope of \GB itself, we specifically provide printer options that produce output compatible with common parsing and plotting software such as \textsf{GetDist} \cite{CosmoMC} and \pippi \cite{pippi}.  We also provide a simple installer for \pippi from within the \GB integrated build system (Sec.\ \ref{cmake}). This allows \GB to effectively produce profile likelihoods, confidence intervals, posterior probability distributions and maximum-posterior-density credible intervals \textit{in situ}, by outsourcing the final step to \pippi.  Bayesian evidences can also be obtained directly from relevant scanners (e.g.\ \multinest), or calculated after the fact with \pippi.  Calculating $p$ values requires the user to make their own \textit{ansatz} for the distribution of the \GB log-likelihood (or other test statistic that they might choose to employ in a \GB scan), and then convert the best fit identified by \pippi to $p$.  Future versions of \scannerbit are planned to include features designed to aid in determining this distribution.

\begin{figure*}[tb]
\centering
\includegraphics[width=0.4\textwidth]{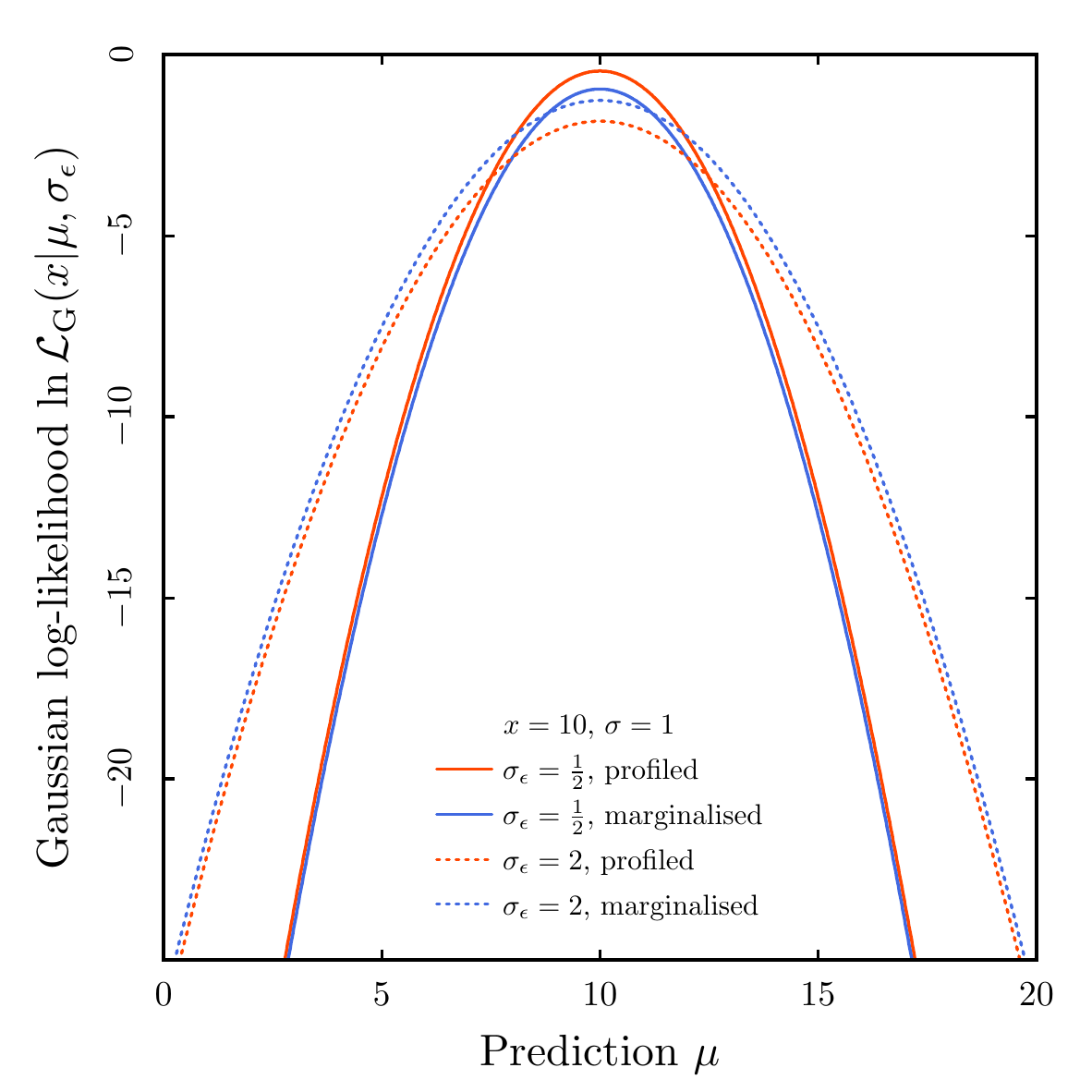}\hspace{0.1\textwidth}%
\includegraphics[width=0.4\textwidth]{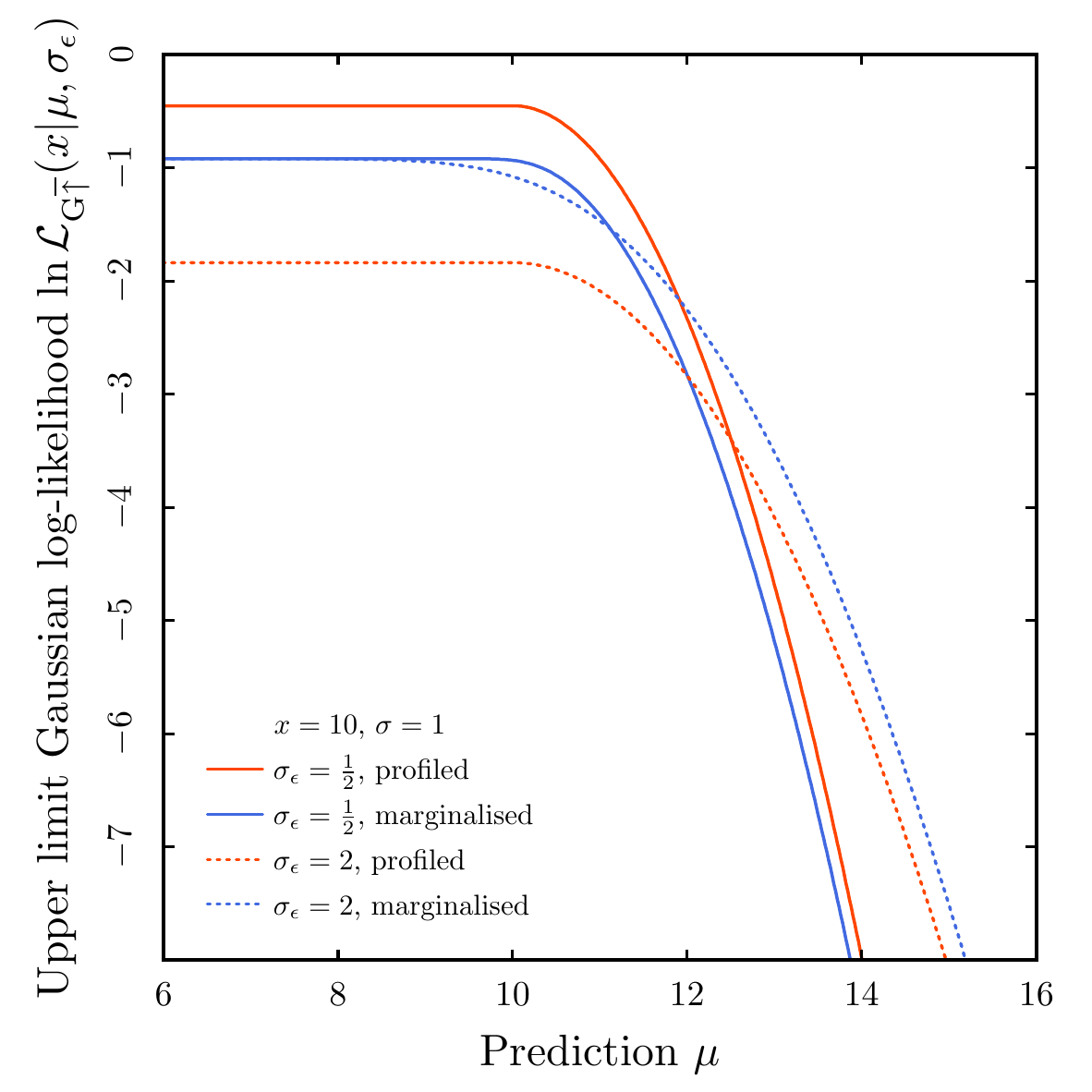}\\
\includegraphics[width=0.4\textwidth]{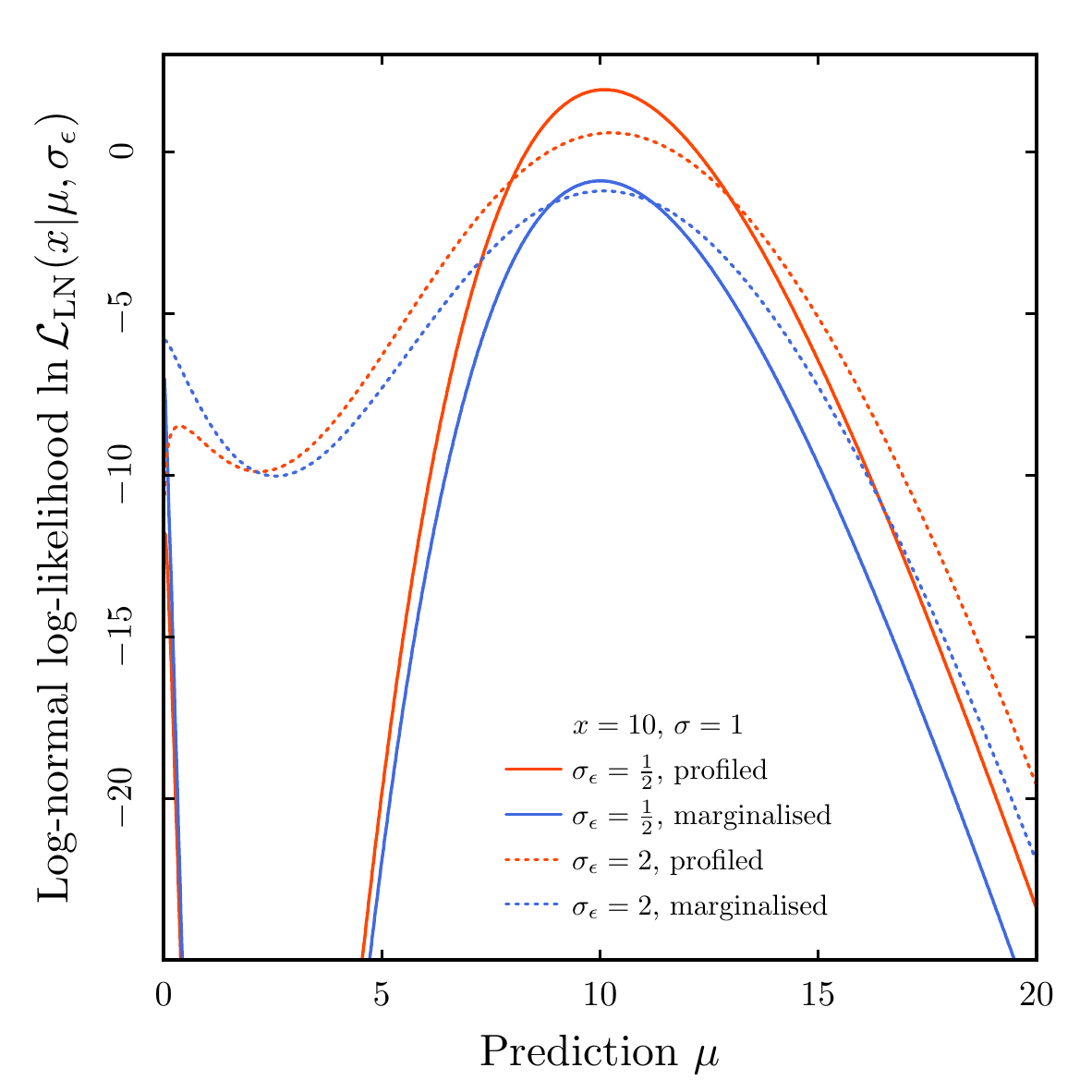}\hspace{0.1\textwidth}%
\includegraphics[width=0.4\textwidth]{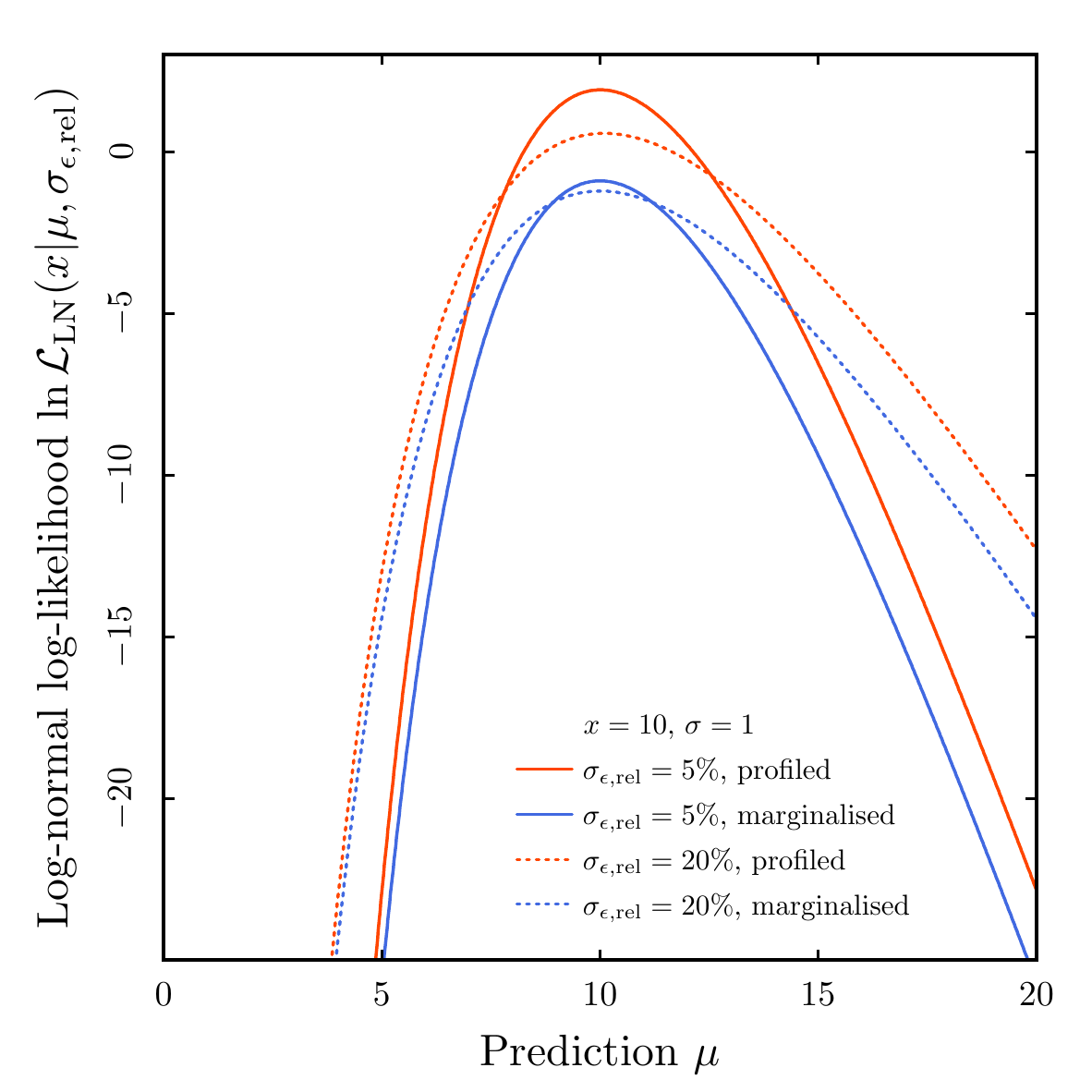}\\
\caption{The different generic likelihood functions available in \GB, described in Sec.\ \ref{stats:likelihoods}: Gaussian (\textit{top left}), Gaussian limit (\textit{top right}) and log-normal (\textit{bottom}). Here we show the log-normal likelihood computed with a fixed \textit{absolute} systematic uncertainty (\textit{bottom left}), and instead with a fixed \textit{fractional} (relative) systematic uncertainty (\textit{bottom right}).  Each curve is computed assuming an observed central value of $x=10$ and standard deviation $\sigma=1$, for two different assumed values of the systematic error. Two potential pitfalls are visible: the profiled upper limit likelihood shows a strong dependence on $\sigma_\epsilon$ at low values of $\mu$, and adopting an absolute systematic uncertainty can introduce additional features in the log-normal likelihood at low $\mu$.}
\label{fig:like}
\end{figure*}

\subsection{Available likelihood forms}
\label{stats:likelihoods}

\GB ships with a number of centrally-implemented, general purpose Gaussian and log-normal likelihood functions.  These can be found in \term{Utils/src/statistics.cpp}.  These are intended for use with simple observables and uncorrelated data, for implementing, e.g., nuisance likelihoods corresponding to well-measured SM parameters (see \cite{SDPBit}).  Module functions responsible for more complicated likelihood calculations typically contain their own implementations of appropriate test statistics, and corresponding translations to a quantity that can be treated as equivalent to $\ln\mathcal{L}$ in a scan (see the indirect detection likelihoods in \cite{DarkBit}, for example).

The centrally-implemented likelihoods come in a number of variants, allowing them to be used for detections, upper limits and lower limits.  They deal with systematic uncertainties (theory errors, experimental systematics, related nuisance parameters, etc) by analytically profiling or marginalising over an assumed distribution for an auxiliary parameter $\epsilon$, which describes departures from a perfect mapping between model parameters and predicted values of observables.  The module author must choose an appropriate central likelihood function to employ when computing a given likelihood.  However, in every module function that uses one of following likelihoods, we choose to implement a boolean \YAML option \yaml{profile_systematics} (default \yamlvalue{false}) that selects at runtime whether systematics will be profiled or marginalised over.\footnote{If the end user so desires, this can even be set differently for different module functions, although the resulting composite likelihood would arguably be inconsistent.}

\subsubsection{Profiled Gaussian}

The basic Gaussian likelihood for data measured with some mean $x$ and standard deviation $\sigma$, given a prediction $\mu$, is
\begin{equation}
\mathcal{L}_\mathrm{G}(x|\mu) = \frac{1}{\sqrt{2\pi}\sigma} \exp\left[-\frac12\frac{(x-\mu)^2}{\sigma^2}\right].
\end{equation}
Here $\mu$ may be a model parameter itself, or some complicated function of the true underlying model parameters.  Taking $\epsilon$ to be an additive offset in $\mu$ induced by some source of error, and modelling its distribution as also Gaussian, centred on zero with standard deviation $\sigma_\epsilon$, the joint likelihood becomes
\begin{equation}
\label{gauss_joint_likelihood}
\mathcal{L}_\mathrm{G} = \frac{1}{2\pi\sigma\sigma_\epsilon} \exp\left[-\frac12\frac{(x-\mu-\epsilon)^2}{\sigma^2} - \frac12\frac{\epsilon^2}{\sigma_\epsilon^2}\right].
\end{equation}
Exactly how to denote $\mathcal{L}$ on the left of this equation depends on whether $\epsilon$ and $\sigma_\epsilon$ are to be interpreted to result from an auxiliary, independent measurement (frequentist), or simply some input systematic, possibly theoretical (Bayesian).  In the former case, $\mathcal{L} = \mathcal{L}(x,\epsilon|\mu)$, and a final form of the likelihood for $x$ alone can be obtained by profiling over the observed value of $\epsilon$.  To do this, we determine the value of $\epsilon$ that maximises $\mathcal{L}(x,\epsilon|\mu)$, by differentiating Eq.\ \ref{gauss_joint_likelihood} to find the root
\begin{equation}
\hat\epsilon = \frac{\sigma_\epsilon^2}{\sigma^2 + \sigma_\epsilon^2}(x-\mu).
\end{equation}
Substituting back into Eq.\ \ref{gauss_joint_likelihood}, the profiled version of the Gaussian likelihood is
\begin{equation}
\label{gauss_prof}
\mathcal{L}_\mathrm{G,prof}(x|\mu) = \frac{1}{2\pi\sigma\sigma_\epsilon} \exp\left[-\frac12\frac{(x-\mu)^2}{\sigma^2 + \sigma_\epsilon^2}\right].
\end{equation}

\subsubsection{Marginalised Gaussian}

If the quantity $\epsilon$ in Eq.\ \ref{gauss_joint_likelihood} is instead interpreted as a direct input from e.g. theory, its Gaussian distribution has the character of a prior and $\mathcal{L} = \mathcal{L}(x|\mu,\epsilon,\sigma_\epsilon)$.  Note that in this case, $\sigma_\epsilon$ has the character of a model parameter (or a quantity derived from the model parameters), indicating that it may vary as a function of the underlying model across the parameter space, independent of any considerations from data.

In this case, the appropriate likelihood for $x$ alone instead comes from marginalising Eq.\ \ref{gauss_joint_likelihood} over the possible values of $\epsilon$, as
\begin{equation}
\mathcal{L}_\mathrm{G,marg}(x|\mu,\sigma_\epsilon) = \frac{1}{2\pi\sigma\sigma_\epsilon} \int^\infty_{-\infty} e^{-\frac{(x-\mu-\epsilon)^2}{2\sigma^2} - \frac{\epsilon^2}{2\sigma_\epsilon^2}}\,\mathrm{d}\epsilon,
\end{equation}
giving
\begin{equation}
\label{gauss_marg}
\mathcal{L}_\mathrm{G,marg}(x|\mu,\sigma_\epsilon) = \frac{1}{\sqrt{2\pi(\sigma^2 + \sigma_\epsilon^2)}} \exp\left[-\frac12\frac{(x-\mu)^2}{\sigma^2 + \sigma_\epsilon^2}\right].
\end{equation}

We compare the marginalised and profiled forms of the Gaussian likelihood for a toy problem with $x=10$ and $\sigma=1$ in the first panel of Fig.\ \ref{fig:like}, assuming $\sigma_\epsilon=0.5$ or $\sigma_\epsilon=2$.

\subsubsection{Profiled Gaussian limits}

The simplest reasonable approximation to the underlying likelihood associated with an upper limit on an observable $\mu$ is to assume flatness below some canonical `observed' or limiting value $x$, and to model the drop-off at $\mu>x$ with a Gaussian of width $\sigma$. This defines the piecewise function
\begin{equation}
\label{gauss_limit_simple}
\mathcal{L}_{\mathrm{G}\overline{\uparrow}}(x|\mu) = \begin{cases}
\frac{1}{\sqrt{2\pi}\sigma}, & \mathrm{if}\ \mu \le x\\
\frac{1}{\sqrt{2\pi}\sigma} \exp\left[-\frac12\frac{(x-\mu)^2}{\sigma^2}\right], & \mathrm{if}\ \mu \ge x.
\end{cases}
\end{equation}
This treatment can be used to directly convert a measured value into a limit likelihood.  An example is the relic density of dark matter $\Omega_\chi h^2$, which has been measured rather precisely, but may not consist exclusively of the dark matter candidate present in any particular BSM theory under investigation.  The same treatment can also be used to implement likelihoods associated with published upper limits, but additional modelling is required to recover the equivalent central value $x$ and falloff width $\sigma$ from a published limit. Typically limits at two different CLs are needed to uniquely determine both $x$ and $\sigma$.

Including an uncertainty from some auxiliary nuisance observable $\epsilon$ proceeds similarly to the pure Gaussian case,
\begin{align}
&\mathcal{L}_{\mathrm{G}\overline{\uparrow},\mathrm{prof}}(x|\mu) = \frac{1}{2\pi\sigma\sigma_\epsilon} \max\left\{\max_{\epsilon\le x-\mu} \exp\left[-\frac{\epsilon^2}{2\sigma_\epsilon^2}\right],\right. \nonumber\\
&\hspace{1.7cm}\left.\max_{\epsilon\ge x-\mu} \exp\left[-\frac{(x-\mu-\epsilon)^2}{2\sigma^2} - \frac{\epsilon^2}{2\sigma_\epsilon^2}\right] \right\}.
\end{align}

Despite the need to carefully piecewise maximise in the different regimes, this leads to the simple result
\begin{equation}
\label{gauss_upper_prof}
\mathcal{L}_{\mathrm{G}\overline{\uparrow},\mathrm{prof}}(x|\mu)=\begin{cases}
\frac{1}{2\pi\sigma\sigma_\epsilon}, & \mathrm{if}\ \mu \le x\\
\frac{1}{2\pi\sigma\sigma_\epsilon} \exp\left[-\frac12\frac{(x-\mu)^2}{\sigma^2 + \sigma_\epsilon^2}\right], & \mathrm{if}\ \mu \ge x.
\end{cases}
\end{equation}
The corresponding expression $\mathcal{L}_{\mathrm{G}\underline{\downarrow},\mathrm{prof}}(x|\mu)$ for a lower limit is identical, except that the inequalities relating $x$ and $\mu$ are reversed.

The simplicity of Eq.\ \ref{gauss_upper_prof} is somewhat beguiling. Incorrectly using this expression when $\epsilon$ and $\sigma_\epsilon$ are interpreted in a Bayesian manner can lead to behaviour of the test statistic that is undesirable in a frequentist analysis.  For example, if $\sigma_\epsilon$ varies over the parameter space, the likelihood function will not actually be flat for $\mu\le x$, despite the fact that the data make no statement about neighbouring values of $\mu$ in this region, and therefore neither should a sensible profile likelihood.  An example of this behaviour can be seen in the second panel of Fig.\ \ref{fig:like}.  In such cases, it is important to carefully decide on the interpretation of $\epsilon$ and $\sigma_\epsilon$ from the outset.  If they cannot be interpreted in a strict frequentist sense, then the marginalised variants of the likelihoods discussed here should be adopted instead, \textit{even} when the final goal of a BSM scan is to produce profile likelihood results.

\subsubsection{Marginalised Gaussian limits}

To produce the marginalised form of Eq.\ \ref{gauss_limit_simple}, we again integrate the joint likelihood over all possible $\epsilon$,
\begin{align}
\mathcal{L}_{\mathrm{G}\overline{\uparrow},\mathrm{marg}}(x|\mu,\sigma_\epsilon) = \frac{1}{2\pi\sigma\sigma_\epsilon} \int_{-\infty}^{x-\mu} \exp\left[-\frac{\epsilon^2}{2\sigma_\epsilon^2}\right]\,\mathrm{d}\epsilon& \nonumber\\
 + \int_{x-\mu}^\infty \exp\left[-\frac{(x-\mu-\epsilon)^2}{2\sigma^2} - \frac{\epsilon^2}{2\sigma_\epsilon^2}\right]\,\mathrm{d}\epsilon&,
\end{align}
leading to
\begin{align}
\label{gauss_upper_marg}
&\mathcal{L}_{\mathrm{G}\overline{\uparrow},\mathrm{marg}}(x|\mu,\sigma_\epsilon) = \frac{1}{2^{3/2}\sqrt{\pi}}\left[\frac{1}{\sqrt{\sigma^2+\sigma_\epsilon^2}} e^{-\frac12\frac{(x-\mu)^2}{\sigma^2+\sigma_\epsilon^2}}\right.\nonumber\\
&\left.\times\,\mathrm{erfc}\left(\frac{\sigma}{\sigma_\epsilon}\frac{x-\mu}{\sqrt{2(\sigma^2+\sigma_\epsilon^2)}} \right) + \frac{1}{\sigma}\mathrm{erfc}\left(\frac{\mu-x}{\sqrt{2}\sigma_\epsilon}\right)\vphantom{\frac{e^{\frac12\frac{(x-\mu)^2}{\sigma^2+\sigma_\epsilon^2}}}{\sqrt{\sigma^2+\sigma_\epsilon^2}}}\right],
\end{align}
where $\mathrm{erfc}(x) = \erf(1-x)$ is the complementary error function. We can now see that
\begin{equation}
\lim_{\mu\to-\infty} \mathcal{L}_{\mathrm{G}\overline{\uparrow},\mathrm{marg}}(x|\mu,\sigma_\epsilon) = \frac{1}{\sqrt{2\pi}\sigma},
\end{equation}
regardless of $\sigma_\epsilon$, and precisely as one would prefer a sensibly-behaved profile likelihood to do.  This behaviour can be seen in the second panel of Fig.\ \ref{fig:like}.

The corresponding marginalised likelihood for a lower limit $\mathcal{L}_{\mathrm{G}\underline{\downarrow},\mathrm{marg}}(x|\mu,\sigma_\epsilon)$ is obtained by making the replacements $x\rightarrow-x$ and $\mu\rightarrow-\mu$ in Eq.\ \ref{gauss_upper_marg}.

\subsubsection{Profiled log-normal}

A log-normal likelihood describes the situation where the distribution of the logarithm of some observation is expected to be Gaussian over repeated experiments. This may occur in cases where, for example, observations must return positive values by construction.  The likelihood takes the form
\begin{equation}
\mathcal{L}_\mathrm{LN}(x|\mu) = \frac{1}{\sqrt{2\pi}\sigma' x} \exp\left[-\frac12\frac{(\ln \frac{x}{\mu} )^2}{\sigma'^2}\right].
\end{equation}
Here $x$ and $\mu$ remain the observed and predicted values of the observable, and the Gaussian distribution for $\ln x$ is centred on $\ln \mu$.  The Gaussian width is $\sigma'$, which is related to $\sigma_\mathrm{rel}$, the relative uncertainty on $x$, as
\begin{equation}
\sigma' \equiv \ln(1+\sigma_\mathrm{rel}) =  \ln(1+\sigma/x).
\end{equation}

This likelihood describes statistical variation in the scale of an observable, and is therefore most prone to the effects of systematics also able to impact that scale.  In this case, $\epsilon$ takes the form of an auxiliary multiplicative source of error, with the corresponding additive offset given by $\ln \epsilon$.  It is therefore appropriate to model the distribution of $\epsilon$ with a log-normal centred on 1.  The corresponding width $\sigma'_\epsilon$ is then given by
\begin{equation}
\sigma'_\epsilon \equiv \ln(1+\sigma_{\epsilon,\mathrm{rel}}) =  \ln(1+\sigma_\epsilon/\mu).
\end{equation}
The joint likelihood is then
\begin{equation}
\label{lognormal_joint_likelihood}
\mathcal{L}_\mathrm{LN} = \frac{1}{\sqrt{2\pi}\sigma'\sigma'_\epsilon x\epsilon} \exp\left[-\frac{(\ln \frac{x}{\epsilon\mu} )^2}{2\sigma'^2} - \frac{(\ln \epsilon )^2}{2\sigma'^2_\epsilon}\right].
\end{equation}
This has its maximum at
\begin{equation}
\hat\epsilon = \exp\left[\frac{\sigma'^2_\epsilon(\ln\frac{x}{\mu} - \sigma'^2)}{\sigma'^2 + \sigma'^2_\epsilon}\right],
\end{equation}
leading to the profiled likelihood
\begin{align}
\label{lognormal_prof}
&\mathcal{L}_\mathrm{LN,prof}(x|\mu) = \nonumber\\
&\frac{1}{x\sqrt{2\pi}\sigma'\sigma'_\epsilon} \exp\left[-\frac12\frac{\left(\ln \frac{x}{\epsilon\mu} \right)^2 + \sigma'^2_\epsilon \ln\frac{x}{\mu} - \sigma'^2\sigma'^2_\epsilon  }{\sigma'^2 + \sigma'^2_\epsilon}\right].
\end{align}

\subsubsection{Marginalised log-normal}
Integrating Eq.\ \ref{lognormal_joint_likelihood} over $\epsilon$ instead of maximising it gives the marginalised log-normal likelihood:
\begin{equation}
\label{lognormal_marg}
\mathcal{L}_\mathrm{LN,marg}(x|\mu) = \frac{1}{x\sqrt{2\pi(\sigma'^2 + \sigma'^2_\epsilon)}} \exp\left[-\frac12\frac{(\ln \frac{x}{\mu} )^2}{\sigma'^2 + \sigma'^2_\epsilon}\right].
\end{equation}

In the lower panels of Fig.\ \ref{fig:like}, we compare the marginalised and profiled forms of the log-normal likelihood, for the same toy problem as discussed previously ($x=10$ and $\sigma=1$).  As in the Gaussian case, the profiled and marginalised versions show very similar behaviour, despite the fact that \textit{unlike} the Gaussian case, they posses somewhat different functional forms.  Here we also show the additional features that can be induced at low $\mu$ if a constant value of the absolute systematic $\sigma_\epsilon$ is employed with the log-normal likelihood, rather than a constant relative uncertainty $\sigma_{\epsilon,\mathrm{rel}}$.

\section{Output}
\label{printers}

Output from \GB scans is handled by the Printer subsystem, which generalises the writing of scan data to disk or any other output medium.  It is designed so that the output format can be chosen at runtime with options in the master \YAML file. Print commands within \GB are issued via a general abstract interface, while the actual writing of data to the chosen output medium is handled by one of several derived classes, known as \doublecross{printers}{printer}.

The actual print commands are automatically issued by \GB whenever it runs a module function, so writers of new module functions need not concern themselves with how to send information to the printer system.  Most users only need to know how to set up a printer via the master \YAML file, and what the format of the output is.  Sec. \ref{printer_setup} covers the \YAML setup.  We deal with the main output format in Sec.\ \ref{output_overview}, and the output formats of specific printers in \ref{output_specific}.

There are three main scenarios that require additional knowledge of the printer system. One is writing a \cross{scanner plugin}, where one must use the printer interface class to output e.g. probability weights or likelihoods. We discuss this briefly in Sec.\  \ref{ascii_output}, but we refer readers to the \scannerbit paper \cite{ScannerBit} for a full exposition.  Another is when a user wishes to make an existing printer emit a new \Cpp type, to e.g.\ allow a new module function returning a custom type to print its result. We deal with this in Sec.\ \ref{print_overloads}.  The final scenario is writing a new printer, for outputting \GB data in a new format. This is a straightforward but quite specialised task, requiring complete knowledge of the class structure of the printer subsystem. The requisite details are left to documentation shipped with the code (found in \term{doc/writing_printers.pdf})

\subsection{Overview of the output format}
\label{output_overview}
Other than in scanner plugin code (see Ref.\ \cite{ScannerBit}), print commands are issued automatically to the \GB printer system.  This occurs after the evaluation of each module function that has been requested for printing from the master \YAML file (see Sec. \ref{printer_setup}).  Nonetheless, it useful to know how this system works when interpreting its output.  The printer system receives print commands via functions with the signature
\begin{lstcpp}
void _print(@\metavar{type}@ const& result,
             const std::string& label,
             const int IDcode,
             const unsigned int MPIrank,
             const unsigned long pointID);
\end{lstcpp}
These contain the following information:
\begin{description}
\item[\CPPidentifierstyle result] The result computed by the module function (printable only if \metavar{type} is registered as printable and has an appropriate function overload defined; see Sec. \ref{print_overloads}).
\item[\CPPidentifierstyle label] A string describing the \cpp{result} of the module function. It typically has the format
 \begin{lsttermalt}
"#*@\metavar{capability}@* @*@\metavar{module}@*::*@\metavar{function}@*"
 \end{lsttermalt}
 where \metavar{capability}, \metavar{module} and \metavar{function} are respectively the \cross{capability}, host module and actual name of the module function that issued the print command. It is left to individual printers to decide what to do with this information (see Secs.\ \ref{ascii_output} and \ref{hdf5_output}).
\item[\CPPidentifierstyle IDcode] A unique integer automatically assigned to each module function for the duration of a scan. This allows printers to identify the origin of each print command without parsing the \cpp{label} string. Generally this number will not be passed on to the output file.
\item[\CPPidentifierstyle MPIrank] The process rank assigned by \mpi. Along with \cpp{pointID} this is needed to identify which parameter space point triggered a given print command.
\item[\CPPidentifierstyle pointID] A unique integer automatically assigned to every parameter combination evaluated by a given \mpi process in a scan. The \cpp{pointID} is not unique across processes, so both \cpp{MPIrank} and \cpp{pointID} need to be used in combination to obtain a globally unique identifier.
\end{description}
These arguments are the totality of information known to the printer at the time of each print command. It is then the job of the printer to assemble this information, from many print commands, into a coherent set of outputs.

Print commands can also be issued by \scannerbit or its plugins. By default, \scannerbit prints the final result returned to it for each model point (i.e.\ the total log-likelihood returned by the \cross{likelihood container}).\footnote{Technically, what is returned to the scanner is actually determined by the \cross{purpose}(s) that the user has associated with their chosen scanner or test function plugin in the master \YAML file (see Sec.\ \ref{stats}).  When using \scannerbit standalone however, \textit{anything} can actually be connected to the scanner as its main objective function, and it will still be printed by default.}  However, scanners will often have other information that they want to record about each model point, and this can be added via manual calls to the print commands.  Details can be found in the \scannerbit paper \cite{ScannerBit}.

In addition to the module function, likelihoood container and scanner plugin outputs sent to the printer, the \cpp{MPIrank} and \cpp{pointID} are also automatically printed for every point. This allows printers the option of writing new data back to previous points.  For example, the \MultiNest scanner plugin computes posterior probability weights for a subset of points, long after the likelihood function is evaluated at those points.  With this setup, such information can be inserted directly into the existing output medium at the position associated with those points, rather than having to write an entirely new output stream, as occurs in the native \MultiNest output. It is up to the individual printers exactly how they handle this; for example, the \textsf{ascii} printer will write out a new file as \MultiNest itself does, but the \textsf{hdf5} printer will automatically update existing \textsf{HDF5} files with new data about old points.

\subsection{Available printers}
\label{output_specific}

Here we give specific details of how print commands are translated into files on disk by the \textsf{ascii} and \textsf{hdf5} printers. This is essential information for interpreting the output of each printer.

\subsubsection{ASCII output}
\label{ascii_output}
The output file produced by the \textsf{ascii} printer (as named by the \yaml{output_file} option; see Sec. \ref{common_printer_setup}) consists of a simple whitespace-separated table of floating point numbers. The table is produced as follows. First, the \GB module functions that are registered for printing issue print commands to the primary print stream, as they are each evaluated, and the result data is stored in a buffer. The print commands contain the \cpp{MPIrank} and \cpp{pointID} (see Sec. \ref{output_overview}) identifying the model point that produced the data.  By monitoring when these identifiers change, the printer detects when the scanner has moved to a new model point. Upon detecting a new point, the buffer begins a new line. Once the buffer is filled with a preset maximum number of lines, it is written to disk.

The structure of the ASCII output table (i.e.\ which data should be assigned to which column) is determined exclusively from the contents of the buffer immediately before the first dump.  This imposes some extra restrictions on the data that the \textsf{ascii} printer can handle. For example, variable-length vectors of data can be printed, but at least one example with the maximum length expected in an entire scan must be sent to the printer before the first buffer dump, otherwise there will be insufficient space allocated in the output table to accommodate the longest such vectors in subsequent dumps.

To interpret the contents of the resulting ASCII file, an accompanying ``info'' file is produced at the time of the first buffer dump.  The info file contains a list of labels identifying the columns of the output data file. If the data file is named \term{output.data}, then the info file will be \term{output.data_info}. When running \GB via \mpi, a separate output file will be produced for each process, with the rank of the host process appended to the root filename. An example info file describing output generated by fitting a normal distribution with \MultiNest\cite{MultiNest}\footnote{See the \scannerbit paper \cite{ScannerBit} for details of the \GB interface to \MultiNest.} is shown below:
\begin{lsttext}
Column 1: unitCubeParameters[0]
Column 2: unitCubeParameters[1]
Column 3: MPIrank
Column 4: pointID
Column 5: LogLikelihood
Column 6: #NormalDist_parameters \
  @NormalDist::primary_parameters::mu
Column 7: #NormalDist_parameters \
  @NormalDist::primary_parameters::sigma
Column 8: #normaldist_loglike \
  @ExampleBit_A::normaldist_loglike
\end{lsttext}
In this example the LogLikelihood (column 5) contains the global log-likelihood used to drive \MultiNest.  It consists of only one component, given in column 8: the log-likelihood returned by the normal distribution log-likelihood function \cpp{normaldist_loglike} from the module \examplebita. Model parameter values are given in columns 6 and 7. The first two columns contain ``unit hypercube'' parameters, which are the raw unit-interval samples produced by \MultiNest, before being transformed into the actual model parameter values by \scannerbit \cite{ScannerBit}. The \term{MPIrank} and \term{pointID} entries contain the model point identification data (Sec. \ref{output_overview}).

Print statements originating from scanner plugins can be issued directly to the main printer --- in which case they will be treated the same as module function output --- or they can be issued to an auxiliary print stream if the data are not synchronised with the likelihood evaluations. Instructions for correctly handling this kind of data when writing scanner plugins are given elsewhere \cite{ScannerBit}.  In the example above, unlike in the the native \MultiNest output format, there are no posterior weights. These are issued to an auxiliary print stream in the \MultiNest scanner plugin, so they end up in a different output file. The auxiliary file is also a plain ASCII table, and it comes with its own info file describing its contents:
\begin{lsttext}
Column 1: Posterior
Column 2: MPIrank
Column 3: pointID
\end{lsttext}

The \term{Posterior} column contains the posterior weights, and the \term{MPIrank} and \term{pointID} contain the point identification data as before. Because \term{MPIrank} and \term{pointID} are shared between output files, they can be used to correlate Posterior weights with other data about the point during post-run analysis. \GB could in principle perform this combination automatically at the end of a run, however it is currently left up to user's preferred post-run analysis tools. Note that the \textsf{hdf5} printer \emph{does} automatically combine the auxiliary print stream data with the primary print stream data, so it is the more convenient format to use when working with auxiliary print data like posterior weights.

\subsubsection{\textsf{HDF5} output}
\label{hdf5_output}
The output file produced by the \textsf{hdf5} printer is set with the \yaml{output_file} option (Sec. \ref{hdf5_printer_setup}).  It contains a separate data record for every output quantity, each of length equal to the number of parameter space points evaluated during the scan. These datasets are located according to the \yaml{group} option (Sec. \ref{hdf5_printer_setup}). The command-line utility \term{h5ls} (included in most \textsf{HDF5} library distributions) can be used to probe the internal layout of an \textsf{HDF5} file. This can be useful for inspecting the names given to each dataset, which are derived from the \cpp{label} supplied via the print commands (see Sec. \ref{output_overview}).  The same information can also be obtained using the \term{probe} command in \pippi \cite{pippi}.

All datasets in the resulting \textsf{HDF5} files are synchronised, meaning that items at index $i$ in every dataset have been obtained from the same model point. Each dataset comes with a second dataset of matching length, containing a flag indicating whether the data at the given index of the host dataset has been identified as valid or not.  The labels for these datasets match their hosts, with \term{_isvalid} appended. For example, if an observable quantity \term{some_obs} was registered as invalid by \GB for that point (perhaps because the result was unphysical), then the entry in the \term{some_obs_isvalid} dataset will be set to $0$ (false). The \term{_isvalid} entries can thus be used as a mask for filtering out invalid or missing entries from the main dataset.  This is done automatically in \pippi --- but a simple example \python script that uses \term{h5py} to inspect the \textsf{HDF5} output of a \GB run serves to illustrate the above concepts:
\begin{lstpy}
import h5py
import numpy as np

#Open the hdf5 file
f = h5py.File("@\metavar{output\_filename}@","r")

#Retrieve the log-likelihoods
logL_label = "@\metavar{group}@/LogLikelihood"
logL = f[logL_label]

#Retrieve flags indicating log-likelihood validity
isvalid_label = "@\metavar{group}@/LogLikelihood_isvalid"
mask = np.array(f[isvalid_label], dtype = np.bool)
print "Successful LogLikelihood evaluations:"
print np.sum(mask)

#Apply flags to print only valid log-likelihoods
print "Valid LogLikelihood values:"
print logL[mask]
\end{lstpy}

Note that the output format described here applies only to the final, combined output of a \GB scan. During a run, the information will be structured differently, and there will be one output file for every \mpi process involved in the scan. This output is combined when scans resume (so that new temporary output can be written), and when they complete. To examine the output of a scan in the format described here before the scan completes, it is necessary to stop the scan and then resume it (see Sec. \ref{resume}) to trigger the combination.

\subsection{Expanding the printable types}
\label{print_overloads}
A module function result can only be printed if its \Cpp type is in the set of printable types. In order for a type to be printable, the printer chosen for a scan must have an appropriate print function overload defined for the type. The internal details of these function overloads must vary with the printer, as they describe how to translate the \Cpp type into the output format specific to each printer.  Here we outline the general requirements.

The process is best illustrated with an example. Suppose one wishes to make the result type \cpp{std::map<std::string,int>} printable via the \textsf{ascii} printer. First, in order for the type to even be potentially printable by any printer, it must be listed in the \cpp{PRINTABLE_TYPES} macro in \term{Elements/include/gambit} \term{/Elements/printable_types.hpp}. Note that commas confuse the macro, so in this example the new type should first be aliased with a typedef, e.g.
\begin{lstcpp}
typedef std::map<std::string,int> map_str_int
\end{lstcpp}

Next, one needs to add a new overload of the \cpp{print} function to the printer class (in this case the \textsf{ascii} printer).  This requires a new declaration to be added to the class \cpp{asciiPrinter}.  This can be acheived automatically by putting the type into one of the two lists of types to be found in \term{Printers/}\term{include/gambit/Printers}\term{/printers} \term{/asciitypes.hpp}:
\begin{lstcpp}
#define ASCII_TYPES                         \
  (std::string)                             \
  // etc

#define ASCII_MODULE_BACKEND_TYPES          \
  (DM_nucleon_couplings)                    \
  (Flav_KstarMuMu_obs)                      \
  (map_str_int) // <--- New printable type.
\end{lstcpp}
Here the type should be added to \term{ASCII_MODULE_BACKEND} \term{_TYPES} if it is defined specifically as a module type or a backend type, and to \term{ASCII_TYPES} otherwise.  Users unsure whether their type is a backend type, module type or some other type should be able to find the answer by studying Secs.\ \ref{backend_types}, \ref{types} and \ref{adding_components}.

The corresponding function definition should then be added to \term{Printers/src/printers/asciiprinter/} \term{print_overloads.cpp}:
\begin{lstcpp}
void asciiPrinter::_print(map_str_int const&
 result, const std::string& label, const int
 IDcode, const unsigned int MPIrank, const
 unsigned long pointID)
{
  std::vector<std::string> names;
  std::vector<double> values;
  names.reserve( result.size() );
  values.reserve( result.size() );
  for (std::map<std::string, int>::iterator
   it = result.begin(); it != result.end(); it++)
  {
    std::stringstream ss;
    ss << label << "::" << it->first;
    names.push_back( ss.str() );
    values.push_back( it->second );
  }
  addtobuffer(values,names,IDcode,MPIrank,pointID);
}
\end{lstcpp}
Note that if the type appears in the \term{ASCII_TYPES} macro above, then the function definition should just go in the main body of \term{print_overloads.cpp}.  If the type is instead part of \term{ASCII_MODULE_BACKEND_TYPES}, the function definition needs to be surrounded by the preprocessor directives \cpp{#ifndef SCANNER_STANDALONE} \cpp{...} \cpp{#endif} in order to retain the ability to use \scannerbit without \GB modules or backends.

Data can be supplied to the \textsf{ascii} printer buffer as a vector of values plus a matching vector of labels, so in this example the input string/integer map is simply converted into two vectors and sent to the print buffer. Of course, to fully understand the detail of the function body above one needs to understand the interior workings of the \cpp{asciiPrinter} class; those details can be found for each printer in the main code documentation (located in the \term{doc} directory).

In general, any expansion of the types printable by a given printer should also involve expanding the types \textit{readable} by the corresponding `inverse printer', which is used by the \textsf{postprocessor} scanner.  See Ref.\ \cite{ScannerBit} for details.

\section{Utilities}
\label{utils}

\subsection{Particle database}
\label{pdb}

The \GB \cross{particle database} provides standardised particle definitions for use throughout the code, in particular for referring to states in\ \GB\ \cpp{Spectrum}, \cpp{DecayTable} and \cpp{ProcessCatalog} objects, which catalogue particle masses, decay and annihilation rates.  It can be found in \term{Models/src/particle\_database.cpp}.

Declaring new particles can be done either in singular form
\begin{lstcpp}
add_particle("~g", (1000021,0) )
\end{lstcpp}
or in sets, as
\begin{lstcpp}
add_particle_set("h0", ((25, 0), (35, 0)) )
\end{lstcpp}
In the first example, the gluino is declared with name \mbox{\lstinline{"~g"}}, PDG code 1000021, and \cross{context integer} 0.  The context integer provides an additional index beyond the PDG code.  This can be used to distinguish different particles that might employ the same PDG code under different circumstances, e.g.\ (s)fermion mass and gauge eigenstates.

In the second example, two new particles are declared, corresponding to the two neutral Higgses in the MSSM.  The names of the new particles are constructed from the string \lstinline{"h0"} and the indices of the individual particles in the set, such that \lstinline{"h0_1"} is created with PDG code 25 and context zero, and \lstinline{"h0_2"} is created with PDG code 35 and context zero.  Essentially any number of particles can be placed together into a set in this manner.

Equivalent versions of both \lstinline{add_particle} and \lstinline{add_particle_set} exist for adding SM particles in particular; these are \lstinline{add_SM_particle} and \lstinline{add_SM_particle_set}.  SM particles are given special treatment and saved as such inside the database, so that filters to e.g. decay final states can be applied according to whether one of the final states is or is not part of the SM.

A special version of \lstinline{add_particle} also exists for defining broad particle classes like `quarks', 'baryons', `mesons', etc,
\begin{lstcpp}
add_generic_particle("quark", (9900084, 0))
\end{lstcpp}
These generic states have rather limited applicability, as they cannot participate in mass spectrum calculations, but they can prove useful for specifying generic decay channels.

Within the rest of \GB, particles defined in the particle database can be referred to in three equivalent ways:
\begin{enumerate}
\item using their full name (e.g.\ \lstinline{"~g"}, \lstinline{"h0_1"}, \lstinline{"h0_2"}, etc)
\item using their PDG-context integer pair (e.g. \{35, 0\})
\item using their short name and set index (e.g.\ \{\lstinline{"h0"}, 2\})
\end{enumerate}
The particle database itself contains various other helper functions for converting between these three conventions, and for converting particles into their corresponding anti-particles.  It can be accessed using the function \mbox{\cpp{ParticleDB()},} which returns a reference to the (singleton) database object.

The particle database in \GB \textsf{1.0.0} contains entries for all SM and MSSM particles, as well as the singlet DM candidate, various significant SM mesons and generic particle classes.

\subsection{Logging}
\label{logs}

The \GB logging system provides a mechanism for writing important messages that occur during a run to disk, so that they can be examined when diagnosing problems or simply trying to understand a scan. Module writers can access the central logging singleton object via the accessor function \mbox{\cpp{logger()},} which can be included via the header \term{Logs/include/gambit/Logs/log.hpp}. Log messages are sent to the logger object via the stream operator \lstinline{<<}. Strings fed into the logger are concatenated until the special object \lstinline{EOM} is received, which marks the end of each log message and causes it to be written to disk. A simple example is:
\begin{lstcpp}
logger() << "Hello world" << EOM;
\end{lstcpp}

Log messages can be assigned tags depending on the nature and origin of the message. Tags can be used for automatically sorting log messages into different output files. For example, the tag \lstinline{LogTags::info} can be attached to a message by inserting it via the stream operator at any point before the \lstinline{EOM} object is received. A list of the available tags along with their string names used in the log output is given below:
\begin{lstcpp}
// Message types
debug    = "Debug"
info     = "Info"
warn     = "Warning"
err      = "Error"
// Flags
fatal    = "Fatal"
nonfatal = "Non-fatal"
// Component tags
def      = "Default"
core     = "Core"
logs     = "Logger"
models   = "Models"
dependency_resolver = "Dependency Resolver"
scanner  = "Scanner"
inifile  = "IniFile"
printers = "Printers"
utils    = "Utilities"
backends = "Backends"
\end{lstcpp}
Note that that namespace qualifier \lstinline{LogTags} is required to access the tags.

If \GB is compiled with \mpi and run with more than one process, the \mpi rank of the process that creates each log file is appended to its filename, separating log messages from different processes into different files.

By default, all log messages are delivered to the files \yaml{runs/}\metavar{yaml\_filename}\yaml{/logs/default.log_}\metavar{rank}, where \metavar{yaml\_filename} is the root name of the \YAML file used to run \GB, and \metavar{rank} is the \mpi rank. There are two ways to change this default path. The first is to specify an override \yaml{default_output_path} in the \yaml{KeyValues} section of the \YAML file (see Sec.\ \ref{keyvalues}).  The second is to specify a new \yaml{prefix} in the \yaml{Logger} section of the \YAML file, which specifies a directory in which to store log files, and overrides any \yaml{default_output_path}. Log messages having a chosen set of tags can then be redirected into files in that directory using the \yaml{redirection} subsection under \yaml{Logger}. For example, to redirect all log messages to the files \term{new_def.log_}\metavar{rank} in the directory \term{/my_dir}, and all messages tagged with both \yaml{Error} and \yaml{Fatal} into the file \term{err_fatal.log_}\metavar{rank} in the same directory, the following \yaml{Logger} setup could be used:
\begin{lstyaml}
Logger:
  prefix: "/my_dir/"
  redirection:
    [Default]     : "new_def.log"
    [Error,Fatal] : "err_fatal.log"
\end{lstyaml}
The tag matching is inclusive, so any message containing the tags \yaml{Error} or \mbox{\yaml{Fatal}} will be directed to the file \term{err_fatal.log}, regardless of what other tags it also has.  Such messages will also go to \term{new_def.log}, seeing as all messages have the \yaml{Default} tag.

By default, messages with the \yaml{Debug} tag will not be logged at all.  The \yaml{Logger} option \yaml{debug} can be used to turn on debug log output, e.g.
\begin{lstyaml}
Logger:
  debug: true
  redirection:
    [Debug]     : "debug_log_messages.log"
\end{lstyaml}
The \yaml{Logger:debug} flag is automatically activated if the central \yaml{KeyValues:debug} flag is set true (cf.\ Sec.\ \ref{keyvalues}).

Messages delivered to the logger from within a \cross{module} are automatically tagged with the name of the module, allowing messages originating from different modules to be easily isolated using the redirection system.

The log system does \textit{not} capture regular print statements sent to \term{stdout} nor \term{stderr}.  This means that any statements printed to the screen in \doublecross{backends}{backend} or modules will appear in \term{stdout} and \term{stderr} as usual. This can be frustrating when working with a massively parallel \mpi job.  We advise users to take advantage of options for sending \term{stdout} and \term{stderr} to separate files for each \mpi process, or tagging outputs with the \mpi rank; these are available in the launcher applications (\term{mpiexec}, \term{mpirun}, etc) of essentially all \mpi implementations, and in some batch schedulers as well.

\subsection{Exceptions}
\label{exceptions}

\GB has separate exceptions for errors, warnings and invalid parameter points, all of which derive from the \Cpp STL \cpp{exception} class.  There is a single
\lstinline{invalid_point_exception} object created at initialisation for use throughout the code, along with a single \lstinline{error} and a single \lstinline{warning} object for each \GB subsystem and each module.  These are accessed by reference with the functions
\begin{lstcpp}
invalid_point();

core_error();
dependency_resolver_error();
utils_error();
backend_error();
logging_error();
model_error();
Printers::printer_error();
IniParser::inifile_error();
DarkBit_error();
ScannerBit_error();
...

core_warning();
dependency_resolver_warning();
utils_warning();
backend_warning();
logging_warning();
model_warning();
Printers::printer_warning();
IniParser::inifile_warning();
DarkBit_warning();
ScannerBit_warning();
...
\end{lstcpp}

Flagging an invalid point is as simple as invoking the \lstinline{raise} method with an appropriate explanation, e.g.
\begin{lstcpp}
invalid_point().raise("Tachyon detected");
\end{lstcpp}
This causes the present parameter combination and the explanation to be logged, and the current module function evaluation to be terminated.

If an invalid point exception is raised during the calculation of the likelihood (or other \yaml{purpose} matching the scanner's requirements, cf.\ Sec.\ \ref{stats}), it short-circuits the likelihood container.  This causes all subsequent calculations in the dependency tree to be skipped, and the point declared invalid.  In this way, by placing the module functions that are most likely to invalidate points earliest in the dependency tree, the dependency resolver can help to optimise a scan by preventing unnecessary calculations from being performed on points that turn out to be invalid for other reasons.

If the invalid point exception is raised during an observable calculation that is not needed for the likelihood, then the likelihood container simply notes that the calculated observable is invalid, and moves on to the next observable, without invalidating the actual likelihood value of the point.

Raising an error or a warning follows in a similar way to an invalid point, but also provides the possibility to provide an additional context string to facilitate future debugging, e.g.
\begin{lstcpp}
DecayBit_error().raise(LOCAL_INFO,"Negative
 width!");
\end{lstcpp}
\GB defines the macro \lstinline{LOCAL_INFO} for this purpose, which unrolls to give a string with the exact file and line number in which it appears.  Users can of course pass different context information if they prefer.

By default, errors are considered fatal and warnings non-fatal.  Fatal exceptions cause a scan to terminate, printing the error message to \mbox{\term{stdout},} whereas non-fatal ones are simply logged and the module function is allowed to continue.  Invalid point exceptions, as well as errors and warnings set to be fatal, all eventually \cpp{throw} themselves in the manner of regular \Cpp exceptions; non-fatal errors and warnings never \cpp{throw}.  Which errors and warnings are considered fatal can be modified from the \yaml{KeyValues} section of the input file, using options such as
\begin{lstyaml}
exceptions:
  dependency_resolver_error: fatal
  dependency_resolver_warning: non-fatal
  core_warning: fatal
\end{lstyaml}

Sometimes, module writers will want to deliberately \cpp{raise} and then \cpp{catch} a \GB exception.  Invalid point exceptions always \cpp{throw} themselves, and if not caught earlier, are always caught and logged by the likelihood container. Module writers who wish to \cpp{raise} and \cpp{catch} invalid point exceptions within their own module functions can therefore safely do so using the regular \cpp{raise} function, under the understanding that any logging of the error is the responsibility of the catching routine.

The optional fatility of \GB errors and warnings makes it impossible to do the same thing with them, however; despite being raised, an error or warning that is deemed non-fatal will never actually be thrown, let alone caught.  \GB errors and warnings therefore also provide \cpp{forced_throw} and \cpp{silent_forced_throw} methods as alternatives to \cpp{raise}.  The \cpp{forced_throw} function raises and logs the exception as usual, but always throws it onwards, regardless of whether or not the exception is deemed fatal.  The silent version does the same, but does no logging.

As throwing exceptions across \omp boundaries constitutes undefined behaviour, \GB exceptions cannot be employed as usual from within nested module functions.  To get around this problem, \GB also includes global threadsafe deferred exception objects \cpp{piped_invalid_point}, \cpp{piped_errors} and \cpp{piped_warnings}.  By calling \cpp{request} from these objects within a nested module function, an exception can be queued up for raising by the nested function's loop manager at the next opportunity.  Developers of loop managers should therefore make a habit of calling \cpp{enquire} (inside \omp blocks) on the piped exception objects at regular intervals to see if any piped exceptions have been requested, and/or \cpp{check} (outside \omp blocks) to raise any queued exceptions.

\subsection{Diagnostics}
\label{diagnostics}

\GB features extensive diagnostic tools, allowing users to quickly check which backends, scanners, modules and models are available at any given time, as well as which module and backend functions offer what capabilities for use in a scan.

\subsubsection{Modules diagnostic}
\label{modules diagnostic}

\begin{lstterm}
gambit modules
\end{lstterm}
\GB lists the modules present and available in the user's current configuration, indicating how many functions each module contains.  Modules that are present on the user's system but have been excluded at configuration time from the compilation of \GB are also listed, but are shown as ditched (see Sec.\ \ref{cmake} for details on the \term{Ditch} process).

\subsubsection{Capabilities diagnostic}
\label{capabilities diagnostic}

\begin{lstterm}
gambit capabilities
\end{lstterm}
\GB lists all capabilities published by module functions, backend functions and backend variables, along with the modules and/or backends in which functions with each capability can be found.

\begin{figure*}[tp]
\centering
\includegraphics[width=\textwidth]{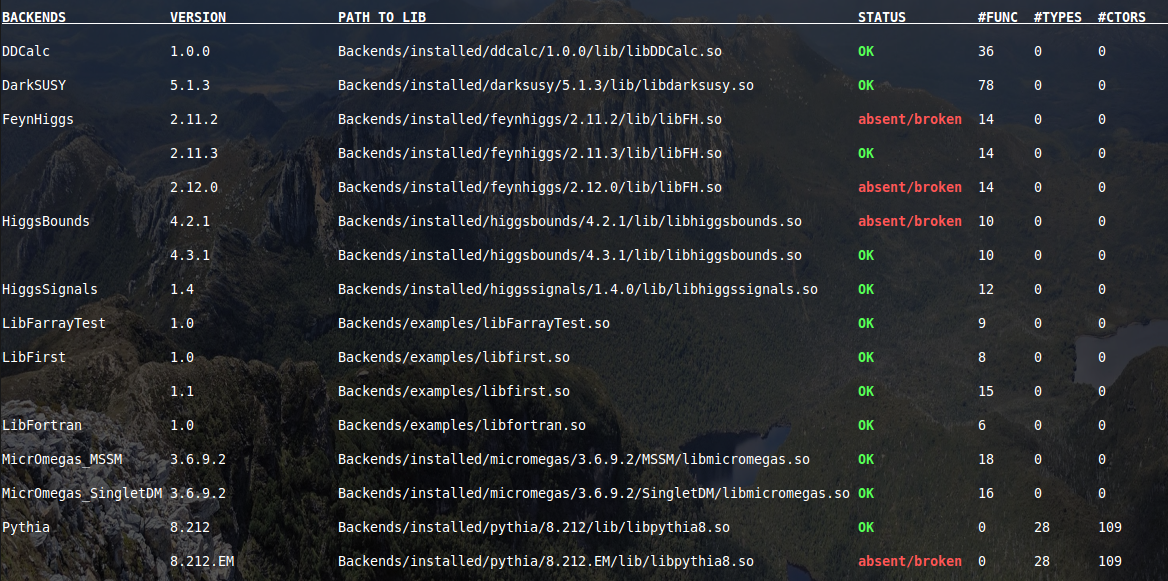}
\caption{Example output of the backends diagnostic mode, showing the statuses and locations of different backend libraries configured for use with \GB.}
\label{fig::backend_diagnostic}
\end{figure*}

\subsubsection{Backends diagnostic}
\label{backends diagnostic}

\begin{lstterm}
gambit backends
\end{lstterm}
\GB lists the backends for which it has frontend interfaces, by backend name and version.  An example is shown in Fig.\ \ref{fig::backend_diagnostic}.

For each version of each backend, the diagnostic shows the path to the shared library containing the backend, the number of functions and variables published to \GB by the frontend interface, the number of classes provided by the backend, and the number of different constructors it provides.  The diagnostic also gives the overall status of the shared library of each backend.  If the library has been located and loaded successfully, the status is \texttt{\color{Green}OK}; if it cannot be loaded or there was an error when loading it, the status is shown as \texttt{\color{red} absent/broken}; if there was a problem finding the necessary symbols for any of the member functions of any of the classes provided by the backend, the status is shown as \texttt{\color{red}bad types}.

In the case of a BOSSed library, any status other than \texttt{\color{Green}OK} causes \GB to disable all module functions that are declared to need classes from that backend.  Refer to the discussion of the rollcall declaration \cpp{NEEDS_CLASSES_FROM} in Sec.\ \ref{boss} for details.

Note that unlike constructor problems, symbol lookup errors for non-constructor backend functions or variables do \emph{not} prevent a backend from presenting status \texttt{\color{Green}OK} overall.  The status of individual functions and variables in a backend \metavar{backend\_name} can be probed using the free-form diagnostic \term{gambit} \metavar{backend\_name}. Symbol errors from non-constructor backend functions and variables cause the individual functions/variables themselves to be disabled, but not the entire backend.

\subsubsection{Models diagnostic}
\label{models diagnostic}

\begin{lstterm}
gambit models
\end{lstterm}
\GB lists the contents of the model database, giving the name of each model, its parent (if any), and the dimensionality of its parameter space.  This diagnostic also produces the necessary files to generate a graph of the model hierarchy (e.g.\ Fig.\ \ref{fig::model_tree}).

\subsubsection{Scanners diagnostic}
\label{scanners diagnostic}

\begin{lstterm}
gambit scanners
\end{lstterm}
\GB lists the names and versions of different parameter samplers for which it has interfaces defined in \scannerbit, and gives a status report on its efforts to load each of their shared libraries.

\subsubsection{Test-functions diagnostic}
\label{test function diagnostic}

\begin{lstterm}
gambit test-functions
\end{lstterm}
\GB lists the names and versions of different objective test functions known to \scannerbit, and gives a status report on its efforts to load each of their shared libraries.

\subsubsection{Priors diagnostic}
\label{priors diagnostic}

\begin{lstterm}
gambit priors
\end{lstterm}
\GB lists its known prior transformations for parameter sampling, giving a brief description of each along with the input file options that it accepts.

\subsubsection{Free-form diagnostics}
\label{free-form diagnostic}

Further information can be found about essentially any component of \GB by simply typing
\begin{lstterm}
gambit @\metavar{component}@
\end{lstterm}
where \metavar{component} is a capability or the name of a module, backend, scanner plugin, test function plugin or model.  The nature of the information subsequently provided depends on the type of component under investigation, but usually includes a short description, a status report and listings of the component's relationship to other components.  Modules come with a list of the functions they contain, including function names, capabilities, types, dependencies and backend requirements.  Backends come with similar information, as well as the individual status of each backend function. Scanners and test functions come with status reports, header and link info, and details of their accepted options.  Models report detailed information about their family tree and the identities of their parameters.  Asking about a capability generates a list of all module and backend functions able to compute that quantity.

\subsection{Type handling}
\label{types}

Dependency resolution works by matching module function capabilities and types to dependencies, and backend function capabilities and types to backend requirements.  The types involved can be \Cpp intrinsic types, \GB intrinsic types, or types associated specifically with a \GB module, model or backend.

Types associated with specific backends are automatically made available to all \GB modules and frontend routines, as all module functions can in principle have a backend requirement filled from any backend.  In contrast, module types are used exclusively by functions associated with that module, and are not available to functions outside the module.  The same is true of model types and their accessibility outside model-associated functions.

Adding a new type is relatively straightforward.  General utility types that will be used throughout \GB, and types with mixed backend, module and/or model character, should be declared in a new header, and that header included from \term{Utils/} \term{include/gambit/Utils/shared\_types.hpp}.

Backend types associated with backend \metavar{x} should be declared in a new header \term{Backends/include/} \term{gambit/Backends/backend\_types/}\metavar{x}\term{\_types.hpp}.  This header will then be automatically included in \term{Backends/} \term{include/gambit/Backends/backend\_types\_rollcall.hpp} by the \GB build system (Sec.\ \ref{cmake}).

Types associated with model \metavar{y} should be declared in \term{Models/include/gambit/Models/models/}\metavar{y}\term{.hpp}, which will then also be picked up and included by the build system, this time in \term{Models/include/gambit/Models/} \term{model\_types\_rollcall.hpp}.

Types for module \metavar{z} should be placed in a new header \metavar{z}\,\term{/include}\term{/gambit/}\metavar{z}\,\term{/}\metavar{z}\,\term{\_types.hpp}, which must then be included from \mbox{\metavar{z}\,\term{/include/gambit/}\metavar{z}\,\term{/}\metavar{z}\,\term{\_rollcall.hpp}}. The build system will automatically include the new header in \term{Elements/include/gambit/Elements/module\_types}\term{\_roll}-\term{call.hpp}.

The above discussion applies not only to new types, but also to typedefs associated with different components of \GB.  One challenge in performing dependency resolution is that types are matched entirely as strings at runtime, meaning that the dependency resolver cannot recognise typedefs \emph{a priori}. To allow it to understand typedefs and treat two types as equivalent for dependency resolution purposes, \GB features a type equivalency database \term{config/resolution\_}\term{type\_equivalency\_}\mbox{\term{classes.yaml}.}  Entries in this file are equivalency classes of different types, such that the dependency resolver considers each type within an equivalency class to be identical to all the others in the same class.

\subsection{Random numbers}
\label{random numbers}

Random numbers in \GB are provided via a threadsafe wrapper to the random number generators of the \Cppeleven STL \cpp{<random>}.  Whether inside or outside an \omp block, single uniform random variates can be obtained by calling
\begin{lstcpp}
double myran = Random::draw();
\end{lstcpp}
\GB seeds and maintains a separate random number generator for each thread, so the resulting deviates are uncorrelated across threads.  The seed for each generator is the sum of the system clock time and the thread index.

The underlying random number generator used by \cpp{Random::draw()} can be configured from the \yaml{KeyValues} section of the input file, as
\begin{lstyaml}
KeyValues:
  rng: @\metavar{chosen\_rng\_engine}@
\end{lstyaml}
where \metavar{chosen\_rng\_engine} can be any of the recognised \Cpp random engines: \cpp{default_random_engine}, \mbox{\cpp{minstd_rand},} \cpp{minstd_rand0}, \cpp{mt19937}, \cpp{mt19937_64}, \cpp{ranlux24_base}, \cpp{ranlux48_base}, \cpp{ranlux24}, \cpp{ranlux48} or \cpp{knuth_b}.\footnote{See \href{http://www.cplusplus.com/reference/random/}{http://www.cplusplus.com/reference/random/} for details.}  It can also be simply \yaml{default}, which selects the \GB default generator; in \GB \textsf{1.0.0} this is the 64-bit Mersenne Twister\ \,\cpp{mt19937_64}.

\subsection{Component databases and documentation}
\label{component_databases}

Although this paper serves as a user and design guide to \GB, as do Refs.\ \cite{DarkBit,ColliderBit,ScannerBit,FlavBit,SDPBit} for each module, \GB also features two additional documentation systems.

The first is a standard \textsf{Doxygen} documentation system, which gives basic information about classes, functions, variables, namespaces and macros defined in \GB.  The documentation can be generated with \term{make docs}, and is also available online at \href{http://gambit.hepforge.org}{http://gambit.hepforge.org}.

The second is a set of descriptive databases, which document individual models, capabilities, scanners, objective test functions and priors.  These are the descriptions that are brought up by the \GB free-form diagnostic (cf.\ Sec.\ \ref{free-form diagnostic}) when querying individual components with e.g.
\begin{lstterm}
gambit DarkBit
gambit NUHM1
gambit MultiNest
gambit SingletDM_spectrum
\end{lstterm}
These can be edited or added to by modifying the text files
\begin{lstterm}
config/models.dat
config/capabilities.dat
config/scanners.dat
config/objectives.dat
config/priors.dat
\end{lstterm}
These files are in fact written in \YAML, albeit much simpler \YAML than the input file.  When adding a new model, scanner, test function or prior, or a module function with a capability that does not already exist in \GB, it is good practice to also add a description of it to one of these files.  If any component in \GB is missing a description in these databases, a warning is raised whenever \term{gambit} is invoked.

\subsection{Signal handling and resuming a scan}
\label{resume}

A \GB scan can be terminated prematurely by sending it an appropriate \posix signal, either SIGINT, SIGTERM, SIGUSR1, or SIGUSR2. Upon receiving one of these signals, \GB will attempt to shut down cleanly, preserving output files and all information required for resuming the scan. The preservation of information required to resume scanning with a particular scan algorithm is the responsibility of \scannerbit, and more specifically each individual scanner plugin. This is described in detail in the \scannerbit paper \cite{ScannerBit}.

To resume a scan, one simply re-launches \GB using the same \YAML file that was used to launch the original scan (making sure that the \term{-r} flag is not present in the argument list; if \term{-r} is present it will cause the scan to restart, deleting existing output). For example, a prematurely terminated scan that was launched using the \YAML file \metavar{myrun.yaml} can be resumed simply by launching \GB as:
\begin{lstterm}
gambit -f @\metavar{myrun.yaml}
\end{lstterm}
where \term{-f} indicates the input file to use (cf.\ Sec.\ \ref{switches}).

\subsubsection{Shutdown behaviour}

There are two possible responses that \GB might make when told to halt a run using a system (\posix) signal. Which one it chooses depends on whether or not the scanner plugin in use can be instructed to stop by setting a \cpp{quit} flag.  A scanner's ability to interpret a \cpp{quit} flag is automatically inferred by \GB, on the basis of whether or not the scanner plugin calls \cpp{like_ptr::disable_external_shutdown()} in its constructor (see Appendix D of Ref.~\cite{ScannerBit}).

First we discuss the case where the scanner can understand a \cpp{quit} flag. In this instance each \GB\ \mpi process will, upon receiving a shutdown signal, take the following actions:
\begin{enumerate}
\item Allow the current likelihood evaluation to complete as normal.
\item Broadcast a stop command via \mpi to all other processes in the job.  This triggers this same shutdown procedure in all other processes, and is necessary in case not all processes receive the original \posix signal.
\item Finalise all printer output.
\item Set the \cpp{quit} flag for the scanner plugin.
\item Pass control back to the scanner plugin.
\end{enumerate}
At this point the \GB core system has completed its shutdown tasks, and assumes that the scanner plugin will do the same. The plugin should then complete its own shutdown tasks and return control to \GB, which will then shut down \mpi and exit the program.

The second case is where the scanner plugin has no way to recognise a \cpp{quit} flag. This is a less-than-ideal situation, as it makes performing a clean shutdown much more difficult, and indeed it is not possible to guarantee that shutdown will succeed in 100\% of cases. However, some third-party scanning algorithms do not provide any mechanism to signal a premature end to a scan, so we have designed the shutdown system to work around this restriction.

In this case the shutdown procedure will operate as follows:
\begin{enumerate}
\item Allow the current likelihood evaluation to complete as normal.
\item Broadcast a stop command via \mpi to all other processes in the job.
\item Enter a custom \mpi barrier, and wait for for all other processes to signal that they have also entered the barrier \label{step:sync}.
\item If the synchronisation in \ref{step:sync} succeeds, skip to step \ref{step:finalise}.
\item If the synchronisation does not succeed within a set time interval, disable all future printer output and return control to the scanner plugin (which cannot be told to stop), and await the next likelihood evaluation request.
\item Upon being requested to evaluate another likelihood, immediately invalidate the model point and return to step \ref{step:sync}.
\item Finalise all printer output \label{step:finalise}.
\item Terminate the program.
\end{enumerate}
The repeated synchronisation attempts are required because the scanner plugin may also be using \mpi; because \GB has no control over how \mpi is used in third-party scanning algorithms, there is a high probability that a deadlock will occur between our synchronisation attempt and a blocking \mpi command (from e.g.\ \cpp{MPI_Barrier}) in the third-party code. We must therefore abandon our synchronisation attempt if takes too long, and return control to the scanning algorithm to allow deadlocks to resolve. However, more blocking calls can easily be initiated in some other process before we attempt to synchronise again, so we have to repeatedly attempt to find a window between these calls in which we can gain control over all \mpi processes simultaneously. Once this succeeds, we can cleanly finalise the output, shut down \mpi, and terminate the program.

It is possible that the synchronisation attempts never succeed. Because of this possibility, \GB will only attempt the procedure a fixed number of times before giving up. In this case, each process will attempt to finalise its output and stop independently. In many cases this will succeed and there will be no problem.  However, if a process has been left in a blocking call by a sampling algorithm, the process will fail to terminate, and will hang until killed by the operating system. This also has the potential to corrupt the printer output for that process.  This is particularly true in the case of \textsf{HDF5} output, as \textsf{HDF5} files are highly vulnerable to corruption if not closed properly.  This can result in data loss, and make a scan impossible to resume. Processes can also hang when running under certain \mpi implementations if \mpi is not finalised correctly, which cannot be done when processes are left to terminate independently.

We remind the reader that this second procedure is not the ideal case. The shutdown should always work smoothly if a \cpp{quit} flag can be set for the scanner plugin, so for large scans that produce valuable data we recommend using scanner plugins that have this feature (e.g.\ \diver; see \cite{ScannerBit} for details).

\section{Configuration and automatic component registration}
\label{cmake}

\GB uses the open-source cross-platform build system \textsf{CMake}\footnote{\url{www.cmake.org}} to configure and build the package. The configuration step identifies the system architecture, available compilers, libraries and \GB components present on a user's system, and creates appropriate makefiles.  \textsf{CMake} and \GB support in-source and out-of-source builds, but we recommend the latter for ease of organisation.  The canonical way to configure and make \GB is therefore
\begin{lstterm}
mkdir build
cd build
cmake ..
make
\end{lstterm}

The build system also incorporates a series of \python \doublecross{harvester scripts}{harvester script} for code generation.  These are used at both configuration and compile time to automatically detect modules, models, backends, printers, scanners, priors, test functions and associated types present in a user's personal \GB configuration.  The harvesters automatically write the header, configuration and \textsf{CMake} files required to register the various components in \GB, and include them in the build.  In this way, users wishing to add a \GB component need only write the source/header files containing the component-specific content they wish to add, and place them in the appropriate folder, relying on the harvester scripts to generate the necessary boilerplate code required to incorporate the component into \GB.

Compilation of \GB module standalone executables is also handled by the same build system, but these are dealt with in Sec.\ \ref{examples}.

\subsection{Adding new models, modules, backends and other components to \GB}
\label{adding_components}

Here we give a quick reference guide to the new files needed when adding new components, along with any notable modifications needed to existing files.  Note that adding any new files to \GB (or moving existing ones) necessitates re-running \textsf{CMake} before rebuilding.
\begin{description}
\item[\textbf{modules}] Add a rollcall header \mbox{\metavar{MyBit}\term{/include/gambit}} \term{/}\metavar{MyBit}\term{/}\metavar{MyBit}\term{\_rollcall.hpp} and source files in \metavar{MyBit}\term{/include}. See Sec.\ \ref{module_declaration} for more details.
\item[\textbf{models}] Add a model declaration header \term{Models/inclu}-\term{de/gambit/}\term{Models/models/}\metavar{my\_model}\term{.hpp}. If needed, add translation function source files to \term{Models/src/models}. See Sec.\ \ref{model_declaration} for more details.  Adding a new model with no parent typically also requires either adding new module functions or verifying that existing ones are safe to use with the new model, and declaring them as such in their rollcall headers. See Sec.\ \ref{module_declaration} for more details.
\item[\textbf{backends}] \textit{We strongly encourage backend authors to go through the following simple steps, to provide official versions of the resulting \GB interface files within the regular releases of their software.}

\hspace{3mm} Add a frontend header \term{Backends/include/gambit/} \term{Backends/frontends/}\metavar{backend\_name}\term{_}\metavar{backend\_version}\term{.hpp}. Add the default location of the backend's shared library to \term{config/backend} \term{_locations.yaml.default}.  If needed, add backend convenience and/or initialisation function source files in \term{Backends/src/frontends}.  If needed, add a backend type header in \term{Backends/include/gambit/Backends/backend_types/} \metavar{backend\_name}\term{\_types.hpp}.  If you want \GB to manage the compilation of the backend, add an entry in \term{cmake/backends.cmake} (see Sec.\ \ref{cmake_backends_scanners}).

\hspace{3mm} If only adding the latest \textit{version} of an existing backend, then the new frontend header, any new frontend source files, the new entry in \term{config/backend_locations.yaml.default} and any new entry in \term{cmake/backends.cmake} can usually just be copied from the previous versions and adapted.  Be sure to update any \cpp{#include} statements in the copied source files to include the new frontend header rather than the previous one.  Any new backend types, or modifications to old backend types, are generally best dealt with by adding the new and revised types to the existing backend types header, and declaring any revised types with entirely new names, in order to avoid breaking the frontend interface to previous versions of the backend.

\item[\textbf{printers}] Add a printer declaration header \term{Printers/include/gambit/Printers/printers/}\metavar{my\_printer}\term{.hpp}. If needed, add source files in \term{Printers/src/printers/}\metavar{my\_printer}.  See Sec.\ \ref{printers} for more details.
\item[\textbf{scanners}] Add a scanner plugin declaration header \term{ScannerBit/include/gambit/ScannerBit/scanners/} \metavar{scanner\_name}\term{/}\metavar{scanner\_name}\term{.hpp}.  Add any additional headers required to the same directory. If needed, add source files to \term{ScannerBit/src/scanners/}\metavar{scanner\_name}.  See Ref.\ \cite{ScannerBit} for more details.  If you want \GB to manage the compilation of the scanner, add an entry in \term{cmake/scanners.cmake} (see Sec.\ \ref{cmake_backends_scanners}).
\item[\textbf{priors}] Add a prior declaration header in \term{ScannerBit/include/gambit/ScannerBit/priors}. If needed, add source files in \term{ScannerBit/src/priors}.  See Ref.\ \cite{ScannerBit} for more details.
\item[\textbf{objective test functions}] Add source files to \term{ScannerBit/src/objectives/test_functions}.  See Ref.\ \cite{ScannerBit} for more details.
\item[\textbf{types}] Exactly what to do depends on which component(s) the type is associated with; see the above entries and Secs.\ \ref{backend_types} and \ref{types} for more information.
\end{description}
When adding any of these components, developers should also add a description of the new component to the relevant component database (see Sec.\ \ref{component_databases}).

\subsection{Building backends and scanners}
\label{cmake_backends_scanners}

Although not strictly necessary for running \GB, we also provide helpful preconfigured methods within the configuration and build system for downloading, configuring, patching (where necessary) and compiling essentially all of the backends and scanners for which \GB has frontend or scanner plugin interfaces.  Although it is straightforward to just manually download and compile backends and scanners as usual, and then enter their shared library locations in custom \term{config/backend\_locations.yaml} and \term{config/scanner\_locations.yaml} files, using the automatic installers in the \GB build system ensures that \GB and all backends and scanners employ consistent compiler and library settings.  As with the \Cpp configuration of \GB itself, for the compilation of backends written in \plainC or \Fortran, \textsf{CMake} automatically searches for the necessary compilers and libraries.  Codes written in \python and other languages can only be backended by \GB \textsf{1.0.0} if they ship with a \plainC API; supporting `native language' backending of such codes is a high priority for future versions.

This system provides a \term{make} target for each installed version of each scanner and backend, along with a corresponding \term{make clean-}\metavar{name} target (where \metavar{name} is the name of the backend or scanner), which calls \term{distclean} or similar in the backend source. Each scanner and backend also gets a target \term{make nuke-}\metavar{name}, which completely erases all downloaded and installed content for the component in question.  The \term{make scanners} target installs and builds the latest versions of all registered external scanning algorithms, and the \term{make backends} target does the same for backends.  All scanners or backends can be cleaned or nuked in one command with the \term{make} targets \term{clean-scanners}, \term{nuke-scanners}, \term{clean-backends} or \term{nuke-backends}.  For the true nihilist, there is also \term{nuke-all}.

Adding a new backend or scanner to the \GB automatic build management system is fairly straightforward.  One adds a new `external project' entry in either \term{cmake/backends.cmake} or \term{cmake/scanners.cmake}, using some of the built-in macros that can be seen demonstrated in those files, for setting up the clean targets and indicating if a given backend requires BOSSing. The minimum information required for a functional entry in either of these files is: the URL from which the package can be downloaded, the MD5 checksum of the download (obtainable for any file via \term{cmake -E md5sum} \metavar{filename}), and basic configure and build commands for the package.  If required, specific build flags can be easily added to whatever \GB passes to the backend.  Custom patches can also be applied.  If a backend should be BOSSed as part of the build process, a \BOSS configuration file must be placed in the \term{Backends/scripts/BOSS/configs} directory, as described in Sec.\ \ref{boss}. The configuration file should be named according to the backend name and safe version, e.g. \term{MyBackend\_1\_2.py}.

\textit{One important vagary of the build system for scanners}: for \GB to properly register a scanner as built and available, it is necessary to re-run \term{cmake} after making the external scanner, and then rebuild \GB.  The most efficient way to get \GB started from scratch with e.g. \textsf{MultiNest} \cite{MultiNest} or \textsf{Diver} \cite{ScannerBit} is therefore
\lstset{style=terminal}
\begin{lstterm}
mkdir build
cd build
cmake ..
make scanners
cmake ..
make
\end{lstterm}
This particular requirement has its roots in the two-step shared library strategy that \scannerbit uses to dynamically load its plugins \cite{ScannerBit}.  This will probably disappear in future versions.

\subsection{Miscellaneous build targets}
\label{misc_build}
\begin{description}
\item \term{make get-pippi} retrieves the latest development version of the analysis and plotting tool \pippi \cite{pippi} from GitHub\footnote{\href{http://github.com/patscott/pippi}{http://github.com/patscott/pippi}}, and places it in the \GB root directory.
\item \term{make docs} builds the \GB doxygen documentation.
\item \term{make clean} removes all compiled and automatically-generated source and header files for \GB itself, but leaves backends and scanners untouched.
\item \term{make distclean} does the same as \term{clean}, but also cleans the \GB doxygen documentation, clears out the \GB scratch directory, and removes all downloaded, installed or compiled backend and scanner content.
\end{description}

\subsection{Configuration options}
\label{config_options}

Here we list the most useful commandline switches for passing to \textsf{CMake} when configuring \GB, by
\begin{lstterm}
cmake -D @\metavar{OPTION\_NAME}@ = @\metavar{value}@
\end{lstterm}
Often none of these is required, but they can be helpful for hinting or forcing \textsf{CMake} to use specific versions of compilers or libraries, or for simply disabling components or features of \GB at the build stage.

\subsubsection{\textsf{CMake} standard variables}

\begin{description}
	\item[\tt{CMAKE\_BUILD\_TYPE}] Sets the build type. Possible values are \term{Debug}, \term{Release}, \term{Release_O3}, \term{MinSizeRel}, \term{RelWithDebInfo} and \term{None}.  The default is \term{None}, which results in the fastest build time, but no debug symbols and the slowest execution.  \term{Release} includes optimisation seetings designed to result in the fastest run time; build time is correspondingly longer.  \term{Release_O3} is a \GB-specific build type that differs from \term{Release} in that it passes \term{-O3} rather than \term{-O2} to the compiler.\footnote{In this sense, \term{Release_O3} in \GB is actually closer to the traditional \textsf{CMake} definition of \term{Release}.  It is not clear that \term{Release_O3} offers any significant advantage over \term{Release}, however, and \term{-O3} may cause instability in some backends.  Use this option with caution.}
	\item[\tt{CMAKE\_CXX\_COMPILER}] Full path to the \Cpp compiler. Alternatively you can specify the environment variable \term{CXX} before invoking cmake.
	\item[\tt{CMAKE\_C\_COMPILER}] Full path to the \plainC compiler. Alternatively you can specify the environment variable \term{CC} before invoking cmake.
	\item[\tt{CMAKE\_Fortran\_COMPILER}] Full path to the \Fortran compiler. Alternatively you can specify the environment variable \term{FC} before invoking cmake.
	\item[\tt{CMAKE\_CXX\_FLAGS}] Extra flags to use when compiling \Cpp source files.
	\item[\tt{CMAKE\_C\_FLAGS}] Extra flags to use when compiling \plainC source files.
	\item[\tt{CMAKE\_Fortran\_FLAGS}] Extra flags to use when compiling \Fortran source files.
\end{description}

\subsubsection{\textsf{CMake} library and \GB-specific variables}

\begin{description}
  \item[\tt{EIGEN3\_INCLUDE\_DIR}] The full path to a local installation of \textsf{Eigen}.  Note that \textsf{Eigen} can be installed automatically from many standard repositories, so a local installation may not be necessary.
  \item[\tt{MPI}] If \term{MPI=OFF}, \mpi is manually disabled even if \textsf{CMake} successfully locates \mpi libraries.  Defaults to \term{ON}.
	\item[\tt{MPI\_INCLUDE}] The full include path of the \mpi distribution to be used (e.g. in case it is not detected automatically).
	\item[\tt{MPI\_LIBRARY}] The full path to the \mpi library file(s) to link against (e.g. in case they are not detected automatically).
	\item[\tt{LAPACK\_LINKLIBS}] The full path to the LAPACK library file(s) to link against (e.g. in case they are not detected automatically).
  \item[\tt{PYTHIA\_OPT}] If \term{PYTHIA\_OPT=OFF} and the Intel compiler is in use, turn off cross-file interprocedural optimisation when compiling the BOSSed \textsf{Pythia} backend (some systems do not have enough memory to perform this optimisation step).  Defaults to \term{ON}.
  \item[\tt{Werror}] If \term{True}, the build system treats all warnings as errors, and halts the build.
	\item[\tt{(D)itch}] Manually selects \GB components to exclude from the build.  Practically anything can be ditched with this command, from modules to models, backends, printers and scanners.  The value should be set to a semi-colon separated list of the beginnings of component names to match for ditching.  For example,
  \begin{lstterm}
cmake -Ditch = "Dark;FeynHiggs_2_11_";
  \end{lstterm}
  would ditch the module \darkbit, all versions of the backend \textsf{DarkSUSY}, and versions 2.11.* of the \textsf{FeynHiggs} backend.  Note that ditching a \GB component does not `clean' any compiled code, so it will not e.g. remove backend shared libraries that have already been compiled.  It \textit{will} however exclude all interfaces to the ditched components the next time \GB is built, making it completely indifferent to the presence or absence of any compiled or uncompiled code associated with those components.
\end{description}

\section{Examples, releases and support}
\label{examples++}

\subsection{Minimal examples}
\label{examples}

\GB ships with a number of different examples.  These include two minimal example modules (\textsf{ExampleBit\_A} and \textsf{ExampleBit\_B}) and some minimal backend examples in \plainC (\textsf{LibFirst}) and \Fortran (\textsf{LibFortran} and \textsf{LibFarrayTest}).  A minimal toy model (\textsf{NormalDist}) can be found declared in \term{Models/include/gambit/Models/models/demo.hpp}.  This file also contains a fully self-contained hierarchy of example models illustrating all the concepts of Sec.\ \ref{models} (note that these are commented out by default, to avoid cluttering the model hierarchy). There is also a matching pair of example \YAML files that use these modules and backends to run rudimentary scans of \textsf{NormalDist} (\term{yaml_files/spartan.yaml}) or the \textsf{CMSSM} (\term{yaml_files/spartan_CMSSM.yaml}).  These two files each contain some simple additional entries, commented out by default, that can be used for experimenting with different printers and scanners.  Most of the features and options outlined in this paper can be found demonstrated in one or another of these example components or scans.

The minimal \term{spartan} example also include a corresponding \term{pip} file (\term{yaml_files/spartan.pip}) for plotting the \textsf{hdf5} results of \term{yaml_files/spartan.yaml} with \pippi \cite{pippi}.  (This file will need to be altered if \term{yaml_files/spartan.yaml} is altered from its default.)

For more complete and realistic examples, users should refer to the full \YAML files corresponding to the MSSM and scalar singlet scans described in Refs.\ \cite{CMSSM,MSSM,SSDM}, which also ship with \GB and can be found in the \term{yaml_files} directory.  These are \term{SingletDM.yaml}, \term{CMSSM.yaml}, \term{NUHM1.yaml}, \term{NUHM2.yaml} and \term{MSSM7.yaml}.

There are also a number of module-specific example \YAML files to be found in the \term{yaml_files} folder: \term{WC.yaml} and \term{FlavBit_CMSSM.yaml} (for \flavbit), \term{ColliderBit_CMSSM.yaml} and \term{ColliderBit_ExternalModel.yaml} (for \colliderbit), \term{DecayBit_MSSM20.yaml} (for \decaybit), \term{PrecisionBit_MSSM20.yaml} (for \precisionbit), \TODO{missing} (for \specbit), \TODO{missing} (for \darkbit), and \TODO{missing} (for \scannerbit).

The full \GB distribution also includes a series of example driver programs that use the different modules as standalone libraries, without the rest of \GB.  Using \GB modules in this manner requires some extra work due to the absence of the dependency resolver and related \GB core components, but allows direct manual control of the functions in a given module, using only a minimal set of \GB components.  In certain cases, using \GB modules as standalone libraries can be a lightweight and even more flexible alternative to employing the full \GB machinery.

The simplest standalone example is \textsf{ExampleBit\_A\_standalone}, found in \term{ExampleBit_A/}\term{examples/}\term{ExampleBit_A} \term{_standalone.cpp}.  Here the driver program carries out breezy versions of many of the tasks performed by \GB in a full-blown scan. It first sets up some files to print log information to, and chooses which model to investigate.  It identifies which module functions in \examplebita it wants to run, then sets their internal options and connects their pipes to other functions within \examplebita, and to relevant backends. It declares and defines an additional \cpp{QUICK_FUNCTION} (cf.\ Sec.\ \ref{declaration_quick_function}) directly in the same file, to fill a missing dependency.  It then chooses what parameter values to run through the resulting pipeline, gathers the results and prints them to \term{stdout}.  Authors of standalone programs have the possibility to intervene in any of these steps, providing the necessary inputs from whatever source they like, or using the outputs directly in whichever manner they prefer.

For many models, the biggest challenge associated with using a module in standalone mode will be fulfilling dependencies on a \GB\ \cpp{Spectrum} object, as these objects are typically created exclusively by \specbit in a regular \GB scan.  For standalone purposes, each model with a \cpp{Spectrum} specialisation in \specbit is expected to also have a correspondingly stripped-down \cross{simple spectrum} defined in \term{Models/include/gambit/Models/SimpleSpectra}.  The simple spectra are essentially mass container objects, devoid of any interface to an actual spectrum generator, which can be used in standalone executables in place of a true \GB\ \cpp{Spectrum}.  The \GB\ \cpp{Spectrum} class and its simple spectrum variants are discussed in more detail in Ref.\ \cite{SDPBit}.

\subsection{Releases}
\label{release}

\GB releases are assigned version numbers of the form \term{major.minor.revision}.  Each version is available from the \GB webpage: \href{http://gambit.hepforge.org}{http://gambit.hepforge.org}.  The code can be downloaded  either directly as a tarball, or accessed through a \term{git} repository, newly forked from the development branch at each minor version update.

As a convenience, for each release we also provide downloadable tarballs of each module, bundled with the minimal number of \GB components required to use it in standalone mode.  For physics modules, these components are the models, backend interfaces, logs and all other utilities except printers.  The required components for \scannerbit are the printers, logs and a smaller subset of the utilities.

\subsection{Support}

Data used in \GB observable and likelihood functions are generally available within regular releases of \GB or relevant backends.  If in any future cases this is not possible for some reason, we will make them available from the main \GB webpage.\footnote{\href{http://gambit.hepforge.org}{http://gambit.hepforge.org}}  Output samples from scans discussed in \GB results papers (such as Refs.\ \cite{CMSSM,MSSM,SSDM}) are also available from the main \GB webpage.

General support information and relevant links are collected in the Support section of the \GB webpage.\footnote{\href{http://gambit.hepforge.org/support}{http://gambit.hepforge.org/support}}  This includes the doxygen documentation for the latest \GB release, a known issues page, and an FAQ dealing mostly with common configuration, compilation and backend questions.

\GB will be supported with regular version updates and revisions.  In general bug fixes will be applied in \term{revision} increments, and new features mostly in \term{minor} version increments.  New \term{major} version increments will be reserved for substantial new features.  After releasing a new \term{major} version, we will continue to support the last \term{minor} version of the superseded \term{major} version with bug fix updates (typically backported from the new \term{major}).

We welcome and encourage bug reports on \GB.  These should be submitted via the \textsf{TRAC} ticket system on the \GB webpage.\footnote{\href{http://gambit.hepforge.org/trac/report}{http://gambit.hepforge.org/trac/report}}  To prevent spam, bug reporters will need to first sign up for an account with \textsf{HEPforge}.\footnote{\href{http://www.hepforge.org/register}{https://www.hepforge.org/register}}

We also welcome enquiries from authors of existing or future backend codes about \GB compatibility; we are willing to work with you to help optimise interoperability of your code with \GB.

Users are also very welcome to suggest contributed code for release in a future version of \GB, particularly new models, observables, likelihood functions, printers, scanners and backend interfaces.  These suggestions will undergo a careful code review before being integrated into the main codebase.  Submitters are expected to pay attention to the coding style of adjacent routines.

\section{Summary}
\label{summary}
\GB is a powerful, general, flexible and extensible tool for phenomenological and statistical analysis of particle theories Beyond the Standard Model.  It includes modules specialised for spectrum and decay calculations, collider, flavour, DM and precision physics, a hierarchical model database of popular BSM theories, flexible interfaces to many of the most popular existing phenomenology codes, extensive statistical and parameter scanning options, and an automatic system for connecting different calculations to their required inputs, outputs and models.  Here we have outlined the main features of the \GB package itself; accompanying papers lay out the details of the individual modules \cite{ColliderBit,FlavBit,DarkBit,SDPBit,ScannerBit} and present first BSM results \cite{CMSSM,MSSM,SSDM}.  The package is fully open source, and can be downloaded from \href{http://gambit.hepforge.org}{gambit.hepforge.org}.

\begin{acknowledgements}
\gambitacknosplus
\end{acknowledgements}


\appendix

\section{Quick start guide}
\label{quickstart}
To configure and build \GB on a machine with $n$ logical cores, retrieve the git repository or the tarball and unpack it, then
\begin{lstterm}
cd gambit
mkdir build
cd build
cmake ..
make -j@\metavar{n}@ scanners
cmake ..
make -j@\metavar{n}@ gambit
\end{lstterm}
To build all backends supported for automatic download:
\begin{lstterm}
make -j@\metavar{n}@ backends
\end{lstterm}
You can also build individual backends with
\begin{lstterm}
make -j@\metavar{n}@ @\metavar{backend\_name}@
\end{lstterm}
and clean them with
\begin{lstterm}
make clean-@\metavar{backend\_name}@.
\end{lstterm}
To see which backends and scanners have been installed correctly, do
\begin{lstterm}
gambit backends
\end{lstterm}
and
\begin{lstterm}
gambit scanners
\end{lstterm}
To run gambit using the included example \YAML files \term{spartan.yaml} or \term{MSSM7.yaml}, do
\begin{lstterm}
gambit -f yaml_files/spartan.yaml
gambit -f yaml_files/MSSM7.yaml
\end{lstterm}
To make a standalone example using any one of the modules:
\begin{lstterm}
make @\metavar{module\_name}@_standalone
\end{lstterm}
and run the resulting executable \metavar{module\_name} \term{\_standalone}.

For more details on the configuration and build options, please see Secs.~\ref{cmake_backends_scanners}--\ref{config_options}.

\section{Supported compilers and library dependencies}
\label{dependencies}
\GB builds and runs under Linux and Mac~OS~X; the architecture is automatically detected by the build system.

\GB is written in \Cpp and requires a compiler that supports a minimal subset of the ISO \Cppeleven standard. For compiling \Fortran backends, a \Fortran compiler is also required. \GB supports GNU\footnote{\url{https://gcc.gnu.org/}} and Intel\footnote{\url{https://software.intel.com/en-us/intel-compilers}} \plainC/\Cpp and \Fortran compilers. The \textsf{Clang}\footnote{\url{http://clang.llvm.org/}} \plainC/\Cpp compiler is not supported due to its lack of historical support for \textsf{OpenMP}.  When newer versions supporting \textsf{OpenMP} become the default in OS X, we will add support for \textsf{clang} in \GB.

The following prerequisite libraries and packages must be installed to configure and to build \GB:
\begin{itemize}
  \item \textsf{gcc}/\textsf{gfortran 4.7.1} or greater, or \textsf{icc}/\textsf{ifort 12.1.0} or greater
  \item \textsf{Python 2.7} or greater (\textsf{Python 3} is not supported)
  \item \python modules \textsf{yaml}, \textsf{os}, \textsf{re}, \textsf{datetime}, \textsf{sys}, \textsf{getopt}, \textsf{shutil} and \textsf{itertools}.  Also \textsf{h5py} if using the \textsf{hdf5} printer with \pippi.
	\item \textsf{Boost}\footnote{\url{http://www.boost.org/}} \textsf{1.41} or greater
	\item GNU Scientific Library (\textsf{GSL})\footnote{\url{http://www.gnu.org/software/gsl/}} \textsf{1.10} or greater
\end{itemize}
Two additional linear algebra libraries are also currently required, but will become optional in future releases (where \flexiblesusy will become a full backend, rather than shipping in the \GB\ \term{contrib} directory):
\begin{itemize}
  \item \textsf{Eigen}\footnote{\url{http://eigen.tuxfamily.org}} \textsf{3.1.0} or greater: required only if using \flexiblesusy from \specbit or \gmtwocalc from \precisionbit
	\item \textsf{LAPACK}\footnote{\url{http://www.netlib.org/lapack/}}: required only if using \flexiblesusy from \specbit or \textsf{MultiNest} from \scannerbit
\end{itemize}
The following are optional libraries and packages:
\begin{itemize}
	\item \mpi: required only if parallelised sampling is desired
  \item \textsf{axel}\footnote{\url{http://axel.alioth.debian.org/}}: if available, will be used to speed up downloads of backends and scanners wherever possible
  \item \textsf{graphviz}\footnote{\url{http://www.graphviz.org/}}: required only if model hierarchy and dependency tree plots are desired
	\item \textsf{HDF5}\footnote{\url{https://www.hdfgroup.org/HDF5/}}: required only if using the \textsf{hdf5} printer
	\item \textsf{ROOT}\footnote{\url{https://root.cern.ch/}} \textsf{5.*}: required if using \textsf{Delphes} from \colliderbit, or \textsf{GreAT} from \scannerbit
\end{itemize}
If any optional package is missing, the build system automatically \term{-Ditch}es the corresponding component or feature relying on the missing package.

Users should note that whilst \GB itself compiles and runs with a wide range of compiler versions, some backend and scanner codes are not compatible with certain newer compilers.  A continually-evolving list of compiler versions tested to date with different backends can be found at \url{http://gambit.hepforge.org/compilers}.  Whilst we obviously cannot assume responsibility for the portability of codes maintained by other members of the community, we are actively working with the authors of the different codes to help improve this situation.

\section{Standard Model definitions}
\label{SMdefs}

The parameters of the \doublecrosssf{StandardModel\_SLHA2}{SM_SLHA2} model are:
\begin{description}
\item[{\footnotesize\tt CKM\_A}]($A$): Wolfenstein parameter defined in \MSbar scheme at scale $m_Z$. Converted into $V_{\rm CKM}$ entries using the 9th-order expansions of Ref.\ \cite{CKMFitter}.
\item[{\footnotesize\tt CKM\_etabar}]($\bar{\eta}$): Wolfenstein parameter defined in \MSbar scheme at scale $m_Z$. Converted into $V_{\rm CKM}$ entries using the 9th-order expansions of Ref.\ \cite{CKMFitter}.
\item[{\footnotesize\tt CKM\_lambda}]($\lambda$): Wolfenstein parameter defined in \MSbar scheme at scale $m_Z$. Converted into $V_{\rm CKM}$ entries using the 9th-order expansions of Ref.\ \cite{CKMFitter}.
\item[{\footnotesize\tt CKM\_rhobar}]($\bar{\rho}$): Wolfenstein parameter defined in \MSbar scheme at scale $m_Z$. Converted into $V_{\rm CKM}$ entries using the 9th-order expansions of Ref.\ \cite{CKMFitter}.
\item[{\footnotesize\tt GF}]($G_\mathrm{F}$):  Fermi coupling, in units of GeV$^{-2}$.
\item[{\footnotesize\tt alpha1}]($\alpha_1$):  First Majorana CP-violating phase of the PMNS matrix, in radians.
\item[{\footnotesize\tt alpha2}]($\alpha_2$):  Second Majorana CP-violating phase of the PMNS matrix, in radians.
\item[{\footnotesize\tt alphaS}]($\alpha_\mathrm{s}(m_Z)^{\MSBar}$):  Strong coupling in \MSbar scheme at scale $m_Z$.
\item[{\footnotesize\tt alphainv}]($\alpha^{-1}_{\mathrm{EM}}(m_Z)^{\MSBar}$):  Inverse electromagetic coupling in 5-flavour \MSbar scheme at scale $m_Z$.
\item[{\footnotesize\tt delta13}]($\delta_{13}$):  Majorana CP-violating phase of the PMNS matrix, in radians.
\item[{\footnotesize\tt mBmB}]($m_b(m_b)^{\MSBar}$): \MSbar mass of the $b$ quark at scale $m_b$, in GeV.
\item[{\footnotesize\tt mCmC}]($m_c(m_c)^{\MSBar}$): \MSbar mass of the $c$ quark at scale $m_c$, in GeV.
\item[{\footnotesize\tt mD}]($m_d(2\,\mathrm{GeV})^{\MSBar}$): \MSbar mass of the $d$ quark at scale of 2\,GeV, in GeV.
\item[{\footnotesize\tt mE}]($m_e$):  Pole mass of the electron, in GeV.
\item[{\footnotesize\tt mMu}]($m_\mu$): Pole mass of the muon, in GeV.
\item[{\footnotesize\tt mNu1}]($m_{\tilde{\nu}_1}$): Pole mass of first left-handed neutrino mass eigenstate, in GeV.
\item[{\footnotesize\tt mNu2}]($m_{\tilde{\nu}_2}$): Pole mass of second left-handed neutrino mass eigenstate, in GeV.
\item[{\footnotesize\tt mNu3}]($m_{\tilde{\nu}_3}$): Pole mass of third left-handed neutrino mass eigenstate, in GeV.
\item[{\footnotesize\tt mS}]($m_s(2\,\mathrm{GeV})^{\MSBar}$):  \MSbar mass of the $s$ quark at scale of 2\,GeV, in GeV.
\item[{\footnotesize\tt mT}]($m_t$):  Pole mass of the $t$ quark, in GeV.
\item[{\footnotesize\tt mTau}]($m_\tau$):  Pole mass of the $\tau$ lepton, in GeV.
\item[{\footnotesize\tt mU}]($m_u(2\,\mathrm{GeV})^{\MSBar}$):  \MSbar mass of the $u$ quark at scale of 2\,GeV, in GeV.
\item[{\footnotesize\tt mZ}]($m_Z$):  Pole mass of the $Z$ boson, in GeV.
\item[{\footnotesize\tt theta12}]($\theta_{12}$):  Solar neutrino mixing angle of the PMNS matrix, in radians.
\item[{\footnotesize\tt theta23}]($\theta_{23}$):  Atmospheric neutrino mixing angle of the PMNS matrix, in radians.
\item[{\footnotesize\tt theta12}]($\theta_{13}$):  Reactor neutrino mixing angle of the PMNS matrix, in radians.
\end{description}

\startglossary

\gitem{backend}\input{"glossary/backend.glossentry"}
\gitem{backend convenience function}\input{"glossary/backend_convenience_function.glossentry"}
\gitem{backend function}\input{"glossary/backend_function.glossentry"}
\gitem{backend initialisation function}\input{"glossary/backend_initialisation_function.glossentry"}
\gitem{backend requirement}\input{"glossary/backend_requirement.glossentry"}
\gitem{backend variable}\input{"glossary/backend_variable.glossentry"}
\gsfitem{BOSS}\input{"glossary/BOSS.glossentry"}
\newcommand{\seecompdatabase}{see Sec.\ \ref{component_databases}}
\gitem{capability}\input{"glossary/capability.glossentry"}
\gitem{child model}\input{"glossary/child_model.glossentry"}
\gitem{conditional dependency}\input{"glossary/conditional_dependency.glossentry"}
\gitem{context integer}\input{"glossary/context_integer.glossentry"}
\gitem{dependency}\input{"glossary/dependency.glossentry"}
\gitem{dependency resolution}\input{"glossary/dependency_resolution.glossentry"}
\gitem{dependency resolver}\input{"glossary/dependency_resolver.glossentry"}
\newcommand{\deptreefig}{Fig.\ \ref{fig::deptree}}
\gitem{dependency tree}\input{"glossary/dependency_tree.glossentry"}
\gitem{friend model}\input{"glossary/friend_model.glossentry"}
\gitem{frontend}\input{"glossary/frontend.glossentry"}
\gitem{frontend header}\input{"glossary/frontend_header.glossentry"}
\gitem{harvester script}\input{"glossary/harvester_script.glossentry"}
\gitem{likelihood container}\input{"glossary/likelihood_container.glossentry"}
\gitem{loop manager}\input{"glossary/loop_manager.glossentry"}
\gitem{model}\input{"glossary/model.glossentry"}
\gitem{model group}\input{"glossary/model_group.glossentry"}
\gitem{module}\input{"glossary/module.glossentry"}
\gitem{module function}\input{"glossary/module_function.glossentry"}
\newcommand{\inifilesec}{Sec. \ref{printer_setup}}
\gitem{printer}\input{"glossary/printer.glossentry"}
\gitem{purpose}\input{"glossary/purpose.glossentry"}
\gitem{nested module function}\input{"glossary/nested_module_function.glossentry"}
\gitem{parent model}\input{"glossary/parent_model.glossentry"}
\gitem{particle database}\input{"glossary/particle_database.glossentry"}
\gitem{physics module}\input{"glossary/physics_module.glossentry"}
\gitem{pipe}\input{"glossary/pipe.glossentry"}
\gitem{quantity}\input{"glossary/quantity.glossentry"}
\gitem{rollcall header}\input{"glossary/rollcall_header.glossentry"}
\gitem{rule}\input{"glossary/rule.glossentry"}
\gitem{safe version}\input{"glossary/safe_version.glossentry"}
\gitem{scanner plugin}\input{"glossary/scanner_plugin.glossentry"}
\gitem{simple spectrum}\input{"glossary/simple_spectrum.glossentry"}
\gitem{test function plugin}\input{"glossary/test_function_plugin.glossentry"}
\gitem{type}\input{"glossary/type.glossentry"}

\finishglossary

\bibliography{R1}

\onecolumn
\section*{}
\vspace{5cm}
\centering\Huge \hypertarget{addendum}{Addendum for \GB \textsf{1.1}}



\section*{Supplementary material}

# Addendum for GAMBIT 1.1

# GAMBIT: The Global and Modular Beyond-the-Standard-Model Inference Tool

## Addendum for GAMBIT 1.1: Mathematica backends, SUSYHD interface and updated likelihoods

The GAMBIT Collaboration: Peter Athron[1,2], Csaba Balazs[1,2], Torsten Bringmann[3], Andy Buckley[4], Marcin Chrząszcz[5,6], Jan Conrad[7,8], Jonathan M. Cornell[9], Lars A. Dal[3], Hugh Dickinson[10], Joakim Edsjö[7,8], Ben Farmer[7,8,a], Tomás E. Gonzalo[3], Paul Jackson[11,2], Abram Krislock[3], Anders Kvellestad[12,b], Johan Lundberg[7,8], James McKay[13], Farvah Mahmoudi[14,15,*], Gregory D. Martinez[16], Antje Putze[17], Are Raklev[3], Joachim Ripken[18], Christopher Rogan[19], Aldo Saavedra[20,2], Christopher Savage[12], Pat Scott[13,c], Seon-Hee Seo[21], Nicola Serra[5], Christoph Weniger[22,d], Martin White[11,2], Sebastian Wild[23]

[1]School of Physics and Astronomy, Monash University, Melbourne, VIC 3800, Australia
[2]Australian Research Council Centre of Excellence for Particle Physics at the Tera-scale
[3]Department of Physics, University of Oslo, N-0316 Oslo, Norway
[4]SUPA, School of Physics and Astronomy, University of Glasgow, Glasgow, G12 8QQ, UK
[5]Physik-Institut, Universität Zürich, Winterthurerstrasse 190, 8057 Zürich, Switzerland
[6]H. Niewodniczański Institute of Nuclear Physics, Polish Academy of Sciences, 31-342 Kraków, Poland
[7]Oskar Klein Centre for Cosmoparticle Physics, AlbaNova University Centre, SE-10691 Stockholm, Sweden
[8]Department of Physics, Stockholm University, SE-10691 Stockholm, Sweden
[9]Department of Physics, McGill University, 3600 rue University, Montréal, Québec H3A 2T8, Canada
[10]Minnesota Institute for Astrophysics, University of Minnesota, Minneapolis, MN 55455, USA
[11]Department of Physics, University of Adelaide, Adelaide, SA 5005, Australia
[12]NORDITA, Roslagstullsbacken 23, SE-10691 Stockholm, Sweden
[13]Department of Physics, Imperial College London, Blackett Laboratory, Prince Consort Road, London SW7 2AZ, UK
[14]Univ Lyon, Univ Lyon 1, ENS de Lyon, CNRS, Centre de Recherche Astrophysique de Lyon UMR5574, F-69230 Saint-Genis-Laval, France
[15]Theoretical Physics Department, CERN, CH-1211 Geneva 23, Switzerland
[16]Physics and Astronomy Department, University of California, Los Angeles, CA 90095, USA
[17]LAPTh, Université de Savoie, CNRS, 9 chemin de Bellevue B.P.110, F-74941 Annecy-le-Vieux, France
[18]Max Planck Institute for Solar System Research, Justus-von-Liebig-Weg 3, D-37077 Göttingen, Germany
[19]Department of Physics, Harvard University, Cambridge, MA 02138, USA
[20]Centre for Translational Data Science, Faculty of Engineering and Information Technologies, School of Physics, The University of Sydney, NSW 2006, Australia
[21]Department of Physics and Astronomy, Seoul National University, 1 Gwanak-ro, Gwanak-gu, Seoul 08826, Korea
[22]GRAPPA, Institute of Physics, University of Amsterdam, Science Park 904, 1098 XH Amsterdam, Netherlands
[23]DESY, Notkestraße 85, D-22607 Hamburg, Germany

**Abstract** In Ref. [1] we introduced the global-fitting framework GAMBIT. In this addendum, we describe a new minor version increment of this package. GAMBIT 1.1 includes full support for Mathematica backends, which we describe in some detail here. As an example, we backend SUSYHD [2], which calculates the mass of the Higgs boson in the MSSM from effective field theory. We also describe updated likelihoods in PrecisionBit and DarkBit, and updated decay data included in DecayBit.

[a]benjamin.farmer@fysik.su.se
[b]anders.kvellestad@nordita.org
[c]p.scott@imperial.ac.uk
[d]c.weniger@uva.nl
[*]Also Institut Universitaire de France, 103 boulevard Saint-Michel, 75005 Paris, France.

# 1 Using Mathematica backends in GAMBIT

For decades, Wolfram Mathematica[1] has been the symbolic computing framework of choice for many physicists. GAMBIT users can now access both public and private Mathematica packages from within GAMBIT modules. This new feature provides a seamless interface to Mathematica, indistinguishable from the existing GAMBIT backend interfaces to codes written in C, C++ and Fortran.

Mathematica is proprietary software. It is the responsibility of the user to acquire a license in order to use the Mathematica framework. The number of Mathematica Kernel instances and subprocesses that can be launched per Mathematica license is limited, and can be found by evaluating the variables `$MaxLicenseProcesses` and `$MaxLicenseSubprocesses` (typically 8 and 16 respectively for a standard Mathematica license).

This addendum describes our implementation of the Mathematica interface. Sec. 1.1 describes the mechanism used to communicate with the Mathematica Kernel from GAMBIT. In Secs. 1.2–1.4 we explain the different aspects of the implementation, consisting of configuration (1.2), registration of backend functions and variables (1.3), and updates to the GAMBIT diagnostics to accommodate Mathematica backends (1.4). In Sec. 1.5, we give a realistic example of a Mathematica backend interfaced to GAMBIT, namely SUSYHD [2], which calculates the mass of the Higgs boson using methods from effective field theory (EFT).

## 1.1 The Wolfram Symbolic Transfer Protocol

The Wolfram Symbolic Transfer Protocol (WSTP[2]), is the communication standard used by Mathematica to communicate with external programs. WSTP provides a two-way communication system, allowing the user to both call external codes from within Mathematica, and to call Mathematica routines from external programs written in many languages (C, C++, Java, etc). GAMBIT 1.1 uses the communication pathway between Mathematica and C++. This allows GAMBIT users to directly call Mathematica functions and variables from GAMBIT modules, in the same way as functions and variables are accessed from C, C++ and Fortran backends in GAMBIT 1.0.

The GAMBIT-Mathematica interface uses the WSTP messaging functions to launch and establish a link to a Mathematica Kernel session, and to send and receive

information from that Kernel. Almost all the messaging functions that we employ use a WSTP link object, of type `WSLINK`, as a handle for communicating with the Kernel. One retrieves such a handle by opening a link to the Kernel with the messaging functions `WSOpenArgcArgv` or `WSOpenString`. The most important messaging functions for our purposes are

```
WSPutFunction(WSlink, function_name, n_args);
WSPutString(WSlink, string_name);
WSPutSymbol(WSlink, symbol_name);
WSPutInteger(WSlink, integer_number);
WSPutReal32(WSlink, float_number);
WSPutReal64(WSlink, double_precision_number);

WSGetString(WSlink, &string_variable);
WSGetInteger(WSlink, &integer_variable);
WSGetReal32(WSlink, &float_variable);
WSGetReal64(WSlink, &double_precision_variable);
```

where `WSlink` is an object of type `WSLINK`. The first argument in each of these must be a C++ object with the appropriate type, e.g. for symbols and functions this is a C++ string, and the argument *n_args* corresponds to the number of arguments of the Mathematica function *function_name*.

To interpret data received from the `WSGet` functions, one must also receive the packets sent by the Kernel. We use `WSNextPacket(WSlink)` to find the next packet head, `WSNewPacket(WSlink)` to skip the rest of the current packet, and `WSError(WSlink)` to check for errors during packet reception.

All these messaging functions live in the header `wstp.h`, native to the Mathematica installation, and will be used by GAMBIT for different purposes, as described below.

## 1.2 Preparing and configuring Mathematica backends

### 1.2.1 Configuration

At `cmake` time, GAMBIT searches for an installation of Mathematica on the user's system, locating the installation directory and defining several `cmake` variables, including the paths to the Mathematica header files and executables. The GAMBIT `cmake` system also searches for the `libuuid` library, required by WSTP. It writes a new header file containing a series of variables relating to the system's Mathematica configuration, including the preprocessor macro `HAVE_MATHEMATICA`, which is later used to enable or disable all code associated with Mathematica backends, according to whether or not the user has Mathematica installed. In order to manually switch off Mathematica support (from GAMBIT 1.1.2 onwards), one may run the `cmake` command as

```
cmake -Ditch=Mathematica ..
```

Mathematica backends are obtained and configured in a similar manner to other backends, by adding relevant entries to `cmake/backends.cmake` and `config/backend_locations.yaml.default`, and then running `make backend_name`. Mathematica backends require neither a configuration nor a build step, only a download link and an installation directory.

### 1.2.2 Frontend header

As described in Sec. 4 of Ref. [1], GAMBIT loads backends at runtime as dynamic libraries, regardless of whether they are written in C, C++ or Fortran, using the POSIX-standard `dl` library. The macro `LOAD_LIBRARY`, used in the corresponding frontend header in GAMBIT, activates the backend library and loads its symbols.

GAMBIT communicates with Mathematica backends in a fundamentally different way to backends written in compiled languages. To redirect compilation flow in the GAMBIT backend system to use the preprocessor directives relevant for Mathematica backends, we define the new macro `BACKENDLANG`. This specifies the language of the backend in question, and can take the values `CC`, `CXX`, `FORTRAN` or `MATHEMATICA`. For example, in GAMBIT 1.1, the frontend header for version 1.2 of a C++ backend called *backend_name* would begin with

```
#define BACKENDNAME backend_name
#define BACKENDLANG CXX
#define VERSION 1.2
#define SAFE_VERSION 1_2
LOAD_LIBRARY
```

### 1.2.3 Backend types

In analogy with the definition of new types and typedefs given for Fortran backends in Sec. 4.4 of Ref. [1], here we provide a list of new types defined in GAMBIT 1.1 for use with Mathematica backends. These have clear correspondences with equivalent types in C and C++ types, and should be adopted whenever one is working with Mathematica backends. These are:

```
MVoid
MInteger
MReal
MBool
MChar
MString
MList<TYPE>
```

where `TYPE` can take any of the other defined types as its template argument. All these types are just convenient type redefinitions of native C++ types and thus can be used in exactly the same manner.

### 1.2.4 Link to the Kernel

Prior to loading the backend library, the connection with the Mathematica Kernel must be established. For this, the macro `LOAD_LIBRARY` initializes a WSTP environment with a call to `WSInitialize`, which returns a handle to the WSTP environment of type `WSENV`. After retrieving this handle, GAMBIT opens a new WSTP connection using the function `WSOpenString`, which launches the Mathematica Kernel and finishes the intialization phase of the connection. These functions are employed in the following way:

```
WSenv = WSInitialize();
WSlink = WSOpenString(WSenv, WSTPflags, &WSerrno);
```

The variable `WSerrno` captures any error that might have occurred during the establishment of the link and `WSTPflags` is a string containing flags that point to the Kernel executable and give the order to initiate the link, e.g. `WSTPflags = "-linkname math -mathlink"` for Linux systems.

If the link is successfully established, the pointer `WSlink` is set to point to the new link between the main program and the Kernel. This handle is used in every subsequent communication with the Kernel.

### 1.2.5 Importing the package

After defining the relevant macros and backend types, loading the Mathematica environment and Kernel, the final task of the `LOAD_LIBRARY` macro is to import the backend package into the Mathematica Kernel. This is achieved by using the pointer to the WSTP link `WSlink` via the following load sequence:

```
WSPutFunction(WSlink, "Once", 1);
WSPutFunction(WSlink, "Get", 1);
WSPutString(WSlink, path);
```

which is equivalent to the call `Get[path]` within a notebook session.

Note that before we load the package, we call the Mathematica function `Once`, which makes sure that the following function will be executed only once in a Kernel session. Here *path* is a string containing the path to the .m file of the backend package; the actual value used is taken from the default GAMBIT backend location file `config/backend_locations.yaml.default` or a user-supplied override.

When loading a package in Mathematica the `Get` function will return information regarding its success. The exact return value will always be package dependent, so is difficult to interpret in general from the GAMBIT interface. Therefore we use the standard library functions

of C++ to check if the package and all required functions and variables exist. This is done via the **GAMBIT** diagnostic system (see 1.4).

### 1.3 Backend functions and variables

The backend system in **GAMBIT**, as described in Sec. 4 of Ref. [1], provides a set of macros for registering functions and variables from backends. In a nutshell, **GAMBIT** extracts a pointer to the function/variable living in the shared library of the backend, and wraps it into a functor object to be handled by the dependency resolver.

When dealing with backends written in **Mathematica**, the above procedure must be modified, as there is no shared library from which to extract the pointer to the function or variable. Instead, we must interface with the functions and variables via the **Mathematica Kernel**. The general strategy for doing this is shown in Figure 1.

#### 1.3.1 Backend functions

Backend functions in **GAMBIT** are defined with the macro `BE_FUNCTION`. This macro creates a pointer to the function in the namespace of the backend, and assigns it a name and capability within **GAMBIT** of the frontend author's choosing. In order to communicate with the **Mathematica Kernel** and provide the same sort of handle to the dependency resolver for **Mathematica** backends, we must define a wrapper function around message calls to the **Kernel**, in such a way as to have the wrapper function operate like a regular C/C++ backend function.

We achieve this by redirecting the macro `BE_FUNCTION` for backends written in **Mathematica**, according to the value of `BACKENDLANG`. The version of `BE_FUNCTION` that gets used for **Mathematica** backends therefore constructs a wrapper function around each **Mathematica** function that the user wishes to access from within **GAMBIT**. The wrapper function handles the submission and receipt of all messages to and from the **Kernel** relating to the function. This includes all function arguments, and the eventual return value of the function. The sequence of messages is:

```
WSPutFunction(WSlink, symbol_name, n);
WSPutVariable(WSlink, arg_1);
WSPutVariable(WSlink, arg_2);
...
WSPutVariable(WSlink, arg_n);
WSGetVariable(WSlink, &val);
```

where *symbol_name* is the name of the function within the **Mathematica** package, *n* is the number of arguments it accepts, *arg_1*, *arg_2*, etc are the arguments themselves,

and *val* is the return value. Here `WSPutVariable` and `WSGetVariable` are overloaded functions, which expand to different type-specific setter and getter functions intrinsic to **WSTP** (mentioned in Sec. 1.1).

There is an important subtlety regarding the names of functions in **Mathematica** backends. **Mathematica** permits non-alphanumeric characters in function names. Many of the backends that are interesting for **GAMBIT** include functions with such names. However, function names in **GAMBIT** are treated as regular strings, so there is no easy way to communicate the actual name of the function to the **Kernel**. To circumvent this issue, names of **Mathematica** backend functions should be declared to **GAMBIT** with all special characters written in terms of their "full names", e.g. $\alpha = $ `\[Alpha]`, but making sure to escape the slash "\" so that C++ can parse it. To illustrate this, we show the signature used for the function $\Delta$`MHiggs` from the backend **SUSYHD** [2], which calculates uncertainties on the Higgs mass:

```
BE_FUNCTION(DeltaMHiggs, MReal,
 (const MList<MReal>&), "\\[CapitalDelta]MHiggs",
  "SUSYHD_DeltaMHiggs")
```

where the first argument is the name of the function within **GAMBIT** with its return type and signature as the second and third arguments. The last two arguments of the `BE_FUNCTION` macro are the symbol associated to the backend function, the **Mathematica** function in our case, and the capability that the function provides.

The generic function call is modified in the case that the symbol name contains non-alphanumeric characters. The symbol name (for example `"\\[CapitalDelta]MHiggs"`) must be sent as a string to the **Kernel**, preceded by a message that transforms it into a valid **Mathematica** expression. This expression matches the string version of the name to the name of the actual backend function, including the correct non-alphanumeric characters. The sequence of messages is

```
WSPutFunction(WSlink, "Apply", 2);
WSPutFunction(WSlink, "ToExpression",1);
WSPutString(WSlink, symbol_name);
WSPutFunction(WSlink, "List", n_args);
WSPutVariable(WSlink, arg_1);
WSPutVariable(WSlink, arg_2);
...
WSPutVariable(WSlink, arg_n);
WSGetVariable(WSlink, &val);
```

which is equivalent to `WSPutFunction(WSlink, symbol_name, n_args)`, in the case that the *symbol_name* contains only alpha-numeric characters.

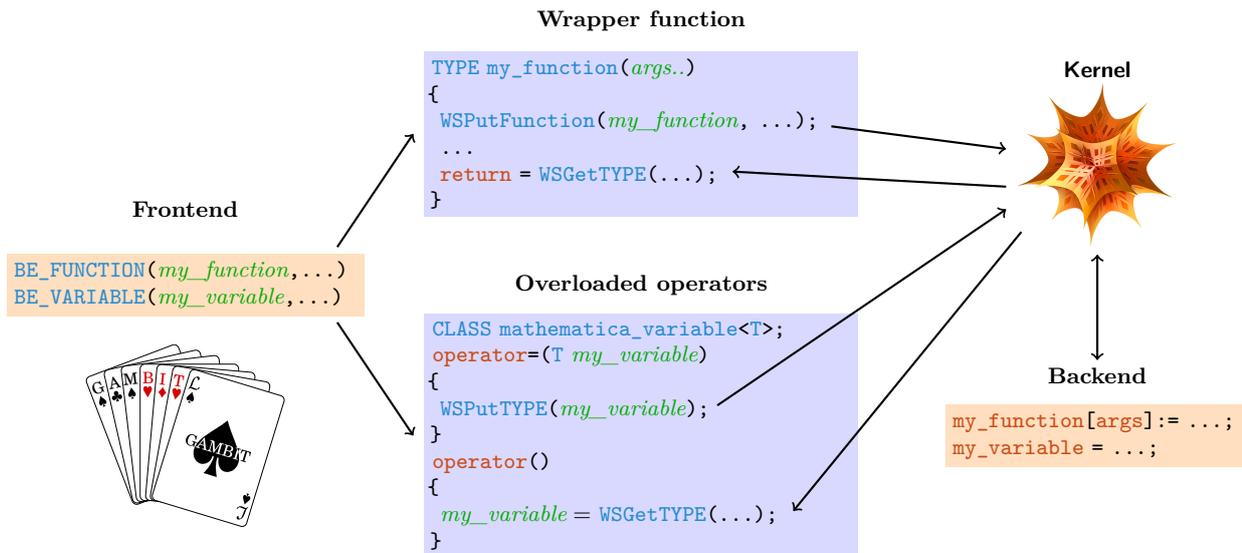

**Fig. 1:** Diagrammatic representation of the communication link between GAMBIT and the Mathematica Kernel, showing the wrappers for backend functions and the variable classes for backend variables.

### 1.3.2 Backend variables

Backend variables in GAMBIT are registered with the preprocessor macro `BE_VARIABLE`. For backends written in C, C++ or Fortran, this macro retrieves a pointer to the variable from the shared library. Through this pointer, module functions can extract and modify values of global variables in backends as needed. However, such an interface cannot be constructed if the backend is written in Mathematica. The WSTP messaging system can only access variables from the Kernel by value, so it is not possible to obtain a pointer via which GAMBIT can modify the value inside the Kernel.

To work around this limitation, the GAMBIT 1.1 backend system creates a templated class `mathematica_variable<TYPE>`, with a private member variable of type `TYPE`. The templated class acts as a wrapper for the required calls to the Mathematica Kernel. As with backend functions, the macro `BE_VARIABLE` is redefined for Mathematica backends. This macro creates an instance of the class `mathematica_variable<TYPE>`, with `TYPE` equal to the type of the variable in the Kernel (see Section 1.2.3). Like backend functions, backend variable names are sent to the Kernel as a string preceded by the `ToExpression` function, instead of simply sending the symbol name via `WSPutSymbol`. This allows access to variables with special characters in their names.

As we show schematically in Figure 1, the `mathematica_variable<TYPE>` class allows variables in the Kernel to be seamlessly read, modified and implicitly cast to variables of type `TYPE`, allowing them to be used directly in C++ expressions. These functionalities are achieved by overloading two operators that act on the `mathematica_variable<TYPE>` class. The first of these is the assignment operator `mathematica_variable&` `operator=(const TYPE &)`, enabling modification of the variables in the Kernel by regular assignment expressions in module functions. The other is the cast operator `operator TYPE const()`, which converts an object of type `mathematica_variable<TYPE>` into its member variable of type `TYPE`, allowing the variable to be used in all C++ expressions where `TYPE` would be permitted.

### 1.4 Diagnostics

GAMBIT already ships with extensive diagnostic tools, providing the user with information about the status of backends, modules, capabilities, scanners and more. We have added a few additional diagnostic checks for Mathematica backends in GAMBIT 1.1.

The first of these simply assesses the status of a Mathematica backend, as typically viewed when using the `gambit backends` diagnostic. GAMBIT 1.1 adds an additional possible status in this readout, triggered in the case that a specific backend requires Mathematica but no acceptable version of it has been found in the system. When backends are loaded at runtime, they each register in the `Backends::backendInfo()` object (described in Sec. 4.6 of Ref. [1]) whether or not they require Mathematica, in the new `BackendInfo` member variable

```
std::map<str, bool> needsMathematica;
```

At runtime, this information is checked against the actual presence (or absence) of Mathematica, and the

status <span style="color:red">Mathematica absent</span> is displayed if it is needed but not found.

GAMBIT will also inform the user of the status of each individual function and global variable in a backend. For Mathematica backends, we achieve this by calling the Mathematica function `NameQ`. This function takes the symbol name (whether a function or a variable) as a string argument, and returns a boolean result indicating if the symbol is currently present in the Kernel. The diagnostic system then assigns a numerical flag to the status of the function or variable:

```
 1: This function is available, but the backend
    version does not match your request.
 0: This function is not compatible with any
    model you are scanning.
-1: The backend that provides this function is
    missing.
-2: The backend is present, but the function is
    absent or broken.
-5: The backend requires Mathematica, but
    Mathematica is absent.
```

### 1.5 Higgs mass calculations with SUSYHD

GAMBIT 1.0 offers three options for calculating the mass of the Higgs boson: FlexibleSUSY [3], SPheno [4, 5] and FeynHiggs [6–11]. The first two are fixed order $\overline{DR}$ calculations of the Higgs pole mass. The third is an on-shell calculation that can be performed at fixed order, or can include some logarithmic corrections to give a hybrid EFT / fixed order calculation [10–12].[3] Here we provide a different approach to computing the Higgs mass, for any MSSM model[4], via the Mathematica package SUSYHD [2].

SUSYHD uses the pure EFT calculation, which resums logarithms through the matching and running procedure, using three-loop Standard Model (SM) renormalisation group equations and two-loop matching, providing a more precise result than fixed-order calculations when $M_{\rm SUSY} \gg m_t$. SUSYHD is the first pure EFT Higgs calculation in GAMBIT, and the calculation remains state-of-the-art. This therefore represents an important new option for calculating the Higgs mass. This calculation is most appropriate when $M_{\rm SUSY} \gg m_t$ because the pure EFT approach neglects terms of order $\mathcal{O}(v_0^2/M_{\rm SUSY}^2)$ (where $v_0$ is the SM Higgs vacuum expectation value), which can be important at the TeV scale and below.

---

[3] Although both FlexibleSUSY [13] and SPheno [14] also include the option of employing a hybrid $\overline{DR}$ calculation, implementing the algorithm in Ref. [13], these options are not yet included in GAMBIT. They will be added in further updates, as will the pure EFT FlexibleSUSY calculation, HSSUSY.

[4] See Sec. 5.4 of [1] for a list of MSSM models in GAMBIT.

SUSYHD includes three main functions: `MHiggs`, $\Delta$MHiggs and `SetSMparameters` with the following signatures

```
MHiggs[shortList_]
ΔMHiggs[shortList_]
SetSMparameters[QMTorgt_,Qα3MZ_]
```

The first two calculate the mass of the lightest Higgs boson and its theoretical uncertainty in the minimal supersymmetric SM (MSSM), starting from a list of $\overline{DR}$ masses at scale $M_{\rm SUSY}$. The third function sets the values of the SM parameters to be used in the calculation.

We use these functions by declaring them to GAMBIT as usual in a frontend header. This SUSYHD frontend header contains the declarations

```
BE_FUNCTION(MHiggs, MReal,
 (const MList<MReal>&), "MHiggs",
 "SUSYHD_MHiggs")
BE_FUNCTION(DeltaMHiggs, MReal,
 (const MList<MReal>&), "\\[CapitalDelta]MHiggs",
 "SUSYHD_DeltaMHiggs")
BE_FUNCTION(SetSMparameters, MVoid,
 (const MReal&, const MReal&), "SetSMparameters",
 "SUSYHD_SetSMparameters")
```

To use SUSYHD from SpecBit to compute the Higgs mass and uncertainty, we define a new module function `SHD_HiggsMass`, which provides the capability `prec_mh`. The declaration of this function in the rollcall header for SpecBit is:

```
#define FUNCTION SHD_HiggsMass
START_FUNCTION(triplet<double>)
DEPENDENCY(unimproved_MSSM_spectrum, Spectrum)
BACKEND_REQ(SUSYHD_MHiggs, (), MReal,
(const MList<MReal>&))
BACKEND_REQ(SUSYHD_DeltaMHiggs, (), MReal,
(const MList<MReal>&))
ALLOW_MODELS(MSSM63atQ, MSSM63atMGUT)
#undef FUNCTION
```

## 2 SpecBit, DecayBit and PrecisionBit updates

GAMBIT 1.1 also includes a simultaneous version update to SpecBit, DecayBit and PrecisionBit [15]. In this section we briefly discuss the changes made to the module functions and structure of the code.

PrecisionBit 1.1 features updated likelihood functions for the first and second generation quark masses, the strong and electromagnetic couplings, the anomalous magnetic moment of the muon, and $\Delta\rho$ (the departure from unity of the ratio of the Fermi couplings associated with $W$ and $Z$ bosons). DecayBit 1.1 contains updates to all SM particle widths, branching fractions and uncertainties. These updates bring the data contained in

| Capability | Function (Return Type): Brief Description | Dependencies | Backend requirements |
|---|---|---|---|
| unimproved_MSSM _spectrum | get_CMSSM_spectrum_FS(Spectrum): Renamed from get_CMSSM_spectrum. Make an MSSM spectrum object with FlexibleSUSY using CMSSM boundary conditions. | SMINPUTS | FlexibleSUSY |
| | get_MSSMatGUT_spectrum_FS(Spectrum): Renamed from get_MSSMatGUT_spectrum. Make an MSSM spectrum object with FlexibleSUSY using soft breaking masses input at the GUT scale. | SMINPUTS | FlexibleSUSY |
| | get_MSSMatQ_spectrum_FS (Spectrum): Renamed from get_MSSMatQ_spectrum. Make an MSSM spectrum object with FlexibleSUSY from soft breaking masses input at (user-defined) scale $Q$. | SMINPUTS | FlexibleSUSY |
| FH_HiggsMasses | FH_AllHiggsMasses (fh_HiggsMassObs): Renamed from FH_HiggsMasses; capability reassigned from prec_HiggsMasses. Higgs masses and mixings with theoretical uncertainties, as computed by FeynHiggs. | – | FeynHiggs |
| prec_mh | FH_HiggsMass (triplet<double>): Mass of the most SM-like neutral Higgs boson and associated uncertainty, as computed by FeynHiggs. | FH_HiggsMasses unimproved_MSSM_spectrum | – |
| | SHD_HiggsMass (triplet<double>): Mass of the most SM-like neutral Higgs boson and associated uncertainty, as computed by SUSYHD. | unimproved_MSSM_spectrum | SUSYHD |
| prec_HeavyHiggsMasses | FH_HeavyHiggsMasses (std::map<int,triplet<double>>): Masses of the three non-SM Higgs bosons and associated uncertainties in a model with two Higgs doublets, as computed by FeynHiggs. | FH_HiggsMasses unimproved_MSSM_spectrum | – |

**Table 1:** New or modified module functions of note in SpecBit 1.1. Functions that use FlexibleSUSY to generate an unimproved_MSSM_spectrum have been appropriately renamed, a new precision SM-like Higgs mass calculator using SUSYHD has been added, and the precision Higgs mass calculation with FeynHiggs has been split into separate functions for the SM-like Higgs and the other three MSSM Higgs bosons.

| Capability | Function (Return Type): Brief Description | Dependencies | Backend requirements |
|---|---|---|---|
| MSSM_spectrum | make_MSSM_precision_spectrum_H(Spectrum): Function to provide an updated MSSM spectrum with precision mass for the most SM-like Higgs boson. | unimproved_MSSM_spectrum prec_mh | – |
| | make_MSSM_precision_spectrum_H_W(Spectrum): Function to provide an updated MSSM spectrum with precision masses for the $W$ boson and the most SM-like Higgs boson. | unimproved_MSSM_spectrum prec_mw prec_mh | – |
| | make_MSSM_precision_spectrum_4H_W(Spectrum): Function to provide an updated MSSM spectrum with precision masses for the $W$ boson and all four Higgs bosons. | unimproved_MSSM_spectrum prec_mw prec_mh prec_HeavyHiggsMasses | – |

**Table 2:** New or modified module functions of note in PrecisionBit 1.1. Precision spectra can now be computed using precision values for 1) the $W$ mass only (as in PrecisionBit 1.0; not shown), 2) the SM-like Higgs mass only, 3) the $W$ mass and the SM-like Higgs, or 4) the $W$ mass and the masses of all four MSSM Higgs bosons. In the latter case, the precision value for the SM-like Higgs mass can be taken from a different source to the precision masses of the other three Higgs bosons.

PrecisionBit and DecayBit up to date with the central values and uncertainties of the 2017 Particle Data Book [16].

The module functions of SpecBit and PrecisionBit have undergone some rearrangement and expansion in order to better accommodate alternative spectrum generators and precision calculators. The updated functions are detailed in Tables 1 and 2.

Functions get_CMSSM_spectrum, get_MSSMatQ_spectrum and get_MSSMatGUT_spectrum of SpecBit 1.0 have been renamed to get_CMSSM_spectrum_FS, get_MSSMatQ_spectrum_FS and get_MSSMatMGUT_spectrum_FS in SpecBit 1.1. This is to explicitly reflect the fact that they make use of FlexibleSUSY for spectrum generation, in contrast to other equivalent functions that use e.g. SPheno (or, in the future, other spectrum generators).

As described in Sec. 4.2.3 of Ref. [15], the Spectrum object computed by SpecBit using a spectrum generator (FlexibleSUSY, SPheno, etc) can be supplemented with more precise calculations of the masses of the $W$ and Higgs bosons. Initial spectrum generation and subsequent precision mass calculations all take place in SpecBit, but PrecisionBit is responsible for combining them to make the final 'precision Spectrum' object.

The computation of the precision mass of the most SM-like Higgs boson in **SpecBit** is now separated from the computation of the precision masses of the three other MSSM Higgs bosons (Table 1). For the SM-like Higgs (capability `prec_mh`), we now provide two module functions: `FH_HiggsMass` and `SHD_HiggsMass`, which provide the prediction from **FeynHiggs** and SUSYHD, respectively. The precision calculation of the masses of the three other MSSM Higgs bosons (capability `prec_HeavyHiggsMasses`) is now provided by the function `FH_HeavyHiggsMasses`, which makes use of results from **FeynHiggs**.

The precision spectrum functions in **PrecisionBit** now allow the mass of the SM-like Higgs boson to be improved independently of the masses of the other three MSSM Higgs bosons (Table 2). This makes it possible to use e.g. SUSYHD for the mass of the SM-like Higgs boson, but **FeynHiggs**, **FlexibleSUSY** or **SPheno** for the other Higgses. Each of the precision spectrum functions in **PrecisionBit** now also recognises a YAML option `allow_fallback_to_unimproved_masses`, which specifies the preferred behaviour when one or more of the requested precision mass calculations returns an invalid value (zero, a negative mass or not-a-number). With this option set to `false` (the default), the parameter combination will be flagged as invalid; if it is set to `true`, failures will instead cause the original masses from the spectrum generator to be retained for the parameter point under investigation. This can be useful for e.g. defaulting to the Higgs mass from **FlexibleSUSY** or **SPheno** for models with Lagrangian mass parameters below the top mass, where the expansion used in SUSYHD breaks down and the backend returns zero for the Higgs mass.

GAMBIT 1.1 also marks the removal of any explicit module function for indicating the PDG code of the most SM-like Higgs boson. This functionality is now provided instead by the internal function `SMlike_higgs_PDG_code`, which takes a high-scale `SubSpectrum` object as input, and can be called from any module function by including the header `gambit/Elements/smlike_higgs.hpp`.

## 3 Updated **DarkBit** likelihoods

DarkBit 1.1 includes an interface to **DDCalc** 1.1, along with corresponding likelihood functions for the XENON1T [17] and 2017 PICO-60 [18] experiments. GAMBIT 1.1 also offers an additional option for turning on one-loop corrections to direct detection cross-sections computed with **DarkSUSY** (`loop`; enabled by default), and switches the default behaviour of the **micrOMEGAs** option `internal_decays` to `false`, causing decay widths and branching fractions to be passed from **DecayBit** by default. More details can be found in the **DarkBit** paper [19].



\end{document}